\documentclass[aps,prb,twocolumn,a4paper,10pt]{revtex4-2}
\usepackage{amsmath}
\usepackage{amsfonts}
\usepackage{amssymb}
\usepackage{bm}
\usepackage[format=hang,indention=-1.25cm,%
justification=RaggedRight,labelformat=parens]{caption}
\usepackage{dcolumn}
\usepackage{enumitem}
\usepackage{graphicx}
\usepackage{hyperref}
\hypersetup{
	pdfstartview=FitH,
	colorlinks=true,
	linkcolor=blue,
	filecolor=cyan,      
	urlcolor=magenta,
	citecolor=blue,
	anchorcolor=black,
}
\usepackage{mathtools}
\usepackage{newtxtext} 
\usepackage[libertine]{newtxmath}
\usepackage{physics}
\usepackage{siunitx}
\usepackage[dvipsnames]{xcolor}
\newcommand{\beq}{\begin{equation}}
\newcommand{\eeq}{\end{equation}}
\newcommand{\bse}{\begin{subequations}}
\newcommand{\ese}{\end{subequations}}
\newcommand{\bea}{\begin{eqnarray}}
\newcommand{\eea}{\end{eqnarray}}
\newcommand{\bem}{\begin{displaymath}}
\newcommand{\eem}{\end{displaymath}}

\newcommand{\nn}{\nonumber}
\newcommand{\bs}{\boldsymbol}
\newcommand{\bc}{\begin{center}}
\newcommand{\ec}{\end{center}}
\newcommand{\bmk}{\bm{k}}
\newcommand{\bmr}{\bm{r}}
\newcommand{\bmq}{\bm{q}}
\newcommand{\tcr}{\textcolor{Red}}

\newcommand{\tcog}{\textcolor{Red}} 

\graphicspath{{inkfigs/}}
\begin{document}
	\title{A primer on twistronics: A massless Dirac fermion's journey to moir\'{e} patterns and flat bands in twisted bilayer graphene}
	\author{Deepanshu Aggarwal, Rohit Narula, and Sankalpa Ghosh}
	\address{Department of Physics, IIT Delhi, Hauz Khas, New Delhi, India}
	
	
	\begin{abstract}
		The recent discovery of superconductivity in magic-angle twisted bilayer graphene has sparked a renewed interest in the strongly-correlated physics of $sp^2$ carbons, in stark contrast to preliminary investigations which were dominated by the one-body physics of the massless Dirac fermions. We thus provide a self-contained, theoretical perspective of the journey of graphene from its single-particle physics-dominated regime to the strongly-correlated physics of the flat bands. Beginning from the origin of the Dirac points in condensed matter systems, we discuss the effect of the superlattice on the Fermi velocity and Van Hove singularities in graphene and how it leads naturally to investigations of the moir\'{e} pattern in van der Waals heterostructures exemplified by graphene-hexagonal boron-nitride and twisted bilayer graphene. Subsequently, we illuminate the origin of flat bands in twisted bilayer graphene at the magic angles by elaborating on a broad range of prominent theoretical works in a pedagogical way while linking them to available experimental support, where appropriate. We conclude by providing a list of topics in the study of the electronic properties of twisted bilayer graphene not covered by this review but may readily be approached with the help of this primer.
	\end{abstract}
	
	\maketitle
	
	\tableofcontents
	
	\section{Introduction}
	The moir\'{e} pattern is a familiar visual effect which manifests when two or more periodic patterns are overlaid, with either a slight rotation or difference in periodicity between them. For more than a century it has been observed in domains as diverse as geometry, optics, wave motion, stress analysis, crystallography, and the psychology of human perception \cite{Amidror,AmidrorRoger2010}. Indeed the recent observation of strongly correlated phenomena in the moir\'{e} pattern-exhibiting twisted bilayer graphene (TBLG), in particular the alternating superconducting and insulating phases  \cite{Cao2018one, Cao2018two, Miller2018}, has been deemed a blockbuster discovery by the popular science press \cite{Freedman2019} and heralding the field of twistronics \cite{Charlie2019} among layered materials and their heterostructures \cite{Ajayan2016}. Theoretical and experimental progress in this direction has already received the attention of several prominent reviews which discuss, $ \mathit{e.g.,} $ the various strongly correlated phases that can be realised \cite{Andrei2020} in magic angle TBLG (MATBLG), the different computational modelling approaches to the various twisted heterostructures \cite{Carr2020}, the emulation of the magic-angle twisting effect by other systems such as cold-atoms, trapped ions and metamaterials \cite{Fu2020}, fabrication techniques and twist-angle dependent properties in TBLG \cite{Nimbalkar2020}, and the experimental advances in twisted graphene moir\'{e} superlattices \cite{Chu2020} and the moir\'{e} pattern in general two-dimensional materials \cite{He_Fang2021}.
	
	The evolution of the weakly correlated phases of the massless Dirac fermions with the conical ultra-relativistic mimicking band structure in pristine monolayer graphene on one hand, to the strongly correlated phases arising from the flat bands with a nearly infinite effective mass in MATBLG on the other, is made possible due to the non-trivial tunability of the band structure of the former, so-called Dirac matter \cite{DiracMatter} subject to periodic potentials created by different superlattices exemplified by the latter. The purpose of this primer is to contribute a self-contained theoretical analysis of the effect of superlattices on the band structure of the Dirac fermions which eventually lead to the flat band physics in magic angle twisted graphene bilayers. To this purpose we have selected a number of interconnected topics, analysing them in detail to provide an origin's story of the flat bands in TBLG at the magic angle. Accordingly, this review is organised as follows.
	
	In section \ref{WeaktoStrong}, we first establish the equivalence 
	between the lattice fermions of gauge theory and the familiar electronic band structure in a condensed matter system. We introduce pristine single-layer graphene (SLG) in the section \ref{Honeycomb} whose massless Dirac fermions act like charge carriers and describe why such charge carriers in SLG are deemed to be \emph{weakly} correlated. We discuss the van Hove singularity (VHS) in SLG in subsection \ref{VHS}, while outlining its significance in affecting the electronic correlations. In \ref{SLZ}, we show how subjecting such massless Dirac fermions to one, and two-dimensional superlattice potentials one obtains new Dirac points which alter the structure of the VHS, the Fermi velocity, and hence the extent of correlation. The foregoing discussion brings us to graphene-based van der Waals heterostructures in section \ref{sec:section2}, which is a natural way of realising such superlattices. We subsequently discuss moir\'{e} superlattices in subsection \ref{ML} viewed as special cases of the van der Waals heterostructures. The first prominent case of such moir\'{e} superlattices with graphene is graphene on hexagonal boron nitride which will be briefly introduced in subsection \ref{GhBN}. Next, the central topic of this review, twisted bilayer graphene, is introduced in section \ref{secTBLG}. We discuss in detail both the commensurate moir\'e pattern in TBLG in subsection \ref{CTBLG} and the incommensurate moir\'e pattern in TBLG in subsection \ref{ICTBLG}. The Bistritzer-MacDonald model which first predicted the flat bands at magic angle in TBLG will be discussed in section \ref{BMmodel}, while a more recent chiral model which explains the origin of the magic angles in TBLG will be briefly reviewed in section \ref{PFB}. We conclude this review by listing a number of important developments in this field that are not covered by this review.

	\section{The transition from weak to strong correlations in Dirac systems}\label{WeaktoStrong}
	The fermions of lattice-gauge theories (henceforth dubbed as the lattice fermions) and the electrons in a crystal familiar to a solid-state physicist both obey lattice translational symmetry, and accordingly their momentum is conserved modulo an integer multiple of the length of the reciprocal vector. Whereas the lattice fermions obey a relativistically-invariant quantum field theory, the electrons of the crystal are sufficiently described by a one-component (non-relativistic) Schr\"{o}dinger equation. The energy eigenvalues of such non-relativistic electrons in a crystal form bands \cite{Ashcroft}. In their seminal 1983 work, Nielsen and Ninomiya \cite{Ninomiya1983} showed the correspondence between the conventional band theory for non-relativistic electrons and the lattice fermion theory. They exploited this resemblance to show that the effective charge carriers in a conventional band theory for gapless semiconductors behave as the massless Dirac fermions or Weyl fermions. In the following description, we shall begin by recapitulating this analogy to introduce the ultra-relativistic Dirac fermions in the conventional band theory of a condensed matter system.
	
	We start with an electron system with a generic Hamiltonian $ H $ obeying the time-independent Schr\"{o}dinger equation in a periodic potential, namely 
	\begin{equation}
		H\Psi(\bm{r}) = E\,\Psi(\bm{r})
		\label{eqn1}
	\end{equation}
	where $ \Psi(\bm{r}) $ are the Bloch wavefunctions. $ \Psi(\bm{r}) $ can be expanded in terms of orthonormal localized functions (Wannier functions) $f_{l}(\bmr-\bm{R}_{n})$, which go to zero exponentially as $ \abs{\bmr - \bm{R}} \rightarrow \infty $, giving
	\begin{equation}
		\Psi(\bmr) = \sum_{l}\sum_{\bm{R}} \phi_{l}(\bm{R})\,f_{l}(\bmr-\bm{R}). 
	\end{equation}
	Here $ l=1,2,\dots, N $ is the band index and $ \bm{R} $ is the lattice translation vector. The substitution of this expansion into (\ref{eqn1}) followed by multiplication by $ f^{*}_{m}(\bmr-\bm{R}')$ and integration over $ \bmr $ gives us
	\begin{widetext}
	\begin{equation}
		\sum_{l,\bm{R}} \phi_{l}(\bm{R})\int \dd{\bmr} f^{*}_{m}(\bmr-\bm{R}')Hf_{l}(\bmr-\bm{R}) = E \sum_{l,\bm{R}} \phi_{l}(\bm{R})\int \dd{\bmr} f^{*}_{m}(\bmr-\bm{R}')f_{l}(\bmr-\bm{R})
	\end{equation}
	Since the Wannier functions are orthonormal, namely $ \int \dd{\bmr} f^{*}_{m}(\bmr-\bm{R}')f_{l}(\bmr-\bm{R}) = \delta_{m,l} \delta_{\bm{R},\bm{R}'}$ and defining $ H_{ml}(\bm{R}'-\bm{R})= \int \dd{\bmr} f^{*}_{m}(\bmr-\bm{R}')Hf_{l}(\bmr-\bm{R})$ we finally get 
	\begin{equation}
		\sum_{l,\bm{R}} H_{ml}(\bm{R}'-\bm{R})\phi_{l}(\bm{R}) = E\,\phi_{m}(\bm{R}'). 
		\label{eqn:matrixham}
	\end{equation}
	The Fourier transformation of both sides of Eq. (\ref{eqn:matrixham}) and some rearrangement finally gives 
	\begin{equation}
		\sum_{l} \iint \frac{\dd{\bmk'}}{(2\pi)^{3}} \frac{\dd{\bmk}}{(2\pi)^{3}} H_{ml}(\bmk') e^{i\bmk'\vdot \bm{R}'} \phi_{l}(\bmk) \sum_{\bm{R}} e^{i\left(\bmk-\bmk'\right)\vdot\bm{R}} \\
		 = E \int \frac{\dd{\bmk}}{(2\pi)^{3}} \phi_{m}(\bmk) e^{i\bmk\vdot\bm{R}'}. 
	\end{equation}
	\end{widetext}
	Insertion of the identity $ \sum_{\bm{R}}e^{i\left(\bmk-\bmk'\right)\vdot\bm{R}} = \sum_{\bm{G}} \delta_{\bmk-\bmk',-\bm{G}}$, where $\bm{G}$ are the reciprocal lattice vectors, yields
	\begin{equation}
		\sum_{l}\sum_{\bm{G}} H_{ml}(\bmk + \bm{G}) \phi_{l}(\bmk) = E\,\phi_{m}(\bmk) \label{matrixHam}
	\end{equation}
	The matrix Hamiltonian in (\ref{matrixHam}) is manifestly the same as the one used in lattice fermion theory  where $H_{ml}(\bmk + \bm{G})$ may contain the essence of relativistic dispersion in an otherwise non-relativistic 
	condensed matter system \cite{Ninomiya1983}. This will be elaborated in the subsequent discussion.
	
	To understand how such a relativistic dispersion comes about, let us note that if the $ i^{\text{th}} $ level $ E_{i}(\bmk) $ and the $ (i+1)^{\text{th}} $ level $ E_{i+1}(\bmk) $ are degenerate at several different points in the dispersion space $ (\bmk,E(\bmk)) $, then the expansion of $ H(\bmk) $ around one of the degeneracy point $ (\bmk_{d},E_{d}  (\bmk_{d})) $ yields
	\bea
		H(\bmk) & =&  H(\bmk_{d}) + \left(\bmk-\bmk_{d}\right)\vdot\grad{H(\bmk)}|_{\bmk=\bmk_{d}} \nonumber \\ 
		&  & \mbox{+} \left(\bmk-\bmk_{d}\right)^{2} \nabla^{2}{H(\bmk)}|_{\bmk=\bmk_{d}} + \dots \label{BlochH}
	\eea
	Particularly, the shifts of the energy, $ E(\bmk) - E(\bmk_{d}) $ to the first order in $ \left(\bmk-\bmk_{d}\right) $ can be determined from the $ 2\times2 $ submatrix $ H^{(2)}(\bmk) $ formed by the $ i^{\text{th}} $ and $ \left(i+1\right)^{\text{th}} $ entries of the $ N\times N $ matrix for H. This effective two-band Hamiltonian $ H^{(2)}(\bmk)$ is given as 
	\begin{equation}
		H^{(2)}(\bmk) = H^{(2)}(\bmk_{d}) + \left(\bmk-\bmk_{d}\right)\vdot\grad{H^{(2)}(\bmk)}|_{\bmk=\bmk_{d}}
	\end{equation}
	where the $ j^{\text{th}}-$component of the derivative term is expressed by the Pauli matrices $ \left(\mathcal{I}_2,\sigma^{\alpha}\right) $ for $ \alpha = 1,2,3 $ and $ \mathcal{I}_2 $ is the $ 2\times2 $ identity matrix, as
	\begin{equation}
		\pdv{H^{(2)}(\bmk)}{k_{j}}|_{\bmk=\bmk_{d}} = a_{j}(\bmk_{d})\mathcal{I}_2 + V_{\alpha}^{j}(\bmk_{d})\sigma^{\alpha} 
	\end{equation}
	where $ \bm{a} $ and $ \bm{V} $ are constant vectors depending on $ \bmk_{d} $. The requirement that the Hamiltonian must be Hermitian, i.e. $ \mathcal{H}^{\dagger}_{\bmk} = \mathcal{H}_{\bmk} $ demands that the functions $ \bm{a},\bm{V} $ must be real. Thus, near $ \bmk=\bmk_{d} $, $ H^{(2)}(\bmk) $ takes the form,
	\begin{equation}
		H^{(2)}(\bmk) = E({\bmk_{d}})\mathcal{I}_2 + \left(\bmk-\bmk_{d}\right)\vdot\bm{a}\,\mathcal{I}_2 + \left(\bmk-\bmk_{d}\right)_{j}V_{\alpha}^{j}(\bmk_{d})\sigma^{\alpha} 
	\end{equation}
	where $E({\bmk_{d}})$ is the energy eigenvalue of the Hamiltonian at the degeneracy point, and the components of $\bm{a}$,  and $V_{\alpha}^{j}(\bmk_{d})$ are all real. 
	
In matrix form, the Hamiltonian may be expressed as:
\begin{equation}
\mathcal{H}^{(2)}({\bmk}) =
\begin{pmatrix}
f^{0}_{\bmk} + f^{3}_{\bmk} & f^{1}_{\bmk} - i\,f^{2}_{\bmk} \\
f^{1}_{\bmk} + i\,f^{2}_{\bmk} & f^{0}_{\bmk} - f^{3}_{\bmk}
\end{pmatrix}
\label{eqn:gentwobandham}
\end{equation}
where the functions $ fs $ are $ f^{0}_{\bmk} = E(\bmk_{d}) + \left(\bmk-\bmk_{d}\right)\vdot\bm{a} $, $ f^{1}_{\bmk} = \left(\bmk-\bmk_{d}\right)\vdot\bm{V}_{1}(\bmk_{d}) $, $ f^{2}_{\bmk} = \left(\bmk-\bmk_{d}\right)\vdot\bm{V}_{2}(\bmk_{d})$, $ f^{3}_{\bmk} = \left(\bmk-\bmk_{d}\right)\vdot\bm{V}_{3}(\bmk_{d})$. The dispersion relation is obtained by diagonalizing the Hamiltonian (\ref{eqn:gentwobandham}) that yields
\begin{equation}
\epsilon_{s}(\bmk) = f^{0}_{\bmk} + s \sqrt{(f^{1}_{\bmk})^2 + (f^{2}_{\bmk})^2 + (f^{3}_{\bmk})^2} \label{dfdispersion}
\end{equation}
and the corresponding eigenstates are
\begin{equation}
\psi_{s\bmk}(\bmr) = \frac{1}{\sqrt{1 + \frac{(f^{1}_{\bmk})^2+(f^{2}_{\bmk})^2}{\left(f^{0}_{\bmk}-\epsilon_{s}-f^{3}_{\bmk}\right)^2}}}
\pmqty{1 \\ -\frac{f^{1}_{\bmk}+i f^{2}_{\bmk}}{f^{0}_{\bmk}-\epsilon_{s}-f^{3}_{\bmk}}}
e^{i\left(\bmk-\bmk_{d}\right)\vdot\bmr}
\end{equation}
where $ s = \pm 1 $ are the indices characterizing the two bands. The function $ f^{0}_{\bmk} $ on RHS does not affect the eigenstates of $ \mathcal{H}^{(2)}_{\bmk} $, instead it just shifts the energy spectrum. Therefore, it is a common practice to define $ \epsilon_{s}(\bmk)- f^{0}_{\bmk} = E_{s}(\bmk) $, such that the energy $ E_{s}(\bmk) $ is measured from the reference level $ f^{0}_{\bmk} $.
	
In a different setting (for example, see \cite{katsnelson2012}), where the Hamiltonian is derived using the tight-binding method, the Hamiltonian can also take a similar form as in (\ref{eqn:gentwobandham}). If we consider that the Hamiltonian is written in the sublattice degrees of freedom denoted as $A$ and $B$ for a bipartite lattice, 
then the diagonal elements represent intra-sublattice couplings (say AA or BB) and off-diagonal elements represent inter-sublattice couplings (AB or BA). Consequently, the second quantized Hamiltonian $ H $ is given as
\begin{equation}
	H = \sum_{\bmk} \Pmqty{a_{\bmk}^{\dagger} & b_{\bmk}^{\dagger}}\mathcal{H}^{(2)}_{\bmk}\Pmqty{a_{\bmk} \\ b_{\bmk}}, \label{diracfermion2}
\end{equation}
where $ a_{\bmk} $ and $ b_{\bmk} $ are the annihilation operators at a particular $ \bmk $. The first and second spinor components would then correspond to the amplitude on the A  and B sublattice, respectively. After expanding the Hamiltonian $ H^{(2)}_{\bmk} $ around the degeneracy point $ \bmk = \bmk_{d} $, the Hamiltonian $ H^{(2)}_{\bmk} $ in (\ref{diracfermion2}) can be written in precisely the same form as (\ref{eqn:gentwobandham}).

Because the linear dispersion is given in (\ref{dfdispersion}) in such condensed matter systems, the low-energy fermionic excitations behave as Dirac particles in contrast to free fermions which obey the usual Schr\"{o}dinger equation and having a quadratic dispersion. 
Defining $ \bmq = \bmk - \bmk_{d} $, the Hamiltonian is written as
\begin{multline}
H^{(2)}(\bmq) = E({\bmk_{d}})\mathcal{I}_2 + \bmq\vdot\bm{a}\,\mathcal{I}_2
+ q_{j}V_{1}^{j}(\bmk_{d})\sigma_{x} \\
+ q_{j}V_{2}^{j}(\bmk_{d})\sigma_{y}
+ q_{j}V_{3}^{j}(\bmk_{d})\sigma_{z}
\label{eqn:hamq}
\end{multline}
Now, according to the definition of time-reversal (TR), $ \mathcal{T}H^{(2)}(\bmq)\mathcal{T}^{-1} = {H}^{(2)*}(-\bmq) $, where $ \mathcal{T} $ is the time-reversal operator, the action of time-reversal operation on (\ref{eqn:hamq}) gives,
\begin{multline}
H^{(2)*}(-\bmq) = E({\bmk_{d}})\mathcal{I}_2 - \bmq\vdot\bm{a}\,\mathcal{I}_2
- q_{j}V_{1}^{j}(\bmk_{d})\sigma_{x} \\
+ q_{j}V_{2}^{j}(\bmk_{d})\sigma_{y}
- q_{j}V_{3}^{j}(\bmk_{d})\sigma_{z}
\label{eqn:eqn2}
\end{multline}
The Bloch Hamiltonian of the system $ H(\bmk) $ given in (\ref{BlochH}) being TR-invariant it implies Kramer's degeneracy, i.e., for each eigenstate $ \psi(\bmk) $ with eigenvalue $ E(\bmk) $, there exists a state $ \mathcal{T}\psi(\bmk) $ with the same eigenvalue. Thus if $ \mathcal{H}^{(2)} $ describes the Hamiltonian near the degenerate point $ \bmk = \bmk_{d} $, then the $ \mathcal{H}^{(2)}(-\bmq) $ must describe the Hamiltonian of its TR-partner at $ \bmk = -\bmk_{d} $. For this to happen the constraints are $ \bm{q}\vdot\bm{a} = 0~\forall~\bmq $, $ E(-\bmk_{d}) = E(\bmk_{d}) $ and the vectors $ \bm{V} $ must also be even functions in $ \bmk_{d} $. The corresponding eigenfunction is then given as
\begin{equation}
\psi'_{s\bmk}(\bmr) = \frac{1}{\sqrt{1 + \frac{(f^{1}_{\bmk})^2+(f^{2}_{\bmk})^2}{\left(f^{0}_{\bmk}-\epsilon_{s}-f^{3}_{\bmk}\right)^2}}}
\pmqty{\frac{f^{1}_{\bmk}-i f^{2}_{\bmk}}{f^{0}_{\bmk}-\epsilon_{s}-f^{3}_{\bmk}} \\ 1}
e^{i\left(\bmk+\bmk_{d}\right)\vdot\bmr}
\end{equation}

In their 1983 work, Nielson and Ninomiya\cite{Ninomiya1983} used such DPs in the band dispersion relation of a prototype condensed matter system to establish that there is no net production of electrons when parallel electric and magnetic fields are switched on. This was termed as the condensed-matter analogue of the (3+1)-dimensional axial anomaly \cite{ABJ1969, ABJ1969Nuovo}. In 1984, Gordon W. Semenoff \cite{Semenoff1984} used this similarity of the fermions between the lattice model of gauge theories and the tight-binding description of electrons in crystals, and proposed a condensed-matter analogue of (2+1)-dimensional electrodynamics. Semenoff identified individual graphite layers (graphene) and hexagonal boron nitride as two possible systems for the realisation of such effects. It was pointed out that contrary to high-energy physics, in such relativistic condensed matter systems, the characteristic velocity that appears in condensed-matter physics is the Fermi velocity and not the speed of light \cite{Novoselov2005nature, CastroNeto2009}.
	
There is a wide range of materials with Dirac fermion low-energy excitations such as high-temperature $d$-wave superconductors \cite{Balatsky2006}, graphene \cite{CastroNeto2009,katsnelson2012}, the surfaces of three-dimensional topological insulators \cite{Hasan2010,Qi2011} and ruby and kagome lattices \cite{Cortes2011}. Regardless of their origin, the materials with the Dirac points in their energy spectrum are referred to as Dirac materials. For a detailed study on Dirac materials, one can refer to the Ref.\cite{Balatsky2014}. In general, the Dirac materials can be classified into three different classes \cite{VanMiert2016} depending on the location of the Dirac points in their Brillouin zone.
\begin{enumerate}[itemsep=0pt]
	\item[(I)] The Dirac point occurs at the high symmetry (HS) points such as graphene,
	\item[(II)] The Dirac points lies along the HS lines such as $ \beta $-graphyne \cite{Malko2012}, square graphynes \cite{LZzhang2015}.
	\item[(III)] The systems that have Dirac cones located at generic points in the BZ such as $ \alpha-$(BEDT-TTF)$_2$I$_3$ \cite{Katayama2006}.
\end{enumerate}
Moreover, the contact, or degenerate $ \bmk- $points among the energy bands are classified as \emph{essential} or \emph{accidental}, according to whether or not one can specify them in advance \cite{Herring,Asano2011}. The point contact is characterized as the \emph{Dirac} point (DP), when two bands split linearly in energy. In Dirac systems with \emph{accidental} (non-essential) degeneracy, a generalized \emph{von Neumann-Wigner} theorem has been proposed which gives the number of constraints on the lattice necessary to have accidental contacts \cite{Asano2011}. Their general treatment also provides a practical procedure to search for the $ \bmk $-points at which the accidental degeneracy takes place.
Keeping in mind the topic of this review, we consider the class (I) systems where the DPs occur at the HS points and are essential. In particular, we shall consider the case of single layer graphene (SLG) and discuss the generic structure of its Hamiltonian and the different symmetries that gives rise to the formation of DPs in its energy spectrum \ref{Honeycomb}.

	\subsection{Honeycomb lattice}\label{Honeycomb}
	
	\begin{figure*}
		\centering
		\includegraphics[scale=0.45]{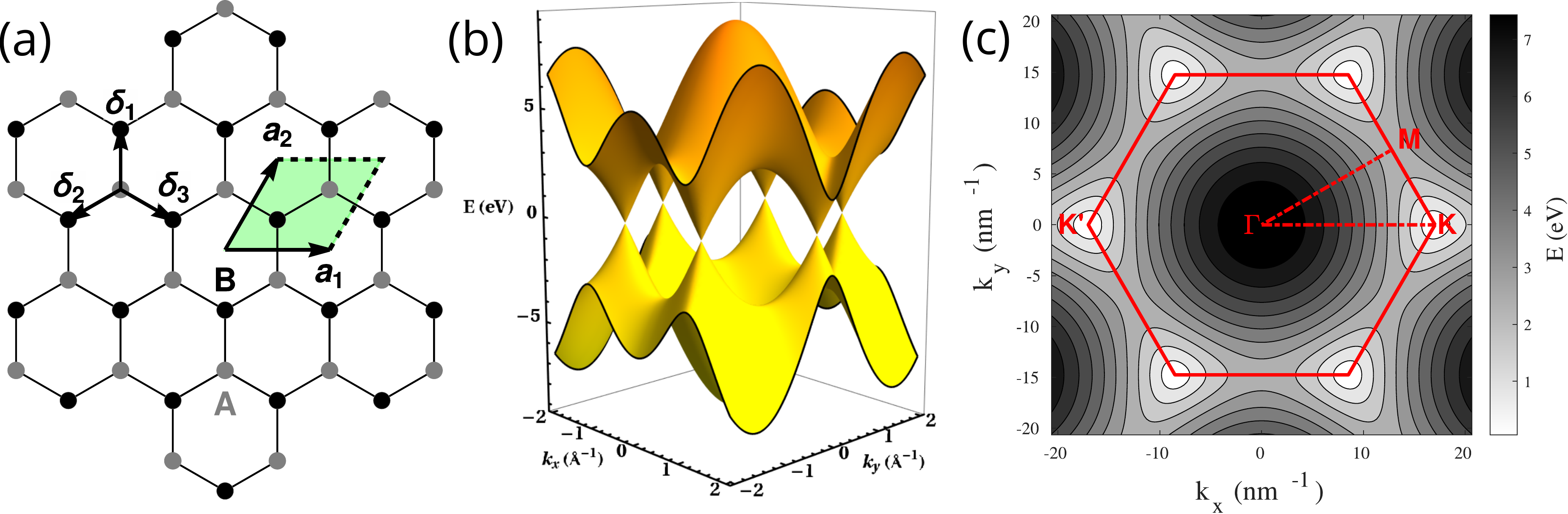}
		\caption{(a) The honeycomb lattice of graphene. The two sublattices are denoted by A (black-points) and B (Gray-points). The two real-space primitive vectors $ \bm{a}_{1} $ and $ \bm{a}_{2} $ are making an unit cell (green rhombus) and contains two atoms of either type A and B. It also shows the three nearest-neighbours vectors $ \bm{\delta}_1 $, $ \bm{\delta}_2 $ and $ \bm{\delta}_3 $ connecting each nearest B-type atom from a given A-type atom. (b) The full dispersion of the pristine SLG using the Hamiltonian (\ref{eqn:slgham}). There are six points (the corners of the first Brillouin zone) where the conduction and valence band make contact forms a hexagon. In the vicinity of these points the bands disperse conically forming the Dirac cones (condensed matter version of light-cone). (c) The energy contour plot of conduction band. The red hexagon marks the FBZ and also label the high-symmetry points, $ \Gamma $: centre of the BZ, $ K $: Dirac point with $ \xi = 1 $, $ M $: Midpoint of the line-joining the two adjacent corners, $ K' $: Dirac point with $ \xi = -1 $. In the close vicinity of Dirac points ($ K $ or $ K' $), the shape of the contours are circular and becomes triangular in shape as one goes away from the Dirac points (the trigonal warping).}
		\label{fig:slgbasic}
	\end{figure*}
	We begin by setting up the notation for SLG, which is a two-dimensional layer of carbon atoms arranged on a honeycomb lattice \cite{Novoselov2004,Novoselov2005pnas} as shown in Fig.\ref{fig:slgbasic}(a). It should be pointed out that pristine SLG has interesting physical properties e.g., large electron mobilities \cite{Geim2007,Novoselov2005nature,Yankowitz2019,Pulizzi2019} and strong mechanical strength \cite{Changgu2008}. It is also used to build Josephson junctions when placed between superconducting contacts \cite{Heersche2007}. 
	SLG can be considered as two interpenetrating triangular lattices (A and B), resulting in a \emph{bipartite}-lattice. 
	The lattice sites $A$ and $B$ for the same carbon atom are inequivalent, and inversion symmetric to each other. 
	
	The direct lattice primitive vectors 
	\beq \bm{a}_1 = a\left(1,0\right),~ \text{and},~\bm{a}_{2} = a\left(1/2,\sqrt{3}/2\right), \label{GUNIT} \eeq connect the center of adjacent hexagons, 
	where $ a = 0.246~\si{\nano\meter} $ is the lattice constant of SLG. The sublattice A and B can be connected from the center of the hexagon by vectors $ 2\left(\bm{a}_1 + \bm{a}_2\right)/3 $ and $ \left(\bm{a}_1 + \bm{a}_2\right)/3 $, respectively. A three-fold rotation symmetry about each lattice point combined with the reflection symmetry relating the two sublattices A and B (Fig.\ref{fig:slgbasic}(a)) leads to the essential degeneracy in the reciprocal space ($ k- $space) where the valence and conduction bands touch each other at the corners of the first Brillouin zone (FBZ) and give rise to the band touching points. At these points the electron dispersion is linear and effectively described by the massless Dirac fermions. The Hamiltonian obtained from the first nearest-neighbour tight-binding model of the $\pi$-electrons at each atomic site is
	\begin{equation}
		H_{\bmk} = -t\sum_{j=1}^{3}\left[\sigma_{x}\cos(\bm{\delta}_{j}\vdot\bmk) - \sigma_{y}\sin(\bm{\delta}_{j}\vdot\bmk)\right]
		\label{eqn:slgham}
	\end{equation}
	where $ t = 2.7~eV $ \cite{Sreich2002} is the tight-binding parameter and $ \bm{\delta}_1 = (a/\sqrt{3})(0,1) $, $ \bm{\delta}_2 = (a/\sqrt{3})(-\cos \ang{30},-\sin \ang{30}) $ and $ \bm{\delta}_3 = (a/\sqrt{3})(\cos \ang{30},-\sin \ang{30}) $ are the three vectors connecting the sublattice $A$ with the nearest-neighbours $B$. It has both time-reversal symmetry ($ \mathcal{T}H_{\bmk}\mathcal{T}^{-1} = H^{*}_{-\bmk} = H_{\bmk} $) and inversion symmetry. Fig.\ref{fig:slgbasic}(b) shows the full band structure. The essential Dirac points are located at the $ \bm{K} = \left(4\pi/3a,0\right) $ and $ \bm{K}' = \left(-4\pi/3a,0\right) $ points located at the corners of the hexagonal Brillouin zone \cite{CastroNeto2009} as depicted in Fig.\ref{fig:slgbasic}(c). The first Brillouin zone (FBZ) of SLG contains two sets of three equivalent $ K $ and $ K' $-points where the $ \pi $-conduction and $ \pi $-valence band touch.
	
	The wave-vector $ \bmk $ in $ H_{\bmk} $ in (\ref{eqn:slgham}) is measured from the $ \Gamma- $point of the Brillouin zone (BZ). Writing the Taylor's expansion of $ H_{\bmk} $ about the $ K $ and $ K' $-point one obtains,
	\begin{widetext}
	\begin{equation}
		H_{\bmk} = -t\left[\sigma_{x}\left(-\xi\frac{\sqrt{3} a k_x}{2} + \frac{a^2}{8}\left(k_{x}^2-k_{y}^2\right)+\dots\right)
		-\sigma_{y}\left(\frac{\sqrt{3} a k_y}{2} +\xi \frac{a^2}{4}k_x k_y + \dots\right)\right]
		\label{eqn:slgHamExpn}
	\end{equation}
	\end{widetext}
	where the quantity $ \xi $ is referred to as the valley index, which takes the values $ 1 $ for $ K $-point and $ -1 $ for $ K' $-point. For small $ k_x $ and $ k_y $, the infinite series (\ref{eqn:slgHamExpn}) is truncated to first order in $ k_{x}, k_{y} $, after which the Hamiltonian (\ref{eqn:slgham}) becomes
	\begin{equation}
		h^{\xi}_{\bmk} = \hbar\,v_F\left(\xi\sigma_{x}k_{x} + \sigma_{y}k_{y}\right)
		\label{eqn:slgHamxi}
	\end{equation}
	where $ \hbar v_F = \sqrt{3}at/2 = 0.58~\si{\electronvolt\nano\meter} $ and $ v_{F} $ is the Fermi velocity which is approximately $ 10^6~\si{\meter\second^{-1}} $. The energy contour in Fig.\ref{fig:slgbasic}(c) in the vicinity of Dirac points ($K$ or $K'$) are circular since the energy dispersion obtained from (\ref{eqn:slgHamxi}) is $ E^2 = \hbar^2 v_{F}^2\left(k^2_{x}+k^2_{y}\right)$. Nevertheless, as one goes away from the Dirac points these circles gradually start to distort and take the shape of triangles since the higher-order terms depend on the direction of $\bmk$ --an effect referred to as \emph{trigonal warping}\cite{Trigonal2000}. In what follows, we shall describe why the electron system in such SLG is considered weakly correlated. 
	 
The effective Hamiltonian $H$ of any such general electronic system that determines its electronic phases is primarily composed of two terms, the kinetic energy term $ H_{\text{K}} $ and the interaction term $ H_{\text{int}} $ which, in turn, contains the terms governing the $e-e$ interactions, the $e$-nuclei interaction or the nuclei -nuclei interactions. The second term is typically responsible for the correlation effects. One thus needs to compare interaction with the kinetic energy to determine whether a system is strongly correlated or weakly correlated. In this context, for a two-dimensional electron gas (2DEG) one can define a dimensionless parameter $r_{s}$ such that $ r_{s}\,a_{0} $ is the radius of a circle whose area is equal to the average area occupied by one electron, namely  $ \frac{1}{n_{e}} = \frac{A}{N} $, where $N$ is the total number of electron and, $ a_{0} = 4\pi\hbar^2\epsilon_{0}/(me^{2}) = 0.529~\si{\angstrom} $ (in S.I. units) is the Bohr radius representing the unit of length. This defines
	\begin{equation}
		\pi r_{s}^{2}a_{0}^{2} = A/N \implies r_{s} = \frac{1}{a_{0}}\left(\frac{1}{\pi n_{e}}\right)^{1/2}. 
	\end{equation}
	For a 2DEG, the Fermi wave-vector $k_{F}$ depends on the carrier density $n_{e}$, and hence on $r_{s}$, as $k_{F} = \left(2\pi n_{e}\right)^{1/2} = \frac{\sqrt{2}}{r_{s}a_{0}}$. 
	In terms of the Fermi wave vector, the total kinetic energy $K$ is given as 
	\begin{equation}
		K = 2A \int_{E<E_{F}} \frac{\dd[2]{k}}{(2\pi)^2} \frac{k^2}{2m} =  \frac{A}{8m\pi}k_{F}^{4}.
	\end{equation}
	The total number of particles $N$ can similarly be written as 
	\begin{equation}
		N = \frac{A}{\pi}\int_{0}^{k_{F}} k \dd{k} =  \frac{A}{2\pi}k_{F}^{2} \implies k_{F} = \left(2\pi n_{e}\right)^{1/2}
	\end{equation}
	Therefore, the kinetic energy per particle in terms of the parameter can be written as 
	\begin{equation}
		K_{E} = \frac{K}{N} = \frac{k_{F}^{2}}{4m} = \frac{1}{2mr_{s}^{2}a_{0}^{2}} \propto \frac{1}{mr_{s}^{2}}
	\end{equation}
	Given that the average Coulomb interaction between electrons can be given as $E^{c}_{\text{int}}= e^{2}/4\pi\epsilon_{0}r_{s}a_{0} $, the ratio between the $ K_{E} $ and the interaction energy is
	\begin{equation}
		\frac{E^{c}_{\text{int}}}{K_{E}} = \frac{2m e^{2}r_{s}^2a_{0}^2
		}{4\pi\epsilon_{0}r_{s}a_{0}}  = 2\hbar^{2}r_{s} \propto r_{s}
	\end{equation}
	When $ r_{s} >> 1 $ (low density) it leads to a strongly correlated system with the energy of interactions dominating the K.E. On the other hand, when $ r_{s} << 1 $ (high density) we have a weakly correlated system with the K.E. dominating over the interactions. For instance, Wigner crystallization \cite{Wigner1934} appears when electrons localize and form a crystal to minimize the potential energy while paying the concomitant kinetic energy cost, which arises from localization as the density of carriers is lowered. Theoretical studies predict that Wigner crystallization in conventional 2DEG occurs at $ r_{s} \approx 37 $ \cite{Tanatar1989}.
	
	In the case of massless-Dirac fermions in SLG, the Fermi wave-vector $k_{F}$ is related to carrier density $n_{e}$ as $\pi n_{e} = k_{F}^{2}$, because of spin and valley degeneracy. 
	The total kinetic energy of the system in the many-body ground state $\ket{0}$ at absolute zero temperature is given by
	\begin{equation}
		E_{k} = \hbar v_{F}\sum_{\lambda,\bmk}\abs{\bmk}\mel{0}{\bm{c}^{\dagger}_{\lambda\bmk}\bm{c}_{\lambda\bmk}}{0}
		= \frac{\hbar v_{F}}{2\pi}\frac{4Ak_{F}^{3}}{3}
	\end{equation}
	where $ \sum_{\lambda} =4 $ is the sum over spin and valley degeneracy, $ A $ is the area of the system, $ \bm{c}^{\dagger}_{\lambda\bmk} $ creates an electron with momentum $ \bmk $ (measured from the Dirac point).
	In terms of the carrier density, the kinetic energy becomes
	\begin{equation}
		E_{\bmk} = \frac{2\sqrt{\pi}}{3}\hbar v_{F}\,A\,n_{e}^{3/2}
	\end{equation}
	In a similar way, the total interaction energy $ E_{\text{int}} $ is calculated as
	\begin{align}
		E_{\text{int}} & = \frac{e^2}{8A\pi\epsilon}
		\sum_{\bmk_{1},\bmk_{2},\bmq} \sum_{\lambda_{1},\lambda_{2}}
		\frac{2\pi}{q} \mel{0}{c^{\dagger}_{\bmk_{1}+\bm{q},\lambda_{1}}c^{\dagger}_{\bmk_{2}-\bm{q}\lambda_{2}}c_{\bmk_{2},\lambda_{2}}c_{\bmk_{1},\lambda_{1}}}{0} \nonumber \\
		& = \frac{\sqrt{\pi}e^{2}}{3\pi\epsilon}A\,n_{e}^{3/2}
	\end{align}
	where $ \epsilon $ is the static dielectric constant for SLG. The ratio of total interaction energy to the total kinetic energy becomes
	\begin{equation}
		\frac{E_{\text{int}}}{E_{\bmk}}
		= \frac{4e^{2}}{h}\frac{1}{\epsilon v_{F}}. 
		\label{eqn:slgratio}
	\end{equation}
	Unlike the case of non-relativistic 2DEG, this ratio is independent of the density of carriers. Therefore, the strength of correlations in a gas of massless Dirac fermions remains unaffected with the variation in the carrier density leading to a situation very different than that in the conventional 2DEG. Using the value of the dielectric constant $ \epsilon = 4.7 $ \cite{Ando2006}, the ratio turns out to be $ E_{\text{int}}/E_{k} \approx 0.32 < 1 $. In pristine graphene, therefore, the kinetic energy dominates over the interactions and, consequently there is weak correlations among electrons. Dahal {\it et al.}\cite{Dahal2006} showed that graphene does not undergo Wigner crystallization to an insulator on reducing the density of carriers. Similarly, Peres {\it et al.}\cite{Peres2005} showed that SLG could not have ferromagnetic phases for realistic values of the parameters. As is evident from the preceding discussion and eq. (\ref{eqn:slgratio}), the ratio $ E_{\text{int}}/E_{k} $ can be increased by reducing the Fermi velocity $ v_{F} $. This can be done by subjecting the charge carriers in SLG to superlattices, as will be discussed in detail in later sections. This will eventually take us to the case of twisted graphene layers which fosters strong correlations in otherwise weakly correlated pristine SLG.
	
	The tuning of correlated states in materials is of great interest to modern condensed matter physicists, and rich physics arises when the position of the Fermi level draws close to the van Hove Singularity \cite{Markiew1997}. 
	Before discussing the role of such superlattices in tuning the interaction effects in graphene, we shall discuss the connection between the Van Hove singularity and the strongly correlated phases.
	
	\subsection{Van Hove Singularity (VHS)}\label{VHS}
	Note that the density of states (DOS) is defined as the number of single-particle states per unit energy range. In general, the expression for DOS reads \cite{Ashcroft}
	\bse
	\begin{align}
		g_n(E) = \int_{C_n(E)} \frac{\dd{l}}{(2\pi)^2}\frac{1}{|\nabla_{\bmk}E_n(\bmk)|} \qquad \left( 2D \right)\\
		g_n(E) = \int_{S_n(E)} \frac{\dd{S}}{(2\pi)^3}\frac{1}{|\nabla_{\bmk}E_n(\bmk)|} \qquad \left( 3D \right)
	\label{eqn:DOS}
	\end{align}
	\ese
	Here `$ n $' indicates the band index, and the integration runs over the constant energy contour $ C_n(E) $ in $2D$ space of $ \left\{ k_x, k_y\right\}$, and constant energy surface $ S_n(E) $ in $3D$ space of $ \left\{ k_x, k_y, k_z \right\} $, respectively. At a $ \bmk- $point, where the denominator in (\ref{eqn:DOS}) vanishes, the $ g_{n}(E) $ blows up and the corresponding $k-$point represents a singularity in the DOS. These singularities are known as the Van Hove singularities \cite{VanHove1953}.
	
	For any scalar field $ f $ of two or more variables $ \left\{ x_i \right\} $ ($ i=1,2,... $), it is said to have a \emph{saddle} point at $ \left( x_{1o}, x_{2o}, ...\right) $ in the space of $ x_1, x_2, ... $ (the stationary point without local maxima or minima) if
	\begin{equation}
		\eval{\pdv{f}{x_i}}_{x_i = x_{io}} = 0 \qq{and} \eval{\pdv[2]{f}{x_i}}_{x_i = x_{io}} = 0 \quad \forall \quad i
	\end{equation}
	In $2D$ materials, the saddle points result in a diverging integrand in (\ref{eqn:DOS}) and make the integral $ g_n(E) $ an improper integral. The DOS diverges logarithmically as $g(E) \propto \ln(W/2\Delta E)$, where $W$ is the bandwidth and $\Delta E = E-E_{\text{vHS}}$. In general, there can also be divergences of the (spin and/or charge) susceptibilities. The saddle points create greatly enhanced density of states (DOS) peaks which are observable in scanning tunnelling spectroscopy (STS) studies \cite{Li2010}.
	
	\begin{figure*}
		\centering
		\includegraphics[scale=0.4]{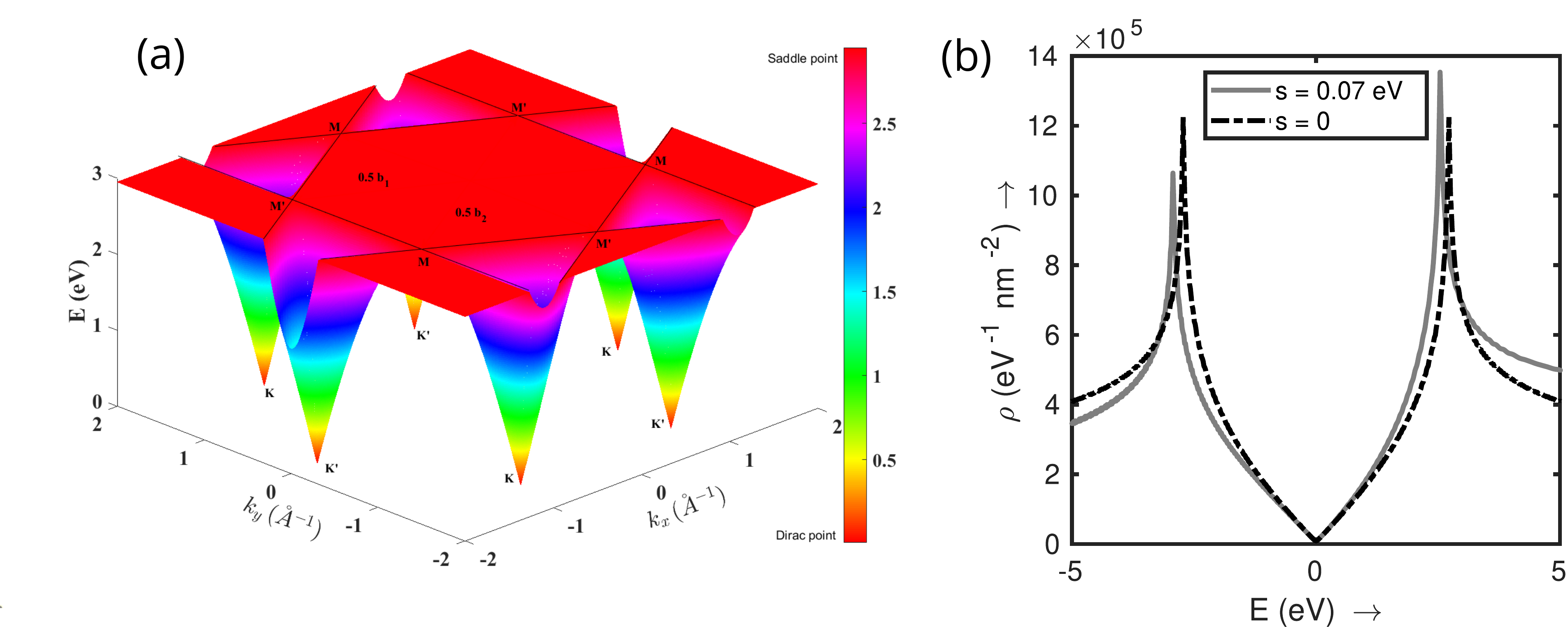}
		\caption{(a) Illustrating the M-points in the SLG band structure where the saddle points occur \cite{Nandkishore2012}. (b) The DOS for SLG with overlap parameter $ s = 0 $ (Gray line) ans $ s = 0.07~\si{\electronvolt} $ (dash-dotted) black line.}
		\label{fig:SLGsaddle}
	\end{figure*}
	
	Of interest to the present discussion is the existence of the Fermi level near the VHS that can lead to magnified interactions among the electrons which result in instabilities and therefore enhances the magnetic and superconducting correlations\cite{Markiew1997,Fleck1997,Rice1975}. In SLG, the energy dispersion from the Hamiltonian (\ref{eqn:slgham}) turns out to be
	\begin{multline}
		E_{s}(k_x,k_y) = \\
		s\,t\sqrt{3+2\cos(k_x a) + 4\cos(k_{x}a/2)\cos(\sqrt{3}k_{y}a/2)}
	\end{multline}
	where $s$ is the band index. The quantities $ \bm{\nabla}E_{s}(\bmk) $ and $ \nabla^2E_{s}(\bmk) $ both vanish simultaneously at the mid-point on the line joining the two adjacent boundaries of FBZ (an $ M $-point) as shown in  Fig.\ref{fig:SLGsaddle}(a). Therefore, the $ M- $point in the SLG band structure is a saddle point that corresponds to the VHS \cite{CastroNeto2009}. Since the saddle points occur both in valence and conduction bands, there are two Van Hove peaks in DOS in valence and conduction band as can be seen in Fig.\ref{fig:SLGsaddle}(b). The two peaks in the conduction and valence bands are located far apart $ (\Delta E = E_{\text{VHS}} - E_{F} \approx 2.7~\si{\electronvolt}) $ with respect to the Fermi level ($ E_{F} = 0 $). Unsurprisingly, therefore, correlation effects are hardly observed in intrinsic graphene. Different methods were implemented to achieve the strong interactions by chemically doping or subjecting the SLG to electrical gating \cite{Novoselov2004} in order to shift the Fermi energy such that interesting correlation phenomena may occur. However, reaching the VHS by either gating or chemical doping is difficult because of the considerable distance $ (\Delta E \approx 2.7~\si{\electronvolt}) $ of VHS from the Dirac point. Despite the difficulties in accessing the VHS in SLG, epitaxial graphene on silicon carbide has been successfully overdoped past the VHS in the conduction band up to a charge carrier density of $ 5.5 \times 10^{14}~cm^{-2} $ \cite{Hrag2020} leading to the observation of exotic ground states driven by many-body interactions \cite{Uchoa2007,Gonzalez2008,McChesney2010}.
	
	The challenges of accessing the VHS in SLG consequently forced the search for exotic many-body effects in other kinds of graphene-like materials. One way to vary the location of VHS is to subject charge carriers in graphene to superlattices. In 1970, Esaki and Tsu proposed the realisation of a novel semiconductor structure \cite{Esaki1970} either by a periodic variation of the doping level in a single material or by a periodic variation of two dissimilar materials, in particular, suggesting a periodic potential \emph{superlattice} (SL) to modify the band structure in semiconductors. The graphene superlattices are formed by applying periodic potentials with periodicities much larger than the lattice constant \cite{Park2008PRL,Park2008nature}. The application of external periodic potentials gives rise to additional Dirac points with renormalised Fermi velocities, with the additional saddle points in the energy bands which lead to the peaks in DOS closer to the Fermi level \cite{Park2008PRL,Park2008nature}. Experimentally, such superlattices have been investigated in monolayer and bilayer graphene $ \mathit{e.g.,} $ the control over the number of superlattice Dirac points by modulating the superlattice potential in SLG \cite{Dubey2013}, the observation of dips in DOS due to the superlattice Dirac points emerging in graphene over hexagonal Boron nitride \cite{Yankowitz2012}, the bilayer graphene superlattices \cite{Killi2011,Killi2012}. In this direction, we review the theory of $1D$ and $2D$ graphene superlattices in the subsequent discussion in some detail. This will help us to understand the physics of van der Waals heterostructures with graphene layers that will be introduced immediately after this discussion.
	
	\subsection{Graphene under external periodic potentials}\label{SLZ}
	A graphene superlattice arises when a periodic external potential, referred to as \emph{superlattice potential}, is applied to the charge carriers of graphene \cite{Park2008PRL,Park2008nature,Barbier2008,Brey2009,Suarez2010,Burset2011,Ortix2012}. This SL potential has been realized by a periodic variation of alloy composition or impurity density introduced during epitaxial growth \cite{Esaki1970}. Since the low-energy spectrum of graphene mimics the massless Dirac fermions, applying the periodic external potential to these quasiparticles alters the electronic spectrum and gives rise to a new set of quasiparticles. We review the effect of this externally applied potential on the band structure, Fermi velocity, DOS, and the location of the VHS. We begin with the Hamiltonian for the massless Dirac fermions under a general periodic potential $ V(\bmr) $, that can be written using (\ref{eqn:slgHamxi}) with $ \xi = +1 $ (K-valley) as
	\begin{equation}
		H = -i\,\hbar v_F \, \bm{\sigma} \cdot \bm{\nabla} + \mathcal{I}_2~V(\bmr)
		\label{eqn:superHamiltonian}
	\end{equation}
	where $ v_F $ is the Fermi velocity of the quasiparticles in graphene in the absence of SL potential, and $ \mathcal{I}_2 $ is a second-order identity matrix. The single-particle eigenstates of the first term in the Hamiltonian $ -i\hbar v_{F}\bm{\sigma}\vdot\bm{\nabla} $ in (\ref{eqn:superHamiltonian}) are given as
	\begin{equation}
		\braket{\bmr}{s,\bmk} = \psi_{s\bmk}(\bmr) = \frac{1}{\sqrt{2A}} \Pmqty{1 \\ s e^{i\theta_{\bmk}}} e^{i\bmk\vdot\bmr}
	\end{equation}
	and the eigenenergies as $ E_{s\bmk} = s \hbar v_{F} \abs{\bmk} $, where $ s $ is the band index, $ A $ is the area of the system and $ \theta_{\bmk} $ is the angle that wave-vector makes with the positive direction of $ x- $axis. Since the applied potential is translationally invariant under lattice translation $ \bm{T} $ of SL, i.e, $ V(\bmr) = V(\bmr + \bm{T}) $, therefore one can write
	\begin{equation}
		V(\bmr) = \sum_{\bm{G}} V_{\bm{G}}~e^{i\,\bm{G}\cdot\bmr}
	\end{equation}
	where $ \bm{G} $ are the reciprocal lattice vectors of the BZ which is referred to as \emph{superlattice} Brillouin zone (SBZ) and $ V_{\bm{G}} $'s are the corresponding complex Fourier components. 
	
	Further, the scattering of a state close to one Dirac point to another does not occur if the SL potential varies slowly over the length scale of lattice constant of graphene \cite{Takeshi1998one,Takeshi1998two,McEuen1999} or, equivalently, the spatial period of the potential is much larger than the nearest-neighbour carbon bond-length in graphene ($ a_{0} = \frac{a}{\sqrt{3}} \approx 0.142~\si{\nano\meter} $). Therefore, the scattering matrix element from a state $ \ket{s,\bmk} $ to $ \ket{s',\bmk'} $ due to applied potential $ V(\bmr) $ can be written as
	\begin{equation}
		\mel{s',\bmk'}{V(\bmr)}{s,\bmk} = \frac{1}{2} \sum_{\bm{G}} \left[1 + s' s e^{i\left(\theta_{\bmk}-\theta_{\bmk'}\right)}\right] V_{\bm{G}}\, \delta_{\bmk',\bmk+\bm{G}}
	\end{equation}
	If the single-particle wave-function $ \psi(\bmr) $ in the Schr\"{o}dinger equation $ H\psi(\bmr) = E\psi(\bmr) $ is expanded as $ \psi(\bmr) = \sum_{s,\bmk} c(s,\bmk) \psi_{s\bmk}(\bmr)$ where $ c's $ are the expansion coefficients, then one can write
	\begin{multline}
		\left(E_{s'\bmk'} - E\right) c(s',\bmk') + \\
		\frac{1}{2} \sum_{s,\bmk}  \sum_{\bm{G}}   \left[1 + s' s e^{i\left(\theta_{\bmk}-\theta_{\bmk + \bm{G}}\right)}\right] V_{\bm{G}}\, c(s,\bmk) = 0
	\end{multline}
	In order to see the behaviour of massless Dirac fermions near the zone boundaries, the energy dispersion has been carried out analytically for a $1D$-potential \cite{Park2008,Park2008PRL,Park2008nature} as given below.
		
	\begin{figure}
		\centering
		\includegraphics[scale=0.38]{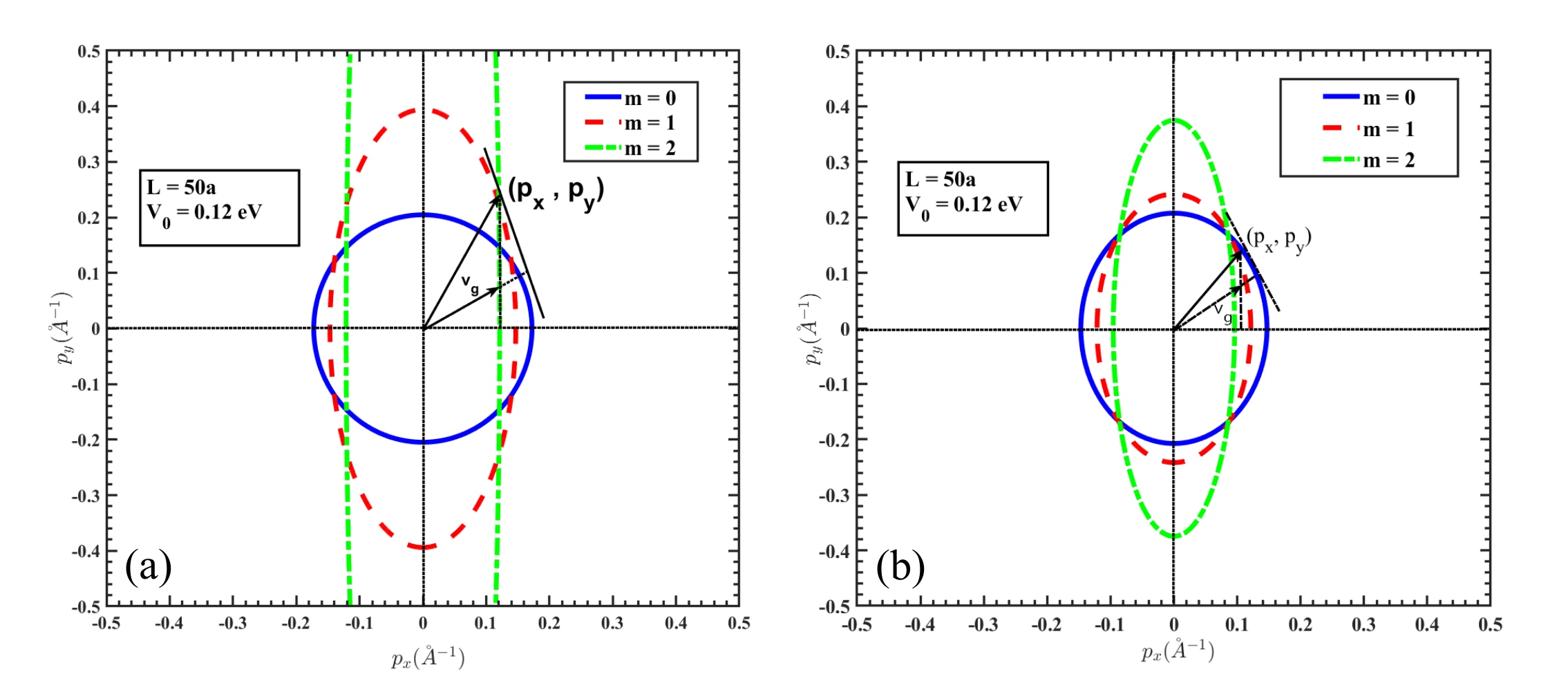}
		\caption{\label{fig4:slgslvel} (a) The equi-energy contours plotted using (\ref{eqn:1dcontoureqn}) for different values of $ m=0,1,2 $ of the newly generated massless Dirac fermions for cosine potential. For $ m=0 $, the blue circle denotes the isotropic Dirac cone. However, as the $ m $ increases, the Dirac cones become anisotropic along the y-axis. (b) Same for the Kr\"{o}nig-Penney-like potential.}
		
	\end{figure}

	\begin{figure*}
		\centering
		\includegraphics[scale=0.34]{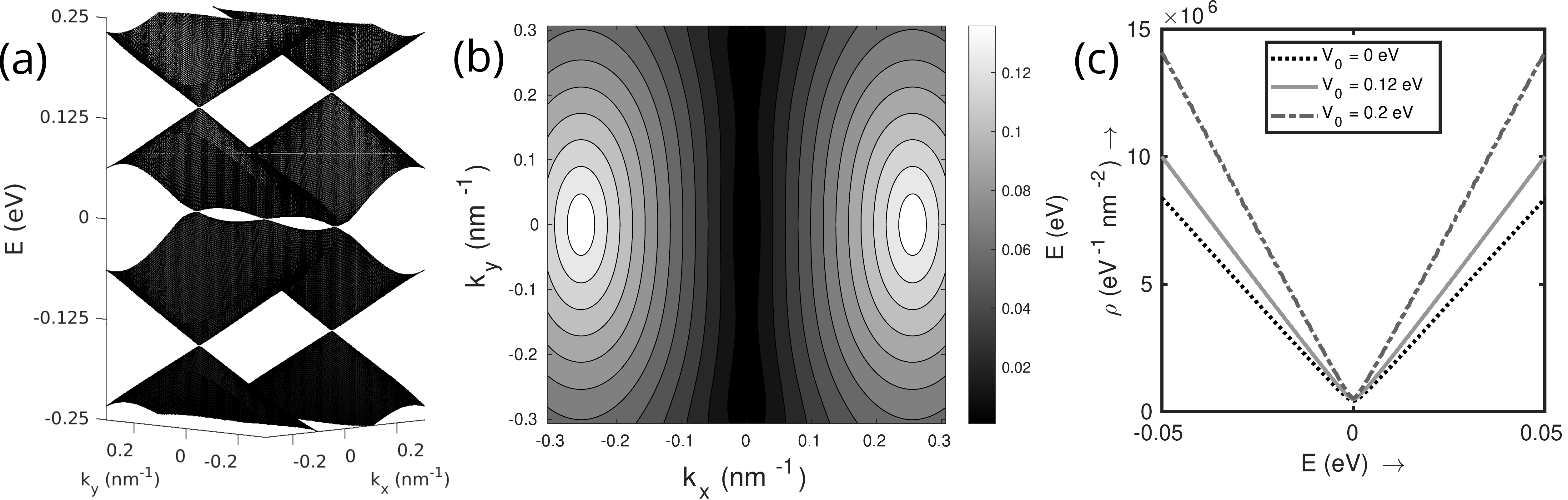}
		\caption{One-dimensional (1D) graphene superlattice formed by a cosine potential periodic along the x-direction with periodicity $ L = 50 a $. (a) It shows the full three-dimensional dispersion of the first two conduction bands and valence bands with the new set of Dirac points at the potential strength $ V_{0} = 0.4~\si{\electronvolt} $. (b) It shows the contour plot corresponding to the first conduction band. The Fermi velocity gets renormalized along the y-direction as the potential is applied along x-axis. (c) It shows the DOS plot at three different potentials $ 0~\si{\electronvolt}, 0.12~\si{\electronvolt} $ and $ 0.2~\si{\electronvolt} $. }
		\label{fig:1dcosbnd}
	\end{figure*}
	\begin{figure*}
		\centering
		\includegraphics[scale=0.38]{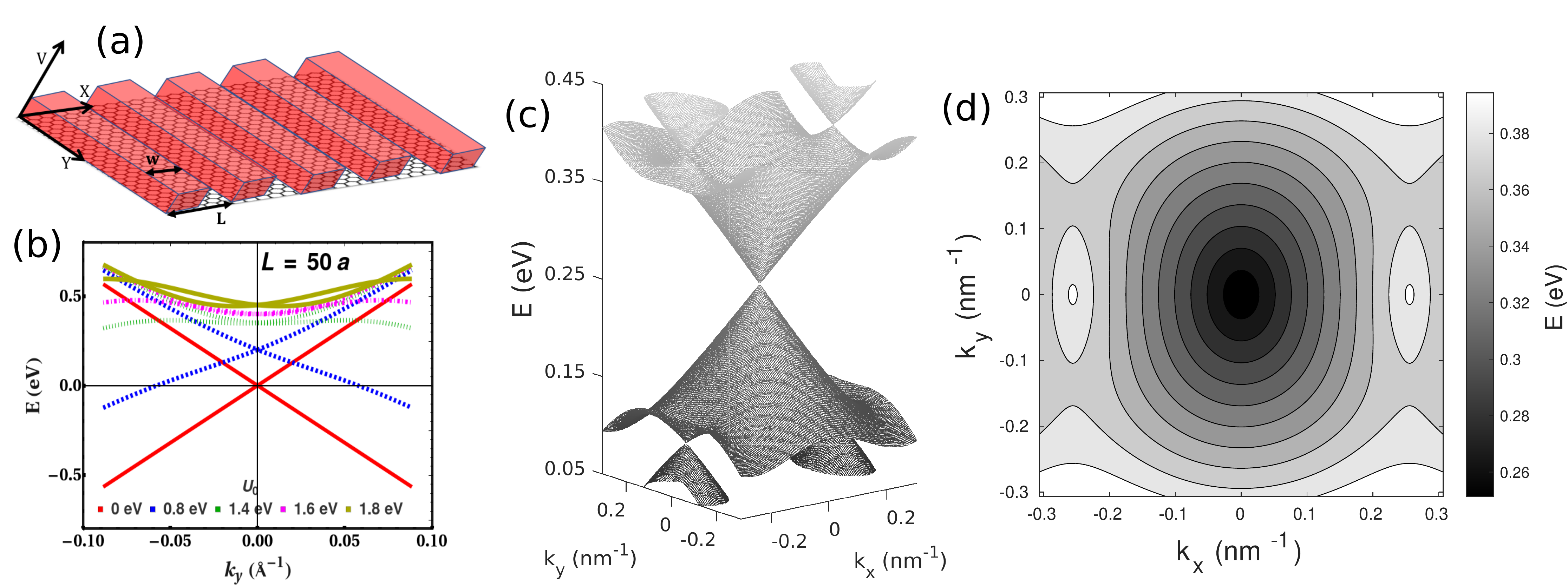}
		\caption{One-dimensional (1D) graphene superlattice formed by a cosine potential periodic along the x-direction with periodicity $ L = 50 a $. (a) It shows the full three-dimensional dispersion of the first two conduction bands and valence bands with the new set of Dirac points at the potential strength $ V_{0} = 0.4~\si{\electronvolt} $. (b) It shows the contour plot corresponding to the first conduction band. The Fermi velocity gets renormalized along the y-direction as the potential is applied along x-axis. (c) It shows the DOS plot at three different potentials $ 0~\si{\electronvolt}, 0.12~\si{\electronvolt} $ and $ 0.2~\si{\electronvolt} $. }
		\label{fig:1dkpband}
	\end{figure*}
	\subsubsection{One-dimensional (1D) potential}\label{1DSL}
	For a generic $1D$-potential $ V(x) $, the Hamiltonian (\ref{eqn:superHamiltonian}) becomes
	\begin{equation}
		H = \hbar\,v_F
		\begin{pmatrix}
			V(x)/\hbar v_F & i\partial_x + \partial_y \\
			i\partial_x - \partial_y & V(x)/\hbar v_F
		\end{pmatrix}.
		\label{eqn:simpcosHam}
	\end{equation}
	The partial differential operators are only present in the off-diagonal terms. A unitary transformation \cite{Talyanskii2001,Novikov2005,Park2008PRL} of the Hamiltonian, $ H' = U_1^{\dagger}\,H\,U_1 $ with 
	\begin{equation}
		U_1 = \frac{1}{\sqrt{2}}
		\begin{pmatrix}
			e^{-i\alpha(x)/2} & -e^{i\alpha(x)/2} \\
			e^{-i\alpha(x)/2} &
			e^{i\alpha(x)/2}
		\end{pmatrix}
		\label{eqn:transmat}
	\end{equation}
	where $ \alpha(x) = \frac{2}{\hbar\,v_F} \int_{0}^{x} V(x')\,\dd{x'} $, brings the $ \partial_x $ in the diagonal. The matrix $ U_1 $ is unitary, namely $U^{\dagger}_{1}\,U_{1} = 1$, as well as unimodular, since $ \text{det}\left(U_1\right) = 1 $. Now, any general unitary unimodular $ 2\cross2 $ or SU(2) matrix can be interpreted as the rotation operation in three dimensions. Thus $ U_{1} $ can be thought of as a rotation by an angle $ \phi $ about the axis $\hat{\bm{n}} = \left(n_x, n_y, n_z\right) $, with the following parameterisation 
	\begin{align}
		\cos(\frac{\phi}{2}) = \frac{1}{\sqrt{2}} \cos(\frac{\alpha(x)}{2});\quad
		n_y = \cot(\frac{\phi}{2}); \nonumber \\
		n_x = n_z = \frac{\sqrt{\frac{1}{2}-\cos^{2}\frac{\phi}{2}}}{\sin(\phi/2)}
	\end{align}
	The transformed Hamiltonian $ H' $ now becomes
	\begin{equation}
		H' = \hbar\,v_F
		\begin{pmatrix}
			-i\,\partial_x & -e^{i\alpha(x)}\,\partial_y \\
			e^{-i\alpha(x)}\,\partial_y & i\partial_x
		\end{pmatrix}. 
		\label{eqn:transformedHamiltonian}
	\end{equation}
	With this transformation, the partial differential operators $ \partial_x $ occurs in the diagonals, while $ \partial_y $ occurs in the in off-diagonal positions with the term $ e^{-i\alpha(x)} $ which contains the effect of the applied potential. The corresponding eigenstates transform, leading to the equation 
	\begin{equation}
		U_1\,H'\,U_1^{\dagger}\,\Phi(x,y) = E\,\Phi(x,y) \implies H'\,U_1^{\dagger}\,\Phi(x,y) = E\,U_1^{\dagger}\,\Phi(x,y)
	\end{equation}
	where $ \Phi(x,y) = \left( \phi_1(x,y) \quad \phi_2(x,y) \right)^T $ is a two-component wave function of the Hamiltonian (\ref{eqn:simpcosHam}), and $ U_1^{\dagger}\,\Phi(x,y) $ are the transformed states. 
	
	The reciprocal lattice vector of the $1D$ SL potential is $ \bm{G}_m = m\,G_0\,\hat{\bm{x}} $ where $ m $ is an integer. We are interested in the behaviour of the quasiparticles in the SBZ boundaries at $ \pm \bm{G}_{m}/2 $, where following two states are used as basis states
	\begin{equation}
		\phi_1 = \Pmqty{1 & 0}^{T} e^{i\left(\bm{p}+\bm{G}_m/2\right)\cdot\bmr}
		,\quad
		\phi_2 = \Pmqty{0 & 1}^{T} e^{i\left(\bm{p}-\bm{G}_m/2\right)\cdot\bmr}
		\label{eqn:basisstates}
	\end{equation}
	This leads us to the representation of  $H'$ in these basis states, namely a $2 \cross 2$ matrix $M$,  
	\begin{equation}
		M_{ij} = \int \dd[2]{\bmr} \phi^{\dagger}_i(\bmr)\,H'\,\phi_j(\bmr). 
	\end{equation}
	Substituting the $ H' $ from (\ref{eqn:transformedHamiltonian}) and the basis states $ \phi_1 $ and $ \phi_2 $ from (\ref{eqn:basisstates}), we get the explicit form of matrix $M$ as 
	\begin{equation}
		M = \hbar\,v_F
		\begin{pmatrix}
			p_x & -i\,f_m\,p_y \\
			i\,f_m^*\,p_y & -p_x
		\end{pmatrix}
		+ \frac{\hbar\,v_F\,m\,G_0}{2}\,\mathcal{I}_2
	\end{equation}
	where $ f_m = \int \dd[2]{\bmr}\,e^{i\alpha(x)} e^{-i\bm{G}_m\cdot\bmr} $ and $ \mathcal{I}_2 $ is the second order identity matrix. This matrix can be further simplified by one more similarity transformation $ M' = U^{\dagger}_2\,M\,U_2 $ to make it analogous to the SLG Hamiltonian, where
	\begin{equation}
		U_2 = \frac{1}{\sqrt{2}}
		\begin{pmatrix}
			1 & 1 \\
			-1 & 1
		\end{pmatrix}
	\end{equation}
	The final effective Hamiltonian that describes quasiparticles at the SBZ boundary now reads 
	\begin{equation}
		M' = \hbar\,v_F
		\begin{pmatrix}
			-p_y\,\text{Im}[f_m] & p_x - ip_y\,\text{Re}[f_m] \\
			p_x + ip_y\,\text{Re}[f_m] & p_y\,\text{Im}[f_m]
		\end{pmatrix}
		+ \frac{\hbar\,v_F\,m\,G_0}{2}\,\mathcal{I}_2. 
		\label{eqn:finalMat}
	\end{equation}
	The corresponding energy eigenvalues are given by
	\begin{equation}
		E_{s}(m,\bm{p}) = s\hbar\,v_F\sqrt{p_x^2 + \abs{f_m}^2p_y^2} + \hbar\,v_F\,m\,G_0/2
		\label{eqn:gendisprel}
	\end{equation},
	respectively. The only difference of the Hamiltonian in (\ref{eqn:finalMat}) \emph{vs.} that in (\ref{eqn:slgham}), other than a constant energy term, is that the group velocity of quasiparticles moving along the $y$-direction has been modified from $ v_F $ to $ f_m\,v_F $ as can be seen from the formula for band velocity below,
	\begin{equation}
		\bm{v} = \frac{1}{\hbar}\,\bm{\nabla}_{\bm{p}}\,E_s(m,\bm{p}) =
		\frac{s\,v_F}{\sqrt{p_x^2 + \abs{f_m}^2\,p_y^2}}\left[ p_x\,\hat{\bm{p}}_x + \abs{f_m}^2\,p_y\,\hat{\bm{p}}_y \right]
		\label{eqn:genfermivel}
	\end{equation}
	Thus, the electronic states near $ \bmk = \bm{G}_m/2 $ are also those of massless Dirac fermions but have a group velocity varying \emph{anisotropically} depending upon the propagation direction.
	For a particular energy value (say $ E_0 $) the equation of the contour is
	\begin{equation}
		\frac{1}{\left(E_0 - \hbar v_F m G_0 /2\right)^2/\hbar^2\,v_F^2}\left( p_x^2 + \frac{p_y^2}{1/\abs{f_m}^2} \right) = 1
		\label{eqn:1dcontoureqn}
	\end{equation}
	Correspondingly, the density of states can be calculated using (\ref{eqn:DOS}) as
	\begin{equation}
		g_s(E) = \frac{1}{s\hbar v_F}\oint_{C_s(E)} \frac{\dd{l}}{(2\pi)^2} \sqrt{\frac{p_x^2 + \abs{f_m}^2\,p_y^2}{p_x^2 + \abs{f_m}^4\,p_y^2}} \label{eqn:gendos}
	\end{equation}
	where $ C_s(E) $ is the constant energy contour at $ E $. The behaviour of massless Dirac fermions at the zone boundary of an SBZ, as seen from Eqs. (\ref{eqn:gendisprel}), (\ref{eqn:genfermivel}), (\ref{eqn:gendos}), 
	is remarkably different from a non-relativistic fermion. To see that, we apply the above theory to the case of 
	the two simple $1D$-potentials, a cosine potential that has a single Fourier component, and a $1D$ Kronig-Penney-like potential which is the fundamental potential for the electrons inside the crystal \cite{kittel2004introduction, Ashcroft}. 
	
	\begin{table*}
	\centering
	\caption{\label{table:slqtys} Tabulates the different quantities with their corresponding expression for the 1D cosine potential and 1D KP potential.}
	\begin{ruledtabular}
	\begin{tabular}{ccc}
		Quantity & Cosine Potential & KP Potential\\
		\hline
		$ V(x) $ & $ V_{0}\cos(G_{0}x) $ &$\begin{cases}
			0 \qquad -L/2 \leq x < -w/2 \\
			U_0 \qquad -w/2 \leq x \leq w/2 \\
			0   \qquad \quad   w/2 < x \leq L/2
		\end{cases} $ \\
		$ \alpha(x) $ & $ \frac{2\,V_0\,\sin(G_0\,x)}{\hbar\,v_F\,G_0} $ & $ \begin{cases}
			1 \qquad \qquad -L/2 \leq x < -w/2 \\
			e^{i\frac{2U_0\,x}{\hbar\,v_F}} \qquad -w/2 \leq x \leq w/2 \\
			1 \qquad \qquad \quad w/2 < x \leq L/2
		\end{cases} $ \\
		$ f_m $ & $ J_m\left(\frac{2V_0}{\hbar\,v_F\,G_0}\right) $ & $ \begin{cases}
			\frac{\hbar\,v_F}{U_0\,L}\sin(\frac{U_0\,w}{\hbar\,v_F}), \quad m = 0 \\
			\frac{1}{\frac{U_0L}{\hbar\,v_F} - m\pi} \sin(\frac{U_0\,w}{\hbar\,v_F} - \frac{m\,\pi\,w}{L}) - \frac{1}{\pi m} \sin(\frac{m\,w\,\pi}{L}), \quad m \ne 0
		\end{cases} $ \\
	\end{tabular}
	\end{ruledtabular}
	\end{table*}
	
	It was already demonstrated that the application of the one-dimensional (1D) superlattice potential leads to the emergence of new Dirac points with a strong anisotropy in the electron velocity around the Dirac point \cite{Park2008PRL,Park2008nature,Brey2009}. 
	The simplest 1D potential with only two Fourier components is the cosine potential
	\begin{equation}
		V(x) = V_0\,\cos(G_0\,x)
	\end{equation}
	where the strength of the potential $ V_0 $ is much smaller than the energy bandwidth of the graphene $ \pi $ orbitals. Periodicity demands, $V_0\,\cos(G_0\,x + G_0\,L) = V_0\,\cos(G_0 x)  $, which gives $ G_0 = 2\pi/L $.
	Now the period $ L $ is much larger than the graphene lattice constant, $\left( L >> a \right)$, and hence SBZ is also referred to as mini-zone (MZ) in the literature \cite{Killi2011}.
	The different quantities which characterise the quasiparticles at the SBZ or MZ are evaluated for the cosine potential, and are listed in Table-(\ref{table:slqtys}). The band structure is plotted in Fig.\ref{fig:1dcosbnd}(a). The energy dispersion for a Dirac point in the presence of cosine potential becomes
	\begin{equation}
		E_s(m,\bm{p}) = s\hbar\,v_F\sqrt{p_x^2 + \abs{J_m\left(\frac{2V_0}{\hbar\,v_F\,G_0}\right)}^2p_y^2} + \frac{\hbar\,v_F\,m\,G_0}{2}
	\end{equation} 
	The extra Dirac points known as \emph{superlattice Dirac points} emerges at $ \bm{G} = \pm G_0/2\,\hat{\bm{x}} $ as shown in the contour plots Fig.\ref{fig:1dcosbnd}(b). The equi-energy contours corresponding to the Dirac point at $ \left(0,0\right) $ have been plotted in the Fig.\ref{fig4:slgslvel}(a) and the analytical expression describing the contour path is
	\begin{equation}
		p_x^2 + \abs{J_m\left(\frac{2V_0}{\hbar\,v_F\,G_0}\right)}^2p_y^2 = \frac{1}{\hbar^2 v_F^2} \left(E - \frac{\hbar\,v_F\,m\,G_0}{2}\right)^2
	\end{equation}
	The ellipticity of these contours rises as $ m $ grows, as shown in Fig.\ref{fig4:slgslvel}(a). Additionally, it displays the group velocity $ \bm{v}_{g} $ at a general point: $\left(p_x, p_y\right)$. The Fermi velocity reduces as the potential $ \left( V_0 \right) $ strength increases, and the other new Dirac cone emerges along the $y$ direction. It is the chirality of the massless Dirac Fermions that protects the band crossing points at the zone boundaries $ \pm G_{0}/2 $ \emph{vs.} the case of parabolic bands of the non-relativistic fermions \cite{kittel2004introduction} where a gap is opened at the zone boundaries due to the periodic external potential. The modification in the DOS due to the $ V(x) $ is also given in Table-(\ref{table:slqtys}). The DOS for different strengths of potential is drawn in Fig.\ref{fig:1dcosbnd}(c) and shows that the VHS on both sides approaches the Fermi level at $ E = 0 $ as the potential strength is increased. Experimentally, such a one-dimensional SL has been realized by applying gate voltages in SLG which leads to a potential barrier of a certain height that can be made periodic \cite{Yankowitz2012,Ponomarenko2011,Pleti2009}. In this direction, a tunable SL has also been proposed where the SL can be tuned by changing the barrier height using a combination of two gates \cite{Dubey2013}.
	
	For a cosine potential, only the two Fourier coefficients are non-zero, leading to the renormalization of one velocity component. A more general model potential is a simple Kr\"{o}nig-Penney (KP) like potential and we shall briefly recapitulate its effect on the band-structure of massless Dirac fermions. As we know, the KP model remains the paradigm for the demonstration of energy bands in periodic crystals. Various KP-like models have been considered $\mathit{e.g.,}$ Magnetic KP model in MLG \cite{Martino2007,Masir2008,Ghosh2009,Ramezani2009}, Dirac electrons in a KP potential \cite{Barbier2009slg}, the KP model in BLG \cite{Barbier2010}. Here, we consider the Dirac electrons in a symmetric KP-like square-well potential $ V(x) $ with height $ U_0 $ and width $ w $ of the potential well that is periodic along the $x$-direction with periodicity $ L $ such that $ L >> a_0 $ as shown in Fig.\ref{fig:1dkpband}(a).

	The vector $ \bm{G}_{m} = m\,G_0~\hat{\bm{x}} $ $ \left(G_{0} = 2\pi/L\right) $ denotes the 1D reciprocal lattice vectors of the \emph{superlattice} Brillouin zone which results from the periodic potential $ V(x) $. Since $ \bmk $ is measured from the Dirac point, the boundaries of the first SBZ lie at $ k_x = \pm~\frac{\pi}{L} $. The Fig.\ref{fig:1dkpband}(b) shows the $2D$ band structure of the Dirac cone centred at the $ \left\{0,0\right\} $ where it shows its deviation as the strength of the potential is increased. The corresponding $3D$ band structure is shown in Fig.\ref{fig:1dkpband}(c), where the new Dirac points emerge at the boundaries $\left( \pm \pi/L \right)$ of the first SBZ ( or MZ). The equienergy contours are plotted in Fig.\ref{fig:1dkpband}(d). The modification of the energy contours is also shown in Fig.\ref{fig4:slgslvel}(b) for different values of $ m $. The new Dirac points lead to the new saddle points in the band structure and, therefore, the VHS in the density of states. In this case, the non-zero Fourier component changes the Fermi level with the change in the strength of the potential. In contrast to the SLG without the potential, the periodic potential causes the VHS to be closer to the Fermi level. Therefore, it provides one of the ways of shifting the singularities near the Fermi level in an otherwise distinct structure.
	
	\begin{figure*}
		\centering
		\includegraphics[scale=0.48]{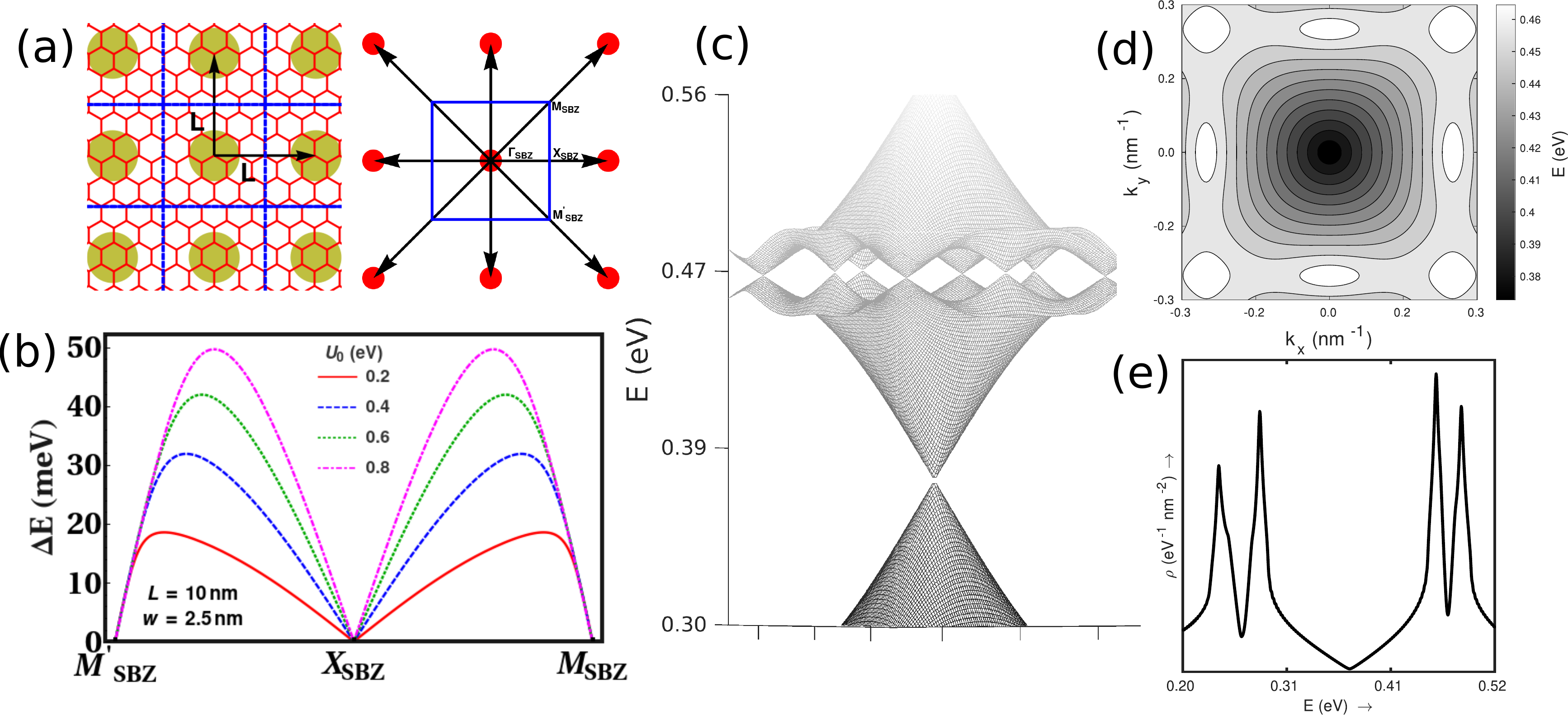}
		\caption{(a) Two-dimensional (2D) graphene SL formed by a muffin-tin \emph{square} potential periodic along the x- and y-direction with translation vector $ \bm{T}(n_1,n_2) = n_1\,L\,\hat{\bm{x}} + n_2\,L\,\hat{\bm{y}} $. The height of the potential barrier is $ U_0 $ inside the circular regions of radius $ w $ and zero outside. Adjacent right figure shows the corresponding reciprocal space where the blue square indicating the superlattice Brillouin zone (SBZ). (b) The difference of the energy ($ \Delta E $) between the first two conduction bands above the Fermi level is plotted against the path $ M'_{SBZ}, X_{SBZ}, M_{SBZ} $. (c) It shows the 3D view of the first two conduction bands with the new set of eight Dirac points occuring at $ \bm{G} = \pm \bm{G}_{1}/2 $, $ \bm{G} = \pm \bm{G}_{2}/2 $ and at the corners of the SBZ. (d) It shows the contour plots corresponding to the first conduction band with elliptical contours indicating the modified Fermi velocity of new Dirac points. (e) It shows the density of states with two symmetrically placed dips.}
		\label{fig:2dsqmtpot}
	\end{figure*}

	\begin{figure*}
		\centering
		\includegraphics[scale=0.48]{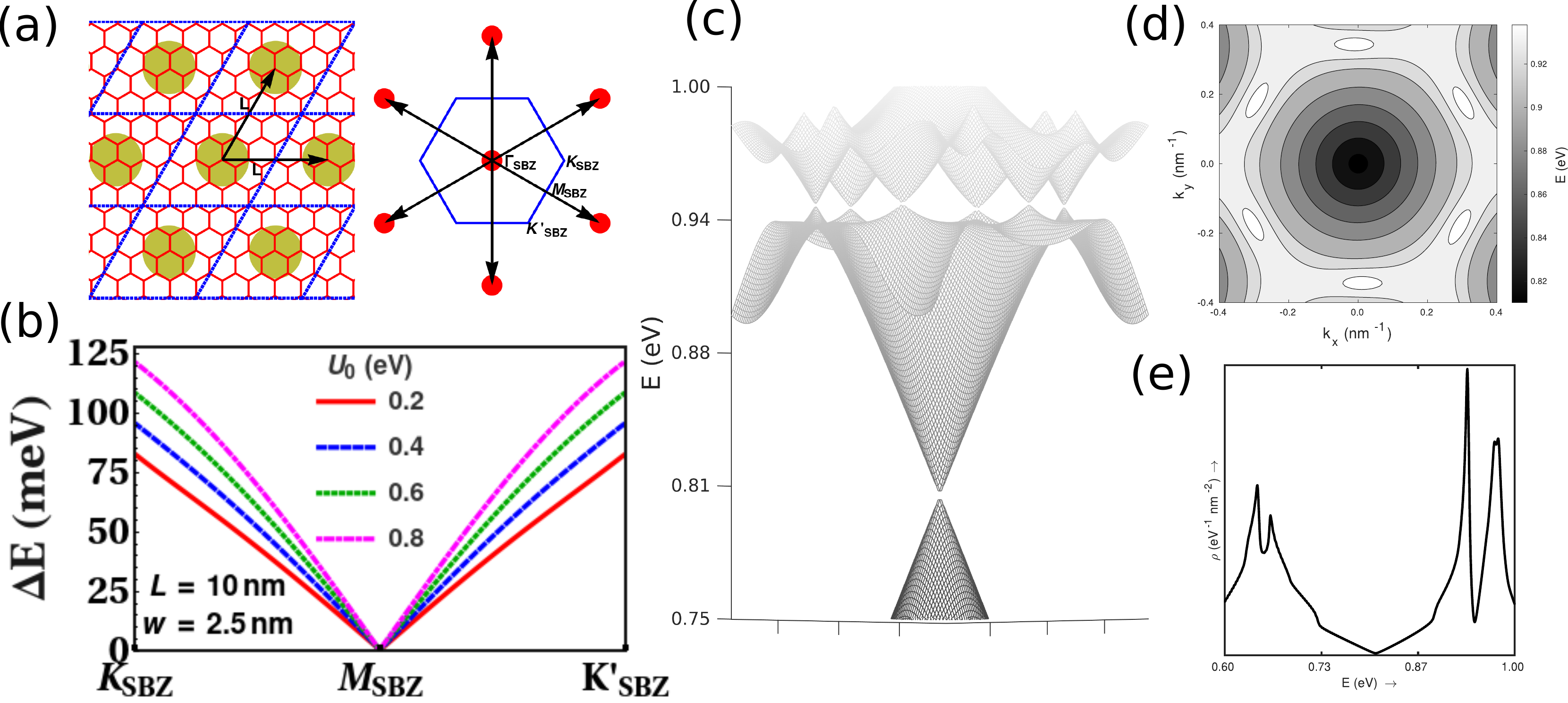}
		\caption{(a) Two-dimensional (2D) graphene SL formed by a muffin-tin \emph{hexagonal} potential periodic along the x- and y-direction with translation vector $ \bm{T}(n_1,n_2) = \left(n_1 + n_2/2\right) L\,\hat{\bm{x}} + \left(\sqrt{3}n_2/2\right)L\,\hat{\bm{y}} $. The height of the potential barrier is $ U_0 $ inside the circular regions of radius $ w $ and zero outside. Adjacent right figure shows the corresponding reciprocal space where the blue hexagon marks the superlattice first Brillouin zone (SBZ). (b) The difference of the energy ($ \Delta E $) between the first two conduction bands above the Fermi level is plotted against the path $ K_{SBZ}, M_{SBZ}, K'_{SBZ} $. (c) It shows the 3D view of the of the first two conduction bands with the new set of six Dirac points  occuring at $ \bm{G} = \pm \bm{G}_{1}/2 $, $ \bm{G} = \pm \bm{G}_{2}/2 $ and $ \bm{G} = \pm \left(\bm{G}_{1}+\bm{G}_{2}\right)/2 $.(d) It shows the contour plots corresponding to the first conduction band with elliptical contours indicating the modified Fermi velocity of new Dirac points. (e) It shows the density of states with two symmetrically placed dips.}
		\label{fig:2dhexmtpot}
	\end{figure*}
	\subsubsection{Two-dimensional (2D) potential}\label{2DSL}
	To understand the connection between van der Waals (VDW) heterostructures which will be introduced in section \ref{sec:section2}, and the non-trivial effect of superlattices on massless Dirac fermion that were analysed in section \ref{SLZ} and \ref{1DSL}, it is important to study the effect of two-dimensional (2D) potential on the massless Dirac fermions. 
	Here we shall briefly review the 2D muffin-tin potentials numerically\cite{Park2008nature,Park2008PRL,Burset2011,Ortix2012} and, consequently, their impact on the electronic properties and the anisotropic nature of the Dirac fermions. For a general $2D$ potential $ V = V(\bm{r}) $, the Hamiltonian reads
	\begin{equation}
		H = i\hbar\,v_F\,\bm{\sigma}\vdot\bm{\nabla} + \mathcal{I}_2\,V(\bmr)
		\label{eqn:genHam2D}
	\end{equation}
	The 2D potential is periodic along the x- and y-direction, $ V(\bmr + n_1\bm{L}_1 + n_2\,\bm{L}_2) = V(\bmr) $, where $ n_1, n_2 \in \mathbb{Z} $ and $ \bm{L}_1 $ and $ \bm{L}_2 $ are the basis vectors. Thus the Fourier expansion of $ V(\bmr) $ can be written as,
	\begin{equation}
		V(\bmr) = \sum_{\bm{G}} V_{\bm{G}}~e^{i~\bm{G}\cdot\bmr}
		\label{eqn:fouexp2d}
	\end{equation}
	where $ \bm{G} = m_1~\bm{G}_1 + m_2~\bm{G}_2 $ with $ m_1,m_2 \in \mathbb{Z} $ are the reciprocal lattice vectors of the reciprocal space associated with the periodicity of the applied potential. A 2D potential that represents an isolated ion within a sphere of radius $r_0$ around each lattice point, and taken to be constant elsewhere \cite{Liberman1967} is the muffin-tin potential. For such potentials the complex Fourier coefficients in (\ref{eqn:fouexp2d}) are given by
	\begin{equation}
		V_{\bm{G}} = \frac{U_0}{L^2} \int_{\mathcal{R}} d\bmr~e^{i\left(m_1\,\bm{G}_1 + m_2\,\bm{G}_2\right)\cdot\bmr} 
	\end{equation}
	where the region $ \mathcal{R} $ is the disk of radius $ w $ of the potential barrier at each lattice site. For a square muffin-tin potential as shown in Fig.\ref{fig:2dsqmtpot}(a), the primitive vectors of the superlattice are $ \bm{L}_1 = L\,\hat{\bm{x}}, \bm{L}_2 = L\,\hat{\bm{y}} $ with $ L >> a_0 $ and the reciprocal lattice vectors are $ \bm{G}_1 = 2\pi/L \, \hat{\bm{x}} \equiv G_0 \, \hat{\bm{x}}, \bm{G}_2 = 2\pi/L \, \hat{\bm{y}} \equiv G_0 \, \hat{\bm{y}}$ also shown in the inset of Fig.\ref{fig:2dsqmtpot}(a). The difference in energies of the first two conduction bands at different heights $ U_{0} $ of the potential is shown in Fig.\ref{fig:2dsqmtpot}(b), which clearly shows the existence of a degenerate point at $ M_{\text{SBZ}} $. The full 3D band structure is shown in Fig.\ref{fig:2dsqmtpot}(c), and a contour plot of the first conduction band is shown in Fig.\ref{fig:2dsqmtpot}(d). The contour plot demonstrates that new Dirac points emerge at the midpoint of the boundaries of SBZ in addition to the Dirac points at the corners of the SBZ. The height of the potential barrier $U_0$ controls the band velocity for both sets of the newly emerged Dirac points. The DOS is plotted in Fig.\ref{fig:2dsqmtpot}(e), where the two dips symmetrically placed about the Fermi level characterize the emergence of new Dirac points.
		
	For a hexagonal potential as shown in Fig.(\ref{fig:2dhexmtpot})a, the real space lattice vectors are given by $ \bm{L}_1 = L \, \hat{\bm{x}}, \bm{L}_2 = L\,\cos(\pi/3)\,\hat{\bm{x}} + L\,\sin(\pi/3)\,\hat{\bm{y}} $, and the reciprocal lattice vectors of the SBZ are $ \bm{G}_1 = 2\pi/L \, \hat{\bm{x}} \equiv G_0 \, \hat{\bm{x}}, \bm{G}_2 = 2\pi/L \, \hat{\bm{y}} \equiv G_0 \, \hat{\bm{y}}$ shown in inset of Fig.\ref{fig:2dhexmtpot}(a). Here too the difference between energies of the first two conduction bands is shown in Fig.\ref{fig:2dhexmtpot}(b). In this case, the new Dirac cone emerges at the corners of SBZ (a hexagon) as shown in Fig.\ref{fig:2dhexmtpot}(c)-Fig.\ref{fig:2dhexmtpot}(d). The electronic states near the SBZ corners are also those of the massless Dirac fermions with their velocity varying \emph{anisotropically}. Ref. \cite{Park2008,Park2008PRL} also showed that there exists an energy window in which only the states of newly generated Dirac fermions can be realized in triangular graphene superlattices (TGSs) where the graphene is subjected to a muffin-tin type of periodic potential. The DOS is shown in Fig.\ref{fig:2dhexmtpot}(e), where the dip on the conduction band side is stronger than on the valence band side.
	
	The behaviour of massless Dirac fermions in graphene under superlattice potentials is so peculiar that it alters the band structure of SLG with extra Dirac points, renormalises the Fermi velocity, and pushes the VHS closer to the Fermi level. They are therefore promising techniques to push graphene physics from a one-body effect-dominated weakly correlated regime to a many-body strong correlation regime. A large number of superlattices are created by stacking or other methods (see section \ref{sec:section2}) 2D layered materials over one another. These layers hold one other by van der Waals forces \cite{Geim2013}. In the following section, we discuss such van der Waals hetero-structures.

	\section{Van der Waals (VDW) heterostructures (a tool for building strong correlations in Dirac systems)} \label{sec:section2}
	In recent years, the van der Waals heterostructures \cite{Geim2013} have attracted a lot of interest due to the ability to tune their electronic properties by changing one or more parameters. In a 2D-material-based superlattice, the interlayer interaction between the layers mediated by van der Waals forces constitutes a key parameter to tune the global properties of the superlattice. Following the work of Y.K.Ryu {\it et al.}\cite{Ryu2019}, one can provide a classification of such 2D-material-based superlattices, which proceeds as the following:
	
	\begin{itemize}[itemsep=0pt]
		\item \textit{Vertical stacking} of dissimilar 2D layers. The basic principle in making a heterostructure is to take, for example, a monolayer, put it on top of another monolayer or few-layer crystal, add another 2D crystal and so on \cite{Dean2010,Ponomarenko2011,Dean2012,Yankowitz2012,Geim2013,Xu2013,Butler2013,Kim2016,Debnath2021}. The resulting stack represents an artificial material assembled in a chosen sequence. Strong covalent bonds provide in-plane stability of 2D crystals, whereas relatively weak, \emph{van-der-Waals-like} forces are sufficient to keep the stack together. Therefore, these structures are referred to as \emph{Van der Waals} heterostructures. Many devices are made by stacking different 2D crystals on top of each other such as isolated monolayers and few-layer crystals of hexagonal boron nitride (hBN), molybdenum disulphide ($ MoS_2 $), other dichalcogenides and layered oxides \cite{Novoselov2005pnas,Xu2013}. Among the above 2D crystals, graphene is likely to be a common component in future van der Waals heterostructures and devices because of its high mechanical strength, and crystalline and electronic quality \cite{Dean2012}. In vertical stacking, the periodic potential is modulated principally by the chemical composition of the layers and/or the lattice mismatch between them.
		
		\item \textit{Intercalation} Multi-layered 2D materials subjected to an intercalation process \cite{Yao2014,Guo2017}. The charge transfer that occurs between the host flakes and the atoms or molecules, the re-structuring of the functionalized layers following the intercalation of the host species, and the chemical composition of the layers can all affect the periodic potential in this kind of superlattice.
		
		\item \textit{Moir\'e Patterns} Moir\'e structures by twisting two 2D material-based layers. When two equivalent (e.g. graphene bilayer) or dissimilar (e.g. graphene/h-BN) adjacent layers are rotated under certain angles, the lattice mismatch and/or the atom rearrangement gives rise to moir\'{e}-pattern superlattices with new lattice periodicities. The twisted bilayer graphene (TBLG) shows the flat bands \cite{Geim2007pt} and is a promising material where the electronic properties are tunable by the relative angle (or \emph{twist} angle) between the layers \cite{Santos2007,Bistritzer2010,Li2010,Suarez2010,Bistritzer2011,Santos2012,Kim2017,Tong2017,Cao2018one,Cao2018two,Xu2018,WuAHM2018,Ochi2018,Tarnopolsky2019,Choi2019,Lian2019,Nuckolls2020,Balents2020,Ledwith2020,Andrei2020,Liu2020,Shen2020,Das2021,Das2021,Mahapatra2017,Chandan2018,Bern12021,Song2021,Bern32021,Lian2021,Bern52021,FangXie2021,Mahapatra2022,Sinha2022,Bhowmik2022,Valagiannaopoulos2022}.
	\end{itemize}
	
	Other types of superlattices are created by strain engineering \cite{Roldan2015,Amorim2016,Lin2016} where a periodic potential can be produced by subjecting a 2D material flake to periodic tensile/compressive strains or by introducing a periodic, controlled mismatch between the lattices of two dissimilar 2D materials. Synthetic superlattices are defined by lithography \cite{Xilin2016}, where the periodic potential can be induced in a 2D material by placing it on a patterned topography or by patterning the material itself with a lithographic technique.
	Each type of superlattice has potential applications. Some of the major discoveries in VDW heterostructures are evidence for moir{\'e} excitons \cite{Tran2019}, resonantly hybridized excitons in moir{\'e} superlattices in VDW heterostructures \cite{Alexeev2019}, giant tunneling magnetoresistance in spin-filter VDW heterostructures \cite{TianSong2018}, intrinsic quantized anomalous Hall effect in a moir\'e heterostructure \cite{Serlin2020}, massive Dirac fermions and the Hofstadter butterfly in a VDW Heterostructure \cite{Hunt2013}, harnessing the photonic local density of states in graphene moir\'e superlattices\cite{Kort2021}, excitonic linewidth approaching the homogeneous limit in ${\mathrm{MoS}}_{2}$-based VDW heterostructures \cite{Cadiz2017}, picosecond photoresponse, \cite{Massicotte2016}, twist-controlled resonant tunnelling in graphene/BN/graphene heterostructures \cite{Mishchenko2014} and optimising graphene's visibility in VDW heterostructures \cite{Menon2019}. Most of the VDW heterostructures are fabricated by encapsulating graphene with other 2D materials \cite{Kretinin2014,Kim2016}.
	
	In this review, since we are primarily concerned with the moir\'e pattern and TBLG, we restrict ourselves to the moir\'e superlattices. In order to understand the electronic properties of VDW heterostructures giving rise to the moir\'e pattern, we first discuss some geometrical properties of the moir\'e systems.
	
	\subsection{Moir\'e superlattices}\label{ML}
	The moir\'e phenomenon exists in disparate fields of arts and science \cite{Amidror}. In condensed matter physics and material science, the advent of scanning tunnelling microscopy (STM) enabled the observation of super-periodic patterns \cite{Tersoff1985,Dean2013}. In graphite, superlattices were discovered in STM studies of its surface \cite{Albrecht1988,Andrei2012} and correctly interpreted as moir\'e patterns caused by misorientation between subsurface graphene layers. For a theoretical interpretation of the moir\'e pattern, one can see refs.\cite{Oster1964,Isaac2009}. When two periodic layers are stacked over one another a periodic pattern (\emph{moir\'e} pattern) emerges if (a) the periodicity of the two layers differs with respect to each other such as graphene over hexagonal Boron Nitride \cite{Yankowitz2012}, and/or (b) one of the layers is rotated and/or translated with respect to another layer such as TBLG \cite{Santos2007}. This moir\'e pattern results because of the interference between the periodicity of the individual layers, and the unit cell of the moir\'e pattern is referred to as the moir\'e cell. In general, if the two sets of real space primitive vectors $ \left\{\bm{a}^{(1)}_1,\bm{a}^{(1)}_2\right\} $ and $ \left\{\bm{a}^{(2)}_1,\bm{a}^{(2)}_2\right\} $ of top and bottom layer, respectively, are connected through a linear transformation $ \mathcal{L} $ \cite{Koshino2012,Koshino2015},
	 \begin{equation}
		 \bm{a}^{(2)}_{j} = \mathcal{L}\,\bm{a}^{(1)}_{j} \quad \forall \quad j = 1,2 \label{MLinT1}
	 \end{equation}
	 then, the primitive vectors of the moir\'e superlattice are given as
	 \begin{equation}
		 \bm{a}^{M}_j = \left[\mathcal{I}_2 - \mathcal{L}^{-1}\right]^{-1}\bm{a}^{(1)}_{j} \quad \forall \quad j = 1,2 \label{MLinT2}
	 \end{equation}
	 where the superscript 'M' stands for the lattice vectors of the moir\'e pattern. The moir\'e periodicity or wavelength is given by $ L_{M} = \abs{\bm{a}^{M}_{j}} $. Corresponding to the moire superlattice, the reciprocal space primitive vectors $ \left\{\bm{b}^{M}_1, \bm{b}^{M}_2\right\} $ are defined as
	 \begin{equation}
		 \bm{b}^{M}_j = \left[\mathcal{I}_2 - \left(\mathcal{L}^{-1}\right)^{\dagger}\right]\bm{b}^{(1)}_{j} \quad \forall \quad j = 1,2
		 \label{eqn:moirerecvec}
	 \end{equation}
	 such that $ \bm{a}^{M}_{i}\vdot\bm{b}^{M}_{j} = 2\pi\delta_{ij} $.  
	 
	 We now discuss the moir\'e system consisting of graphene over the well-known hexagonal boron nitride (hBN) substrate. Indeed, this moir\'e structure has provided interesting experimental results well before TBLG was experimentally explored in detail. The hBN offers minimal interlayer coupling to the overlaid graphene sheet, and therefore it was demonstrated that the effect of this substrate could be modelled as a periodic external potential ( superlattice potential) to the massless Dirac fermions \cite{Yankowitz2012,Yankowitz2014}. In the following discussion, we look at the effect of this periodic substrate potential on the quasiparticles of SLG before moving to the TBLG where the interlayer coupling among the two layers strengthens as the twist angle is decreased.
	
	\begin{figure*}
		\centering
		\includegraphics[scale=0.45]{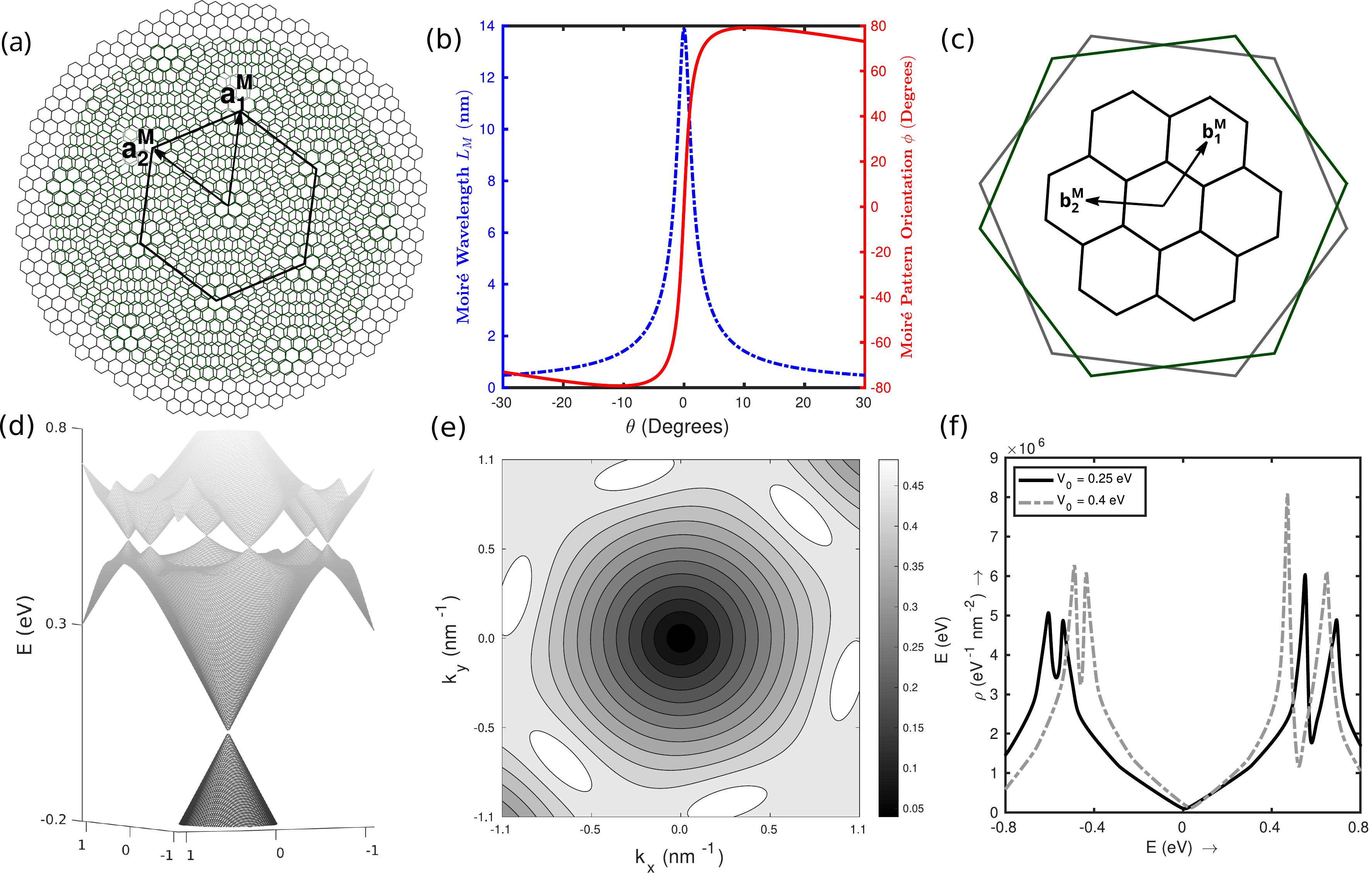}
		\caption{(a) The schematic of the moir\'e pattern in G/BN at the rotation angle of $ 8^{\circ} $. (b) The left y-axis is showing variation of moir\'e wavelength with the relative rotation angle $ \theta $ and the right y-axis shows the relative misorientation of moir\'e pattern with graphene layer. (c) It shows the reciprocal space BZ of SLG (Dark green), hBN (Gray) and the BZ corresponding to moir\'e primitive vectors. (d) The 3D band structure of G/BN and clearly shows the superlattice Dirac points in addition to the Dirac point at zero energy. (e) A contour plot where the white elliptical contours are clearly showing the superlattice Dirac points with modified Fermi velocity perpendicular to the reciprocal lattice vectors $ \bm{b}^{M}_{1} $,$ \bm{b}^{M}_{1} $ and $ \bm{b}^{M}_{1} + \bm{b}^{M}_{2} $. (f) It shows the DOS plot for two different strengths of potentials $ V_{0} = 0.25~\si{\electronvolt} $ and $ V_{0} = 0.4~\si{\electronvolt} $ at the angle $ \theta = \ang{4} $. The two symmetrically placed dips on either side of the Fermi level indicates the occurrence of superlattice Dirac points.}
		\label{fig:hBN1}
	\end{figure*}
	\subsection{Graphene on hexagonal Boron Nitride (G/BN)}\label{GhBN}
	SLG over hBN can be prepared either by transfer \cite{Dean2010} or growth \cite{Yang2013} techniques. Experimentally, it has been found that the SLG on boron nitride (BN) is very flat and that its low-energy electronic states are often very weakly perturbed by the substrate \cite{Dean2010,Xue2011,Decker2011}. However, these fabrication procedures for creating the graphene on BN result in a random rotational orientation between the graphene and hBN lattices. Moreover, a moir\'e pattern is formed both in perfectly aligned  ($\theta = 0$) due to the slightly larger lattice constant of hBN than graphene \cite{Giovannetti2007}, and misaligned ($\theta \ne 0$) configurations \cite{Yankowitz2012,Wallbank2015} where $ \theta $ is the angle through which graphene is rotated with respect to the BN layer. The underlying moir\'e potential due to hBN forms a large moir\'e superlattice, consequently resulting in a mini Brillouin zone (MBZ) which gives rise to minibands in the electronic spectrum \cite{Wallbank2013}. Here, we review the problem to show the emerging superlattice Dirac points due to the moir\'e potential. As we pointed out already, in contrast to the problem of a free electron in a periodic potential where a band gap opens at the corners of the first Brillouin zone (FBZ), the chiral nature of quasiparticles in graphene prevents the opening of band gap at the corners of the superlattice BZ \cite{Yankowitz2012,Yankowitz2014}. This will now be verified in a real VDW heterostructure that has a moir\'e pattern. To this end we work out the moir\'e pattern of G/BN primitive vectors. The lattice constant of BN is larger than that of SLG, and the graphene layer is rotated counter-clockwise by $ \theta $ with respect to the hBN layer. Therefore, following (\ref{MLinT1}) the primitive lattice vectors $ \left\{\bm{a}'_{j}\right\} $ of BN and $ \left\{\bm{a}_{j}\right\} $ of SLG are then related through the linear transformation \cite{Moon2014hbn,Koshino2015iop}
	\begin{equation}
		\bm{a}'_{j} = \left[\left(1+\epsilon\right)\mathcal{R}(-\theta)\right]\bm{a}_{j} \quad \forall \quad j = 1,2
	\end{equation}
	where $ \epsilon = 0.0172 $ is the mismatch in the lattice constants, and $\mathcal{R} $ is the 2D rotation matrix about the z-axis. The sublattice A atom (boron atom for hBN) in a unit cell of SLG (hBN) is located at $ 2\left(\bm{a}_1 + \bm{a}_2 \right)/3 $ $ \left(2\left(\bm{a}'_1 + \bm{a}'_2 \right)/3\right) $ of the cell, and the positions of the sublattice B atoms (nitrogen atom for hBN) in the unit cell are $ \left(\bm{a}_1 + \bm{a}_2 \right)/3 $ and $ \left(\bm{a}'_1 + \bm{a}'_2\right)/3 $ for SLG and hBN, respectively. The primitive lattice vectors of the moir\'e superlattice can be obtained from (\ref{MLinT2}) as 
	\begin{equation}
		\bm{a}^{M}_{j} = \left[\mathcal{I}_2 - \frac{1}{1+\epsilon}\mathcal{R}(\theta)\right]^{-1}\bm{a}_{j} \quad \forall \quad j = 1,2
	\end{equation}
	The corresponding moir\'e pattern is shown in Fig.\ref{fig:hBN1}(a) at an angle $ \theta = \ang{8} $ with the moir\'e primitive vectors. The moir\'e wavelength $ L_{M} $ is then given as $L_{M}= \abs{\bm{a}_1^{M}} = \abs{\bm{a}_2^{M}}$, and comes out to be
	\bse
	\begin{equation}
		L_{M} = \frac{(1+\epsilon)\,a}{\sqrt{2\,(1+\epsilon)\,(1-\cos \theta)+\epsilon^2}}
	\end{equation}
	where $ a $ is the lattice constant of graphene. The moir\'e wavelength $ L_{M} $ is plotted against the orientation $ \phi $ in  Fig.\ref{fig:hBN1}(b). The moir\'e pattern is also misoriented by an angle $ \phi $ with respect to the graphene layer and depends on the relative rotation angle $ \theta $ through the expression
	\begin{equation}
		\tan \phi = \frac{\sin \theta}{(1+\epsilon)-\cos \theta}
	\end{equation}
	\ese
	The orientation of the moir\'e pattern with respect to the graphene layer is also shown in Fig.\ref{fig:hBN1}(b). Using (\ref{eqn:moirerecvec}) the reciprocal space primitive lattice vectors can be obtained as 
	\begin{equation}
		\bm{b}_{j}^{M} = \left[\mathcal{I}_2 - \frac{1}{1+\epsilon}\mathcal{R}(-\theta)\right]\bm{b}_{j} \quad \forall \quad j = 1,2, 
	\end{equation}
	where $ \left\{\bm{b}_{j}\right\} $ are the reciprocal lattice vectors of graphene sheet. The BZ formed by the reciprocal lattice vectors $ \bm{b}^{M}_{j} $ is shown in Fig.\ref{fig:hBN1}(c) at an angle $ \ang{8} $.
	
	Since we are interested in the emerging superlattice Dirac points in the energy spectrum, the influence of the hBN can be modelled by an effective periodic potential with the exact symmetry as the observed moir\'e pattern \cite{Xue2011,Decker2011,Marchini2007,Parga2008}, and therefore the Hamiltonian is given as \cite{Yankowitz2012}
	\begin{equation}
		H = i\hbar\,v_F\,\bm{\sigma}\cdot\bm{\nabla} + \mathcal{I}_2\,V_{0}\,\sum_{\alpha =1,2,3} \cos(\bm{b}^{M}_{\alpha}\cdot\bm{r})
	\end{equation}
	where $ \bm{b}^{M}_{3} = \bm{b}^{M}_1 + \bm{b}^{M}_2 $ and $ V_{0} $ is the strength of the potential. The band structure is plotted in Fig.\ref{fig:hBN1}(d) with a corresponding contour plot in Fig.\ref{fig:hBN1}(e). The superlattice Dirac points emerge at energies $ E = \pm \hbar v_F \abs{\bm{b}^{M}_{\alpha}}/2 $, where $ \bm{b}^{M}_1 $ is a reciprocal lattice vector in SBZ. Here too the VHS comes nearer to the Fermi level due to the SL Dirac points, as can be seen in Fig.\ref{fig:hBN1}(f). Here, we discussed the external or substrate potential as weak perturbations to the massless Dirac fermions, and the electron-electron interactions are entirely ignored. The electron-electron interactions in graphene superlattices are considered in \cite{Song2013,Woessner2015}. SLG over hBN is discussed to demonstrate the emerging superlattice Dirac points in the electronic spectrum and their impact on the VHS. For a more in-depth study, we refer the reader to recent reviews on SLG over hBN \cite{Yankowitz2014,Wang2017hbn,Yankowitz2019}. In the subsequent section \ref{secTBLG}, we shall discuss the primary topic of this review, which is twisted bilayer graphene (TBLG).

	\begin{figure*}
		\centering
		\includegraphics[scale=0.45]{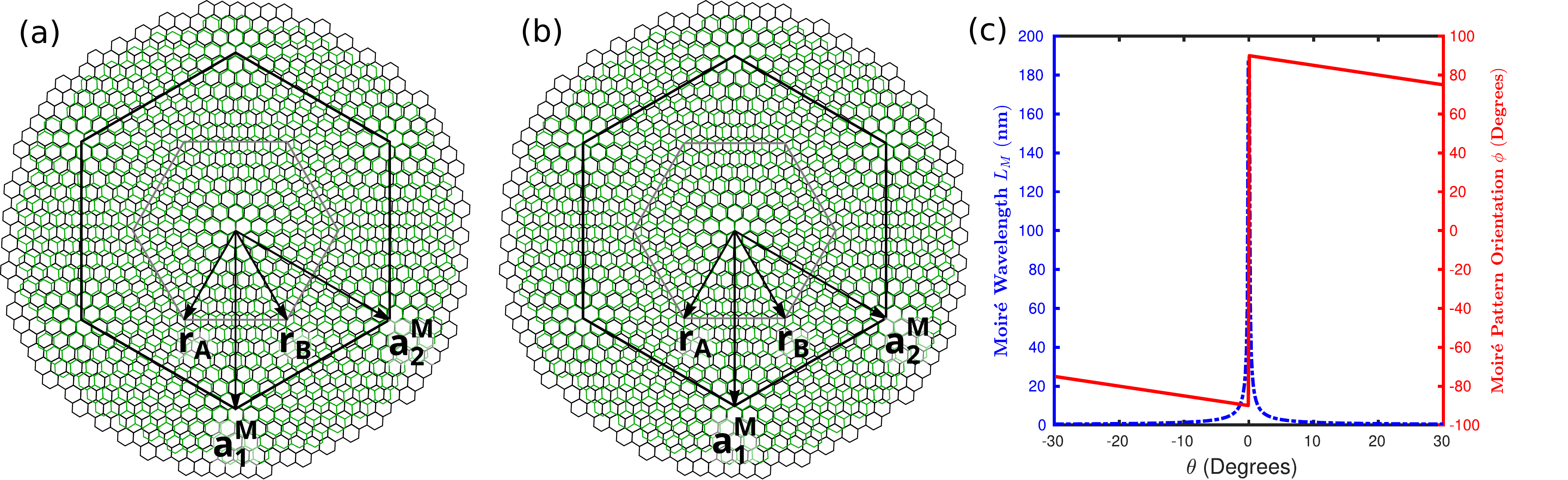}
		\caption{\label{fig:TBLG1} A real space moir\'e-pattern in TBLG at (a) $ \theta = \ang{5} $ and (b) $ \theta = \ang{5.086} $. The moir\'e primitive vectors $ \bm{a}^{M}_{1} $ and $ \bm{a}^{M}_{2} $ are shown by black arrows. The black hexagon is formed by connecting AA-rich regions and forms the moir\'e unit cell, and $ L_M = \abs{\bm{a}^{M}_{1}} = \abs{\bm{a}^{M}_{2}} $ is the corresponding side length of the hexagon referred to as the \emph{moir\'e period} or \emph{moir\'e wavelength}. The two vectors $ \bm{r}_{A} $ and $ \bm{r}_{B} $ are connecting the centre of the moir\'e cell to the local AB and BA-stacking regions, respectively. (c) The variation of moir\'e period $ L_M $ with the twist angle $ \theta $.}
	\end{figure*}

\section{Twisted bilayer graphene}\label{secTBLG}
In twisted bilayer graphene, two layers of graphene are stacked over one another as in $AB$ bilayer graphene, followed by a relative rotation which leads to a moir\'e pattern as shown in Fig.\ref{fig:TBLG1}(a). The localized "AA"- and "AB"-rich regions give rise to alternating bright and dark fringes analogous to interference patterns in optics or tapestry. A unit cell of the moir\'e pattern is created by connecting these local "AA"- or "AB"-regions. The moir\'e periodicity $ L_{M} $ is the distance between the two nearest "AA"- or "AB"- regions. Assuming that the top layer is rotated anticlockwise by an amount $ \theta /2 $ and the bottom layer is rotated clockwise by the same amount such that the relative angle between them is $ \theta $, the relation between two sets of primitive lattice vectors $ \left\{\bm{a}^{(T)}_1,\bm{a}^{(T)}_2\right\} $ of top-layer and $ \left\{\bm{a}^{(B)}_1,\bm{a}^{(B)}_2\right\} $ of bottom-layer, is given by
\begin{equation}
 \bm{a}^{(T)}_{j} = \mathcal{R}(\theta)\,\bm{a}^{(B)}_{j}
\end{equation}
where $ \mathcal{R}(\theta) $ is a 2D rotation matrix about z-axis. Using (\ref{MLinT2}) and (\ref{eqn:moirerecvec}), the primitive vectors of the real space moir\'{e} pattern
\begin{equation}
 \bm{a}^{M}_{j} = \left[\mathcal{I}_2 - \mathcal{R}^{-1}(\theta)\right]^{-1} \bm{a}^{(B)}_{j} \\
\end{equation}
 and of the reciprocal space are
 \begin{equation}
	 	\bm{b}^{M}_{j} = \left[\mathcal{I}_2 - \left(\mathcal{R}^{-1}(\theta)\right)^{T}\right] \bm{b}^{(B)}_{j}
	 	\label{eqn:moiregenrec}
 \end{equation}
for all $ j = 1,2 $. The moir\'{e} periodicity $ L_M $ is plotted against the twist angle $ \theta $ in Fig. \ref{fig:TBLG1}(c). The moir\'{e} periodicity ($ L_{M} $) is much larger than the lattice constant of graphene for smaller twist angles. Consequently the moir\'{e}-cell contains a very large number of atoms. This apparent moir\'{e} periodicity is an approximate periodicity in a sense that the corners of the hexagon made by joining the head of the vectors $ \pm\bm{a}^{M}_{1} $,$ \pm\bm{a}^{M}_{2} $ and $ \pm\left(\bm{a}^{M}_{1}-\bm{a}^{M}_{2}\right) $ (Fig.\ref{fig:TBLG1}(a)) may not align with the same atomic configuration compared to the atomic configuration at the point about which rotation is performed. Thus, the number of atoms in one hexagon may differ from another. In TBLG, the underlying structure is exactly periodic only at a certain number of twist angles. At these angles, a hexagon made by the primitive vectors contains precisely the same number of carbon atoms in any other hexagon (Fig.\ref{fig:TBLG1}(b)). These are commensurate structures and do not occur in G/BN moir\'{e} pattern. In the next subsection \ref{CTBLG}, we therefore begin with the detailed theory of such exact periodic (commensurate) structures.

In TBLG, the position of two VHSs in valence and conduction bands can be varied by tuning the twist angle between the individual layers \cite{Li2010,Andrei2020} without introducing any defects, chemical doping or by electrical gating. Therefore, the VHS can be accessible to the electrons by simply tuning the twist angle. This suggests that the superconducting and correlated-insulating states may also be tunable by a single parameter (twist angle).  
In the absence of the interlayer coupling between the graphene layers, the band structure consists of four Dirac cones from the two layers and two valleys. The interlayer interaction between the layers produces avoided crossings at the intersection of the Dirac cones in the absence of interlayer tunnelling. This leads to saddle points in the band structure. At larger twist angles, the Dirac cones are widely separated, and the low-energy state in one layer is only weakly influenced by tunnel coupling to the adjacent layer.

\subsection{Commensurate TBLG}\label{CTBLG}

\begin{figure*}
		\centering
		\includegraphics[scale=0.45]{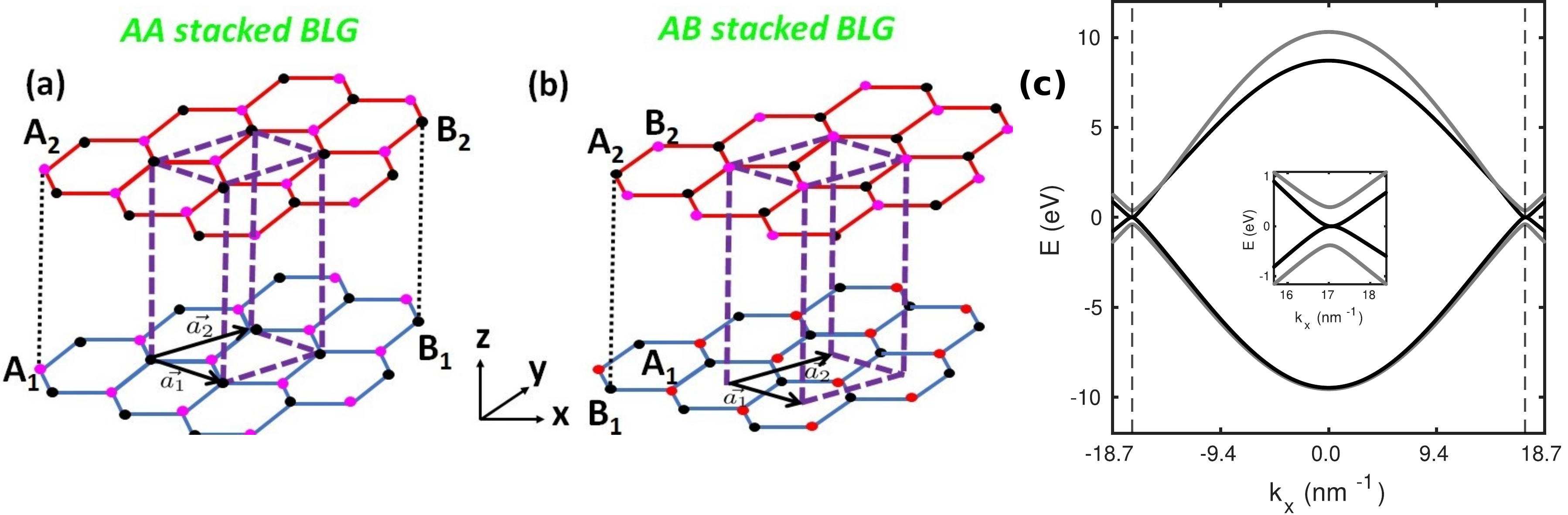}
		\caption{\label{fig:bilayer} (a) AA-stacked bilayer graphene. (b) AB- or Bernal-stacked bilayer graphene. (c) The band structure is plotted along the x-axis and passes through both the valleys $ K' $ (left gray dashed line parallel to the y-axis) and $ K $ (right gray dashed line). The inset shows the zoom-in of the bands near the $ K $-valley.}
	\end{figure*}

The commensuration problem in TBLG has been studied in several papers \cite{Santos2007,Santos2012,Shallcross2008,Shallcross2010}. We shall now study the theory of this crystallographic problem in detail. The crystalline nature or long-range ordering is preserved in two isolated graphene layers stacked over one another at a certain but countably infinite set of twist angles. When this occurs the periodicity is restored with a large unit cell compared to the unit cell of constituting layers. Based on the coinciding sublattice sites, there are two configurations: (a) AA-Bilayer graphene, where each sublattice site in the top layer coincides with the lower layer as shown in Fig.(\ref{fig:bilayer}a), and (b) AB-Bilayer graphene (Bernal stacking) where the layers are arranged so that one of the atoms from the lower layer $ B1 $ is directly below an atom, $ A2 $. In this second case, from the upper layer ('dimer' sites) and the other two atoms, A1 and B2, do not have a counterpart on the other layer which is directly above or below them (‘non-dimer’) as shown in Fig.(\ref{fig:bilayer}b) \cite{EMcCann2013}. These configurations are commensurate, and we must find other twist angles where the periodicity restores itself again.
	
	Starting with Bernal-stacked bilayer graphene, the commensurability is defined such that the stacking repeats in any direction, \emph{i.e.,} some arbitrary ($ n_1',n_2' $) site B of the rotated layer lies just above the site A ($ n_1,n_2 $) of the lower layer. Mathematically,
\begin{equation}
	n_1 \bm{a}^{(1)}_1 + n_2 \bm{a}^{(1)}_2 = n_1' \bm{a}^{(2)}_1 + n_2' \bm{a}^{(2)}_2
	\label{eqn:commveccond}
\end{equation}
	where $ \bm{a}_1^{(2)} = \mathcal{R}(\theta) \bm{a}^{(1)}_1 $ and $ \bm{a}_2^{(2)} = \mathcal{R}(\theta) \bm{a}^{(1)}_2 $. Here $\bm{a}_{1,2}$ are the lattice vectors of graphene and are defined in (\ref{GUNIT}). The superscripts indicate the layer indices. In other words, it tells us that there is an identical pair (A-B) which can be reached from the origin using both the unrotated basis vectors ($ \bm{a}^{(1)}_1 $, $ \bm{a}^{(1)}_2 $) and rotated basis vectors ($ \bm{a}^{(2)}_1 $, $ \bm{a}^{(2)}_2 $). With the help of definition (\ref{GUNIT}), the equation (\ref{eqn:commveccond}) can equivalently be written as \cite{Shallcross2008,Shallcross2010},
	\begin{equation}
		\bm{n}	=
		\begin{pmatrix}
			\cos \theta - \frac{\sin \theta}{\sqrt{3}} & -\frac{2}{\sqrt{3}} \sin \theta \\ \frac{2}{\sqrt{3}} \sin \theta & \cos \theta + \frac{\sin \theta}{\sqrt{3}}
		\end{pmatrix}
		\bm{n}'
		\label{eqn:commmateqn}
	\end{equation}
	where $ \bm{n} = \pmqty{n_{1} & n_{2}}^{T} $ and $ \bm{n}' = \pmqty{n'_{1} & n'_{2}}^{T} $. This is a linear system of Diophantine equations in four unknowns $ n_1, n_2, n_1'~\&~n_2' $, and maps one integer pair $ \left(n'_1,n'_2\right) $ to another $ \left(n_{1},n_{2}\right) $. Therefore the necessary and sufficient condition for integer solutions is that the matrix elements assume rational values only \cite{Shallcross2008}. Writing $ \cos \theta = y $ and $ \sin \theta/\sqrt{3} = x $, where $ x,y \in \mathbb{Q}^{+} $, the trigonometric identity $ \sin^2 \theta + \cos^2 \theta = 1 $ gives,
	\begin{equation}
		3 x^2 + y^2 = 1
		\label{eqn:trigident}
	\end{equation}
	\begin{figure}
		\centering
		\includegraphics[scale=0.5]{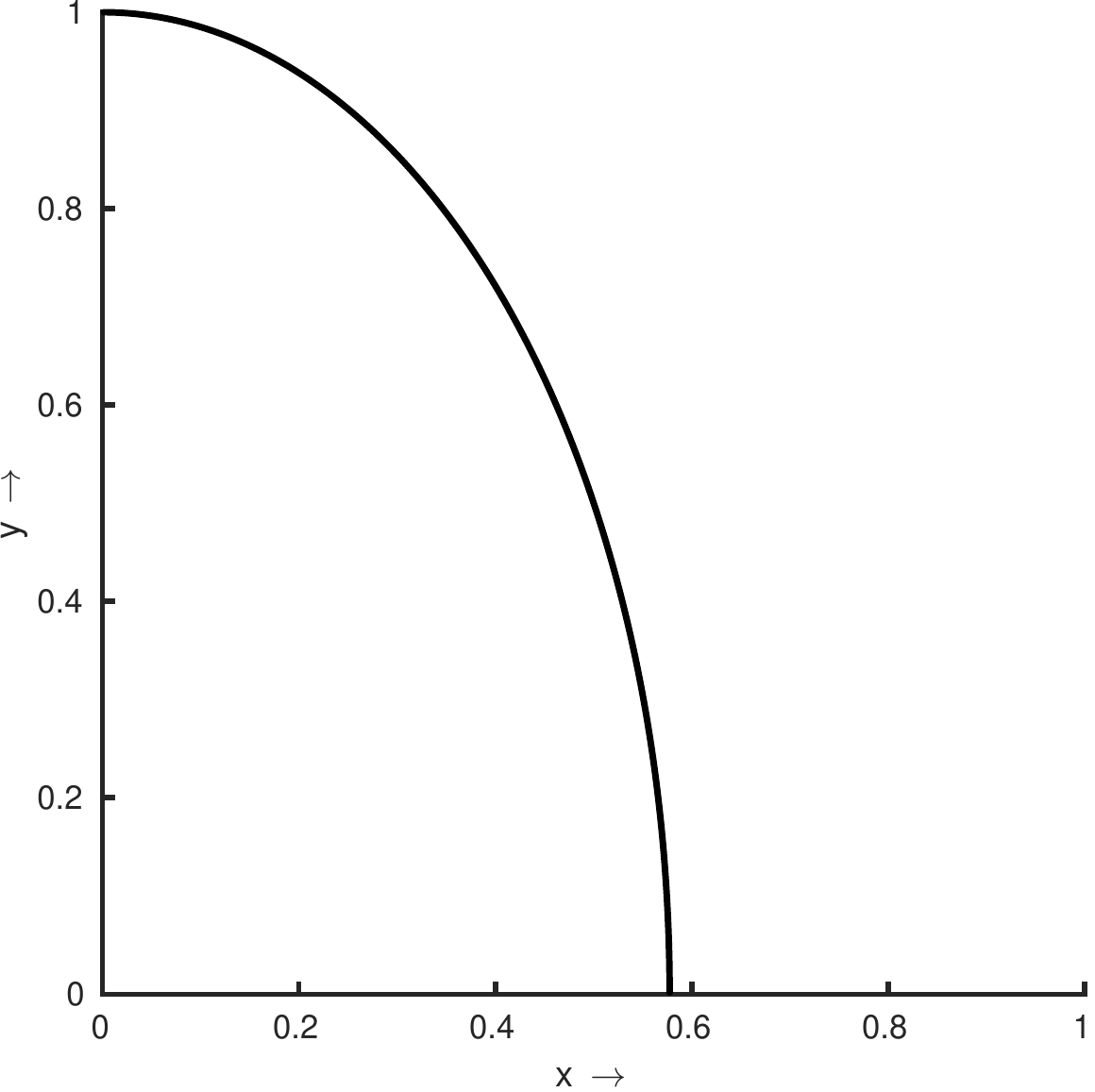}
		\caption{\label{fig:ellipse} The ellipse generated by the equation $ 3x^2 + y^2 = 1 $. Only those points for which $ x,y \in \mathbb{Q}^{+} $ are allowed solutions.}
	\end{figure}
	This is known as a Pythagorean Diophantine equation of the second degree. The solution of this equation proceeds analogously to the case of Pythagorean triplets. There is thus a one-to-one mapping between solutions of (\ref{eqn:trigident}) and rational points on the ellipse $ 3x^{2} + y^{2} = 1 $ shown in Fig.(\ref{fig:ellipse}). This leads to the following solutions for $ x $ and $ y $,
	\begin{equation}
		x = \frac{2pq}{3q^2 + p^2} \qq{and}
		y = \frac{3q^2 - p^2}{3q^2 + p^2}
	\end{equation}
	where $ q,p \in \mathbb{N} $. From these equations, one can immediately find the set of rotation angles leading to commensurations,
	\begin{equation}
		\cos(\theta) = \frac{3q^2 - p^2}{3q^2 + p^2}
		\label{eqn:cosinecommcond}
	\end{equation}
	\begin{figure*}
			\centering
			\includegraphics[scale=0.44]{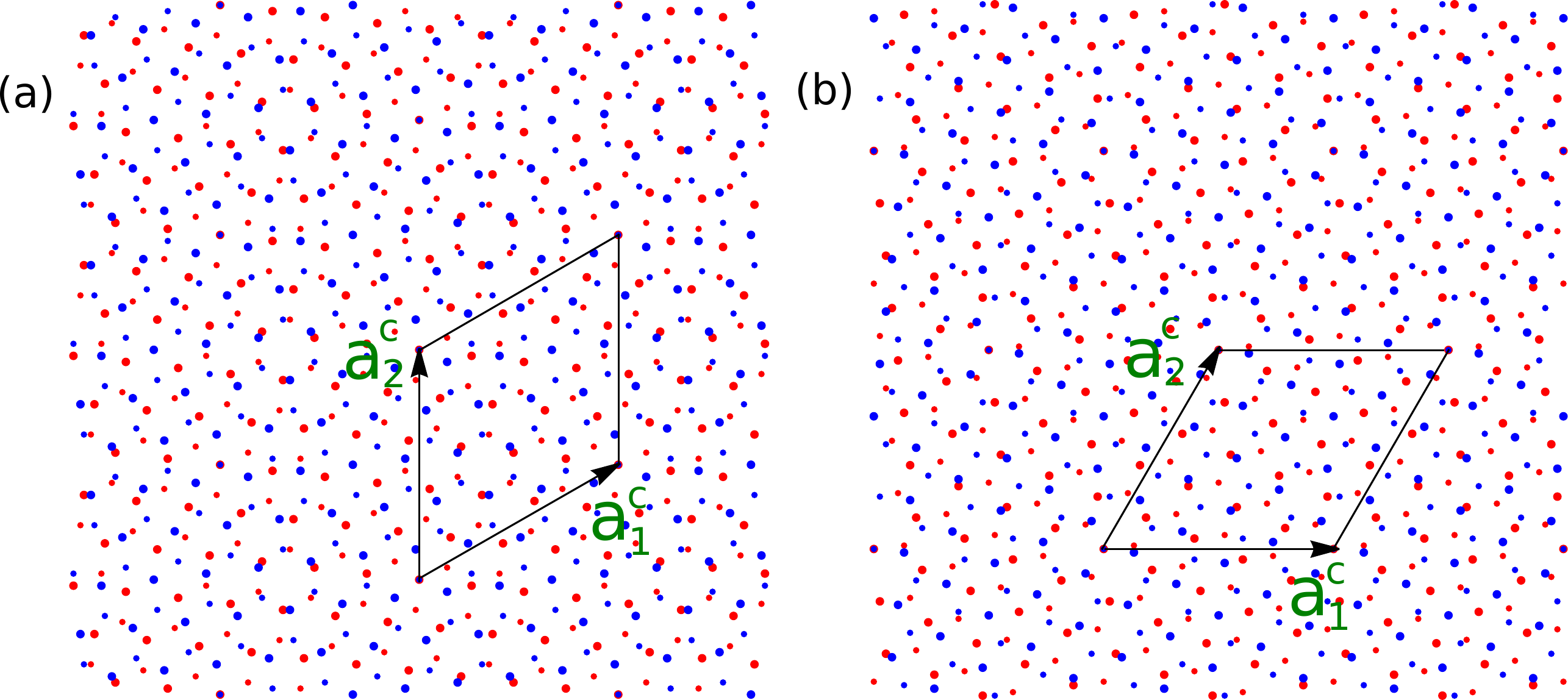}
			\caption{\label{fig:commstrs}(a) Commensurate TBLG at $ \theta = 13.1736^o $. The two black arrows mark the primitive lattice vectors $ \bm{a}^{c}_{1} $ and $ \bm{a}^{c}_{2} $ of the commensurate structure and the corresponding unit cell is shown by the rhombus. This commensurate structure is SE-even. (b) The commensuration pair at twist angle $ \theta = 46.8264^o $. The primitive lattice vectors are $ \bm{a}^{c}_{1} $ and $ \bm{a}^{c}_{2} $. This commensurate structure is SE-odd.}
		\end{figure*}

	For $ q > p \geq 0 $, this formula produces rotation angles that lie in	the range $ \theta \in \left[\ang{0}, \ang{60}\right]$. All other rotation angles are equivalent due to the symmetry of the hexagonal lattice. One requires the corresponding primitive vectors of the commensuration lattice. Substituting $ x $ and $ y $ back into (\ref{eqn:commmateqn}) leads to the coupled linear Diophantine equation $ \bm{n} = \mathcal{T}(x,y)~\bm{n}' $ with the transformation matrix $ \mathcal{T} $ as
	\begin{equation}
		\mathcal{T}(x,y) = \pmqty{y-x & -2x \\ 2x & y + x}
	\end{equation}
	The solution follows by a similarity transform of $ \mathcal{T} = \mathcal{S}\mathcal{T}^{D}\mathcal{S}^{-1} $, where $ \mathcal{S} $ is the matrix of eigenvectors of $ \mathcal{T} $ and $ \mathcal{T}^{D} $ is the diagonal matrix of eigenvalues. Then, one can write
	\begin{equation}
		\mathcal{S}^{-1}\,\bm{n} = \mathcal{T}^{D}\,\mathcal{S}^{-1}\,\bm{n}'
		\label{eqn:commmateqn2}
	\end{equation}
	Since the eigenvectors of $ \mathcal{T} $ are independent of $ x $ and $ y $ and therefore $ \mathcal{S} $ is also independent of $ x $ and $ y $. The $ x $ and $ y $ dependence and, in turn $ q $ and $ p $ are present in the diagonal matrix $ \mathcal{T}^{D} $. The eigenvalues of $ \mathcal{T} $ are $ \lambda_{1} = -\left(p+i\sqrt{3}q\right)/\left(p-i\sqrt{3}q\right) $ and $ \lambda_{2} = -\left(p-i\sqrt{3}q\right)/\left(p+i\sqrt{3}q\right) $.  Using them one gets from (\ref{eqn:commmateqn2})
	\begin{equation}
		\pmqty{3q+p & 2p \\
			-2p & 3q-p}\bm{n} = \pmqty{3q-p & -2p\\
			2p & 3q+p}\bm{n}'
    \end{equation}
	Therefore, one can write
	\bse
	\begin{align}
		\bm{n} = \pmqty{3q-p \\ 2p}\alpha + \pmqty{-2p \\ 3q+p}\beta \\
		\bm{n}' = \pmqty{3q+p \\ -2p}\alpha + \pmqty{2p \\ 3q-p}\beta
	\end{align}
	\ese
	where $ \alpha $ and $ \beta $ are constants such that the components of $ \bm{n} $ and $ \bm{n}' $ are integers. The primitive vectors of the commensuration turn out to depend on a parameter $ \delta = 3/\text{gcd}(p,3) $ \cite{Shallcross2010}.
	\begin{itemize}[itemsep=0pt]
		\item For the case where $ \delta = 1 $, one get
		\bse
		\begin{align}
			\bm{a}^{c}_{1} & = \frac{1}{\gamma}\left[\left(3q+p\right)\bm{a}^{(1)}_{1} - 2p\,\bm{a}^{(1)}_{2}\right] \\
			\bm{a}^{c}_{2} & = \frac{1}{\gamma}\left[2p\,\bm{a}^{(1)}_{1} + \left(3q-p\right)\bm{a}^{(1)}_{2}\right]
			\label{eqn:primvectscomm}
		\end{align}
		\ese
		\item For the case where $ \delta = 3 $, one get
		\bse
		\begin{align}
			\bm{a}^{c}_{1} & = \frac{1}{\gamma}\left[\left(-q-p\right)\bm{a}^{(1)}_{1} + 2q\,\bm{a}^{(1)}_{2}\right] \\
			\bm{a}^{c}_{2} & = \frac{1}{\gamma}\left[2q\,\bm{a}^{(1)}_{1} + \left(-q+p\right)\bm{a}^{(1)}_{2}\right]
		\end{align}
		\ese
	\end{itemize}
	where $ \gamma = \text{gcd}\left(3q+p,3q-p\right) $. The condition (\ref{eqn:cosinecommcond}) leads to the characterization of each commensurate cell to a pair of integers $ q,p $, which is denoted by $ \left(q,p\right) $ throughout the article. A special case occurs in the limit $ q/p \rightarrow 0 $, the twist angle $ \theta \rightarrow 0 $ generates the initial stacking configuration, which is the Bernal stacked bilayer graphene. The conventional unit cell consists of four atoms, labelled $ A1, B1 $ on the lower layer and $ A2, B2 $ on the upper layer. In the tight-binding description of BLG, one considers the $ p_z $ orbitals on the four atomic sites in the unit cell, labelled as $ A1,B1,A2,B2 $. The properties of electrons in the vicinity of the $\bm{K}$ points are described by a $ 4 \times 4 $ Hamiltonian, which contains only linear terms in the momentum $ k $ \cite{EMcCann2013}.
	\begin{equation}
		H = 
		\begin{pmatrix}
			\epsilon_{A1} & v\pi^{\dagger} & -v_4 \pi^{\dagger} & v_3\pi \\
			v\pi & \epsilon_{B1} & \gamma_1 & -v_4\pi^{\dagger} \\
			-v_4\pi & \gamma_1 & \epsilon_{A2} & v\pi^{\dagger} \\
			-v_3\pi^{\dagger} & -v_4\pi & v\pi & \epsilon_{B2}
		\end{pmatrix}
		\label{eqn:blgHamiltonian}
	\end{equation}
	where $\pi = \hbar(k_x+ik_y)$ and $ \pi^{\dagger} = \hbar(k_x-ik_y) $ and the effective velocities, $ v_3=\frac{\sqrt{3}a\gamma_3}{2\hbar} $ and $ v_4 = \frac{\sqrt{3}a\gamma_4}{2\hbar} $. Fig.(\ref{fig:bilayer}c) shows the band structure of the Bernal-stacked bilayer graphene. In the sublattice basis $ \left\{ A1, B1, A2, B2 \right\} $, there are four bands: a pair of conduction bands and a pair of valence bands. Near the \tcr{$ K $}-point, the two bands touch each other and are contributed by the non-dimer sites, while one conduction band and one valence band are split away from zero energy by an energy of the order of the interlayer coupling $ \gamma_1 $ due to the dimer sites. The transition of the electron from non-dimer sites via the dimer sites leads to the 'mass' term, and therefore the dispersion of the Bernal-stacked bilayer near the Dirac point (\tcr{$ K $}-point) is parabolic \cite{EMcCann2013}.

 	As an example for the cases with $\delta=1$, we consider the commensurate structure $ \left(q,p\right) = \left(5,1\right) $ at an angle $ \theta = \ang{13.1736} $. The corresponding structure is shown in Fig.\ref{fig:commstrs}(a) with the unit cell, and the primitive lattice vectors are calculated from (\ref{eqn:primvectscomm}). Using the primitive vectors, the number of atoms $ N $ enclosed by the commensurate cell can be calculated as \cite{Santos2007,Shallcross2008,Shallcross2010}
 	\begin{equation}
 		N = 4 \frac{\abs{\bm{a}^{c}_1 \times \bm{a}^{c}_2}}{\abs{\bm{a}_1 \times \bm{a}_2}} =
 		\frac{12}{\delta}\frac{1}{\gamma^2} \left(3q^2 + p^2\right)\label{eqn:commNatoms}
 	\end{equation}
 	where $ 4 $ is multiplied as there are four atoms in a unit cell of Bernal stacked bilayer graphene. The size of the unit cell in the case of Bernal stacked bilayer graphene is the smallest (enclose only four atoms) compared to the size of the unit cell in any other commensurate structures as $ N \ge 4 $.
 	
 	E.J. Mele \cite{Mele2010} showed that the twisted bilayers with rotation angles $ \theta $ and $ \bar{\theta} = 60^{\circ} - \theta $ are \emph{related} and referred to as \emph{commensuration pair}. The simplest trivial example of such a pair occurs for $ \theta = 0 $ and $ \bar{\theta} = 60^{\circ} $ corresponding to \emph{AB}(Bernal) stacking. These commensuration pairs are distinguished by their sublattice symmetries \cite{Mele2010}. A symmetry operation which exchanges the sublattice is referred to as a sublattice exchange (SE) operation, namely
 	\begin{equation*}
 		\text{A1} \longleftrightarrow \text{B1} \qquad \text{A2} \longleftrightarrow \text{B2}. 
 	\end{equation*} 
 	Twice, the operation of the sublattice parity restores the structure to its original form, and consequently, the eigenvalues are -1 (odd parity) and 1 (even parity). In this context, a commensuration is SE-even if the primitive cell contains an A and a B sublattice site in each layer that are coincident with atomic sites in the neighbouring layer, such as AA-stacking in bilayer graphene. On the other hand, a commensuration is SE-odd if only one sublattice site in the primitive cell is covered, such as the Bernal stacked bilayer graphene. Fixing the rotation centre of the twist at an atom site guarantees that there will be at least one coincident site. Accordingly, the Fig.\ref{fig:commstrs}(a) is SE-even, as the each of the sublattice $ A_1 (or~A_2)$, $ B_1 (or~B_2) $ and the center of the hexagon coincides. On the other hand, its commensurate partner in Fig.\ref{fig:commstrs}(b) is SE-odd, as only the sublattice $ A_1 (A_2) $ coincide.
	Although one can always find a unit cell in commensurate structures, they usually contain {\it one-too-many} atoms, making {\it ab initio} calculations time-prohibitive. Owing to the progress in computational power, however, several works detailing the atomistic simulations of these systems have been carried \cite{Suarez2010,Trambly2010,Ohta2012,Uchida2014,NNTnam2017,Zhang2018,Larson2020,Long2022}. Particularly the reference \cite{Suarez2010} was one of the first such works which suggested the existence of flat bands in slightly twisted bilayer graphene and its connection with high-temperature superconductivity.
	Nevertheless, constructing a long wavelength description, or in other words, the continuum model provides key simplifications over the microscopic lattice models and reproduce the energy spectrum with reasonable accuracy with the electronic spectrum calculated using {\it ab initio} large-supercell tight-binding Hamiltonians or density functional theory (DFT) based calculations. Particularly, the continuum models of TBLG should be able to reproduce the energy spectrum as a function of their relative twist angle $ (\theta) $, which captures the nature of single-particle states near the Fermi level. Surprisingly, the benefits of continuum models in TBLG include the applicability of the Bloch theory of bands and the reduction of the problem's two dimensionless parameters—the twist angle and the interlayer to intralayer hopping ratio—to a single one \cite{Balents2019}. In the following subsection \ref{contimod}, we review the first continuum model for a commensurate TBLG by Santos {\it et al.}\cite{Santos2007} and their salient results.
	\begin{figure}
		\centering
		\includegraphics[scale=0.8]{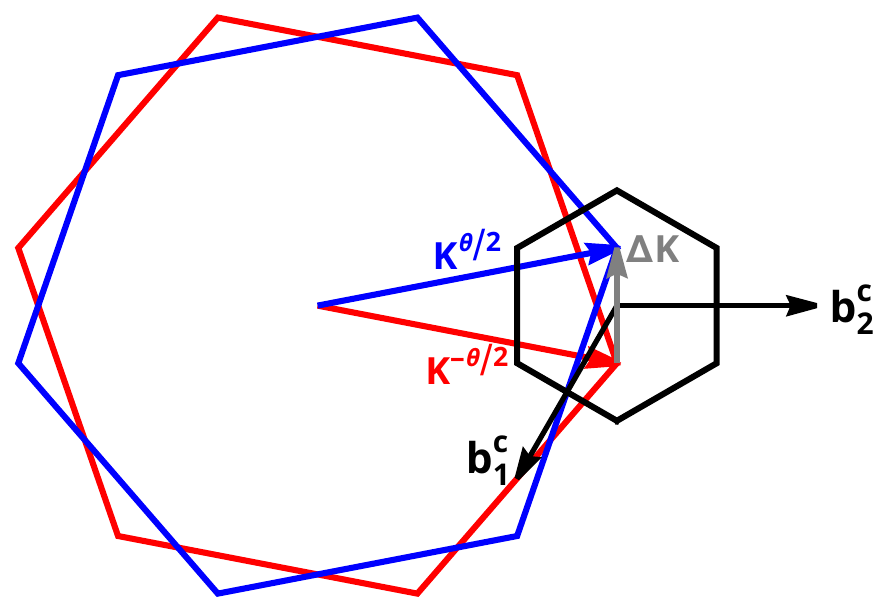}
		\caption{\label{fig:santosBZ}The red and blue hexagons show the BZ of two rotated SLG at $ -\frac{\theta}{2} $ and $ \frac{\theta}{2} $, respectively. The two vectors $ \bm{K}^{\frac{\theta}{2}} $ and $ \bm{K}^{-\frac{\theta}{2}} $ points the corresponding Dirac points at the right valley $ \xi = 1 $ and the vector $ \Delta\bm{K} = \bm{K}^{\frac{\theta}{2}} - \bm{K}^{-\frac{\theta}{2}} $ points the shift between them. The two primitive reciprocal space lattice vectors corresponding to the real-space commensurate TBLG are $ \bm{b}^{c}_{1} $ and $ \bm{b}^{c}_{2} $ such that $ \bm{a}^{c}_{i}\vdot\bm{b}^{c}_{j} = 2\pi\delta_{ij} $ for $ i,j = 1,2 $ and the black hexagon centred at the middle point of $ \Delta\bm{K} $ shows the BZ of the real space commensurate TBLG.}
	\end{figure}
	\subsection{The Continuum model by Santos {\it et al}}\label{contimod}
	In the previous section, we noted that the small twist angle among the graphene bilayers leads to large-period primitive cells. For smaller twist angles, the low-energy electronic spectrum of TBLG is dominated by the electronic states of SLG. The Hamiltonian for TBLG has the form \cite{Santos2007,Santos2012}
	\begin{equation}
		\mathcal{H} = H_1 + H_2 + H_{\perp} \label{Hconti1}
	\end{equation}
	where $ H_1 = \hbar\,v_F\bm{\sigma}^{\xi}_{-\theta/2}\vdot\bm{k} $ is the intralayer Hamiltonian of layer-1 (L1) and $ H_2 = \hbar\,v_F\bm{\sigma}^{\xi}_{\theta/2}\vdot\bm{k} $ is of layer-2 (L2), where $ \bm{\sigma}^{\xi}_{\theta} = e^{i\sigma_z\theta/2}\left(\xi\sigma_x,\sigma_y\right)e^{-i\sigma_z\theta/2} $ for valley index $ \xi $. $H_{\perp}$ represents the interlayer coupling Hamiltonian and will be deliberated upon in the subsequent discussion. For the rest of the derivation in this section, the calculation is done using the right-valley unless stated otherwise.

The low-energy electronic structure is obtained by knowing the behaviour of the interlayer coupling among the layers. The interlayer coupling $ H_{\perp} $ is modelled by retaining hopping from each sublattice site $ \alpha \left( = A_1, B_1 \right) $ in L1 to the closest sublattice sites $ \beta \left( = A_2, B_2 \right) $ of L2. If $ \bm{\delta}^{\alpha\beta}(\bmr_{i}) $ be the horizontal in-plane displacement from an atom $ \alpha $ of L1 to the closest atom $ \beta $ in L2 at position $ \bmr_{i} $, then the functional $ t_{\perp}\left[ \bm{\delta}^{\alpha\beta}(\bmr_{i}) \right] \equiv t^{\alpha\beta}_{\perp}(\bmr_{i}) $ denotes the position-dependent hopping between $ p_z $ orbitals. The second quantized form of $ H_{\perp} $ is
	\begin{equation}
		H_{\perp} = \sum_{i,\alpha,\beta} t^{\alpha\beta}_{\perp}(\bmr) \,c^{\dagger}_{\alpha}(\bmr_i) \, c_{\beta}(\bmr_i + \bm{\delta}^{\alpha\beta}(\bmr_i)) + h.c.
		\label{eqn:secquantlop}
	\end{equation}
	where $ c^{\dagger} $ and $ c $ are creation and destruction operator respectively. To write the momentum space representation of $ H_{\perp} $, the on-site operators are written as
	\bse
	\begin{align}
		c^{\dagger}_{\alpha}(\bmr_i) & =\sqrt{v_{c}}\, \psi_{1,\alpha}(\bmr_{i})\,e^{i\bm{K}^{-\theta/2}\vdot\bmr_{i}} \\
		c_{\beta}(\bmr_i + \bm{\delta}^{\alpha\beta}(\bmr_i)) & = \sqrt{v_{c}}\, \psi_{2,\beta}(\bmr_{i})\,e^{i\bm{K}^{\theta/2}\vdot(\bmr_{i}+ \bm{\delta}^{\alpha\beta}(\bmr_{i}))}
	\end{align}
	\ese
	where $ \psi $ are the field operators and $ \psi_{j,\beta}(\bmr_{i} +  \bm{\delta}^{\alpha\beta}(\bmr_{i})) \approx \psi_{j,\beta}(\bmr_{i}) $ for $ j=1,2 $ as the field operator varies slowly over the lattice length scale. Here $v_{c}$ is the volume of the unit cell. Using the above operator representations in Eq. (\ref{eqn:secquantlop}), one gets
	\begin{multline}
		H_{\perp} = v_{c} \sum_{i,\alpha,\beta} t^{\alpha\beta}_{\perp}(\bmr_{i}) e^{i\bm{K}^{-\theta/2}\vdot\bmr_{i}} \\ \times e^{i\bm{K}^{\theta/2}\vdot(\bmr_{i}+ \bm{\delta}^{\alpha\beta}(\bmr_{i}))}
		\psi^{\dagger}_{1,\alpha}(\bmr_{i})
		\psi_{2,\beta}(\bmr_{i}) + \text{H.c.}
		\label{eqn:intlayerlop}
	\end{multline}
	\begin{figure*}
		\centering
		\includegraphics[scale=0.45]{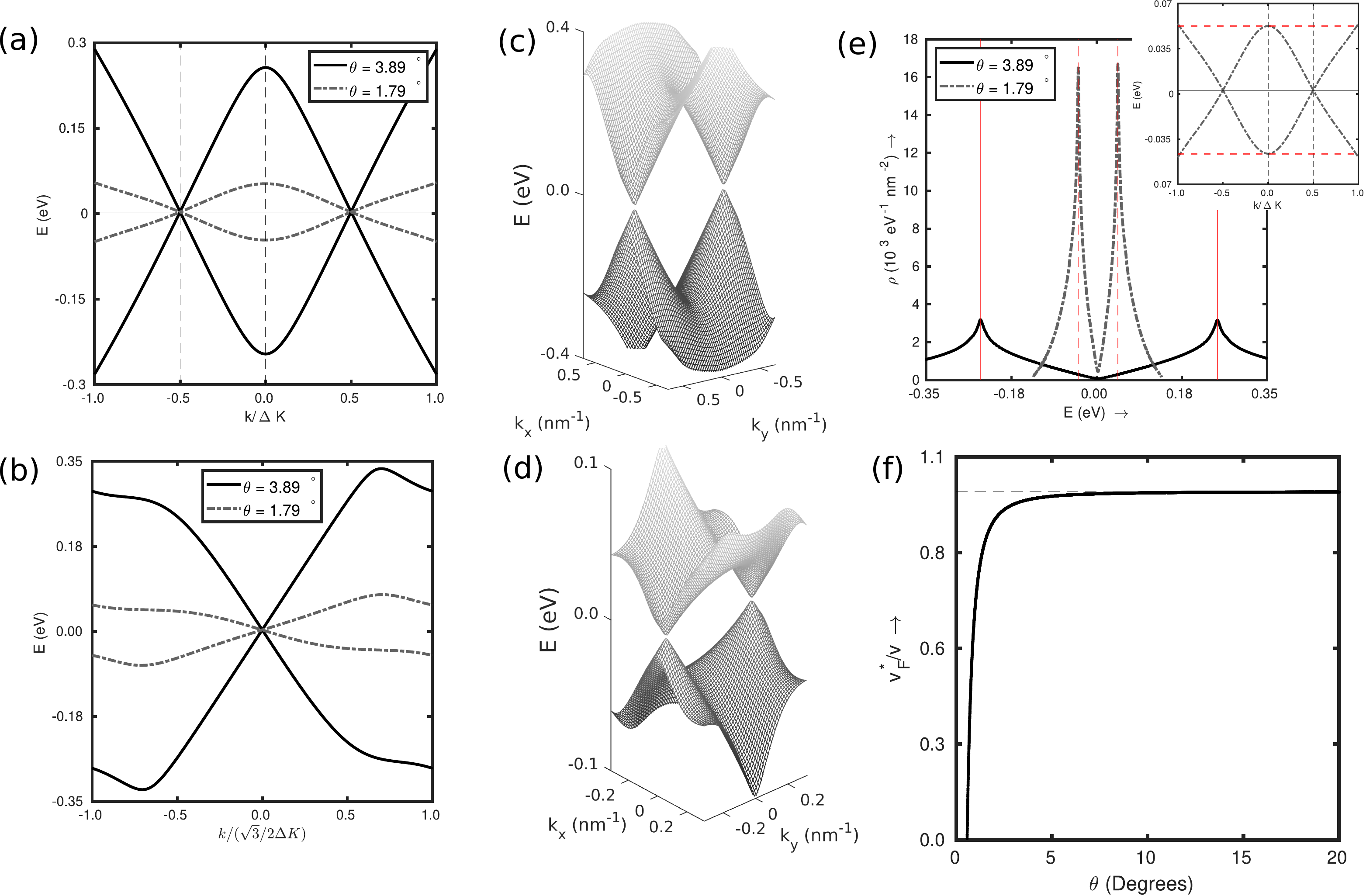}
		\caption{\label{fig:santos}(a) The band structure along y-axis passing through both the Dirac points at $ \pm \Delta\bm{K}/2 $ at twist angles $ \ang{3.89} $ and $ \ang{1.79} $. (b) The band structure along x-axis passing through through the Dirac point  at $  \Delta\bm{K}/2 $ at twist angles $ \ang{3.89} $ and $ \ang{1.79} $. (c)-(d) The 3D band structures showing the first valence and conduction bands at $ \theta = \ang{1.79} $ and $ \theta = \ang{3.89} $,respectively. (e) The DOS for two angles at $ \theta = \ang{3.89} $ and $ \theta = \ang{1.79} $. The red dashed lines parallel to y-axis marks the VHS for twist angle $ \ang{1.79} $ and the dashed red lines in inset marks the maximum of conduction band (saddle point) and minimum of valence band (saddle point) where the VHSs occur. (f) The figure shows the renormalized Fermi velocity for the commensurate TBLG at different commensurate angles. For very small angles, the quantity $ \frac{\tilde{t}_{\perp}}{v_F~\Delta K} $ is no longer small and the perturbation theory breaks down.}
	\end{figure*}
	As shown in Fig.(\ref{fig:santosBZ}), the two vectors $ \bm{K}^{\theta/2} $ and $ \bm{K}^{-\theta/2} $ in the interlayer Hamiltonian (\ref{eqn:intlayerlop}) join the centres of the BZ of each layer to the corresponding Dirac points at the right valley $ \xi = 1 $. The vector $ \Delta\bm{K} = \bm{K}^{\theta/2} - \bm{K}^{-\theta/2} $ points the shift between the tips of these vectors. For commensurate structures, the function $ t^{\alpha\beta}_{\perp}(\bmr_{i})\, e^{i\bm{K}^{\theta/2}\vdot\bm{\delta}^{\alpha\beta}(\bmr_{i})} $ for a given $ \theta $ is translationally invariant under $ \bmr_i \rightarrow \bmr_i + \bm{a}^{c}_{1,2} $, since the displacement $ \bm{\delta}^{\alpha\beta} $ at each $ \bmr_i $ remains same in each of the commensurate cell. The $\bm{a}^{c}_{1,2}$ are defined in (\ref{eqn:primvectscomm}). 
	This periodicity allows one to expand it in terms of a Fourier series as 
	\begin{equation}
		t^{\alpha\beta}_{\perp}(\bmr_{i})\,e^{i\bm{K}^{\theta}\vdot\bm{\delta}^{\alpha\beta}(\bmr_{i})}
		= \sum_{\bm{b}^{c}} \tilde{t}^{\alpha\beta}(\bm{b}^{c})\, e^{i\bm{b}^{c}\vdot\bmr_{i}}. \label{periodichopping}
	\end{equation}
	Here $ \bm{b}^{c} $ are the reciprocal lattice vectors of the reciprocal space corresponding to the real space commensurate structure, and the Fourier coefficient $ \tilde{t}^{\alpha\beta}(\bm{b}^{c}) $ is given by
	\begin{equation}
		\tilde{t}^{\alpha\beta}(\bm{b}^{c}) = \frac{1}{v_{c}} \int_{v_{c}} \dd[2]{\bmr_{i}} t^{\alpha\beta}_{\perp}(\bm{r}_{i})\,e^{i\bm{K}^{\theta}\vdot\bm{\delta}^{\alpha\beta}(\bmr_{i})}\,e^{-i\bm{b}^{c}\vdot\bmr_{i}} \label{FTTC}. 
	\end{equation}
	A relative rotation by an angle $ \theta $ between the two layers shifts the crystal momenta of their closest Dirac points by $ \Delta \bm{K} = \bm{K}^{\theta/2} - \bm{K}^{-\theta/2} $ as shown in Fig.(\ref{fig:santosBZ}), with magnitude $ \Delta K = 2K_{D} \sin(\theta/2) $. Here $K_{D}$ is the magnitude of the wave vector from the centre ($ \Gamma $) to one of the corners of FBZ in unrotated SLG. Writing the momentum-space representation of field operators $ \psi $ and measuring the momentum from the middle point of $ \Delta\bm{K} $ as shown in Fig.(\ref{fig:santosBZ}), the interlayer tunnelling Hamiltonian $ H_{\perp} $ now becomes
	\begin{multline}
		H_{\perp} = \sum_{\alpha,\beta}\sum_{\bmk,\bm{b}^{c}} \tilde{t}^{\alpha\beta}(\bm{b}^{c})
		\phi^{\dagger}_{1,\alpha}(\bmk+\bm{b}^{c})
		\phi_{2,\beta}(\bmk) + \text{H.c.}
		\label{eqn:intlayerksp}
	\end{multline}
	where the field-operators $ \phi_{1,\bmk,\alpha} \equiv \psi_{1,\bmk +\Delta\bm{K}/2} $ and $ \phi_{2,\bmk,\beta} = \psi_{2,\bmk - \Delta\bm{K}/2} $. The full second quantized Hamiltonian $ \mathcal{H}$ introduced in (\ref{Hconti1}) now becomes
	\begin{widetext}
	\begin{multline*}
		\mathcal{H} =  \sum_{\bmk,\alpha,\beta} \left[
		-\hbar v_{F} \phi^{\dagger}_{1,\bmk,\alpha}\bm{\sigma}^{\xi}_{-\theta/2,\alpha\beta}\vdot \left(\bmk + \frac{\Delta\bm{K}}{2}\right)\phi_{1,\bmk,\beta}
		-\hbar v_{F} \phi^{\dagger}_{2,\bmk,\alpha}\bm{\sigma}^{\xi}_{\theta/2,\alpha\beta}\vdot\left(\bmk - \frac{\Delta\bm{K}}{2}\right)\phi_{2,\bmk,\beta}
		+ \left(\sum_{\bm{b}^{c}} \tilde{t}^{\alpha\beta}_{\perp}(\bm{b}^{c}) \phi^{\dagger}_{1,\bmk+\bm{b}^{c},\alpha}\phi_{2,\bmk,\beta}+ H.c.\right)\right] 
	\end{multline*}
	 For the valley index $ \xi = 1 $, the above Hamiltonian contains only four states $ \ket{\bmk,1} $ of L1 which couples to the three states $ \ket{\bmk,2} $,$ \ket{\bmk + \bm{b}^{c}_{1},2} $, and $ \ket{\bmk + \bm{b}^{c}_{1} + \bm{b}^{c}_{2},2} $ of L2, and is given as
	\begin{equation}
		H_{\bmk} = 
		\pmqty{\hbar v_{F}\bm{\sigma}_{-\theta/2}\vdot\left(\bmk + \frac{\Delta\bm{K}}{2}\right) & S_{1} & S_{2} & S_{3} \\
			S^{\dagger}_{1}  & \hbar v_{F}\bm{\sigma}_{\theta/2}\vdot\left(\bmk - \frac{\Delta\bm{K}}{2}\right) & 0 & 0 \\
			S^{\dagger}_{2} & 0 & \hbar v_{F}\bm{\sigma}_{\theta/2}\vdot\left(\bmk - \frac{\Delta\bm{K}}{2} - \bm{b}^{c}_{1}\right) & 0 \\
			S^{\dagger}_{3} & 0 & 0 & \hbar v_{F}\bm{\sigma}_{\theta/2}\vdot\left(\bmk - \frac{\Delta\bm{K}}{2} - \bm{b}^{c}_{1} - \bm{b}^{c}_{2}\right)} \label{Hconti2}
	\end{equation}
	\end{widetext}
	where $ S_{1},S_{2} $ and $ S_{3} $ are the interlayer tunneling matrices as listed in Table-(\ref{table:inttunnmat}).  
	
	To gain more insight into the Hamiltonian (\ref{Hconti2}), in Fig.\ref{fig:santos}(a) the corresponding bands are drawn along the y-axis passing through both the Dirac points at $ -\Delta\bm{K}/2 $ and $ \Delta\bm{K}/2 $. Whereas in Fig.\ref{fig:santos}(b), the same are drawn along the x-axis passing through the Dirac point at $ \Delta\bm{K}/2 $.
	Figs.\ref{fig:santos}(c)-(d) show the 3D view of the valence and conduction bands in the neighbourhood of both the Dirac points at $ \Delta\bm{K}/2 $ and $ -\Delta\bm{K}/2 $ at angles $ \theta = \ang{3.89} $ and $ \ang{1.79} $, respectively. The DOS is plotted in Fig.\ref{fig:santos}(e), the dashed black line shows the DOS for $ \theta = \ang{1.79} $ where the inset shows the 2D band structure in which the maximum and minimum of conduction and valence bands at $ k/\Delta K = 0 $ are marked using the red, dashed lines since those points are the saddle points. The difference between the energies at which the singularities occur in the DOS at $ \ang{1.79} $ is $ \Delta E_{\text{VHS}} \approx 67.1~\si{\milli\electronvolt} $. On the other hand, the solid black line shows the DOS at twist angle $ \theta = \ang{3.89} $ with the van Hove singularities are marked by solid, red lines. The difference between the energies at which the singularities occur in the DOS at $ \ang{3.89} $ is $ \Delta E_{\text{VHS}} \approx 420.2~\si{\milli\electronvolt} $. Experimentally, the DOS was observed by Li et al. \cite{Li2010} in 2010, where they reported that the difference $ \Delta E_{\text{VHS}} $ increases as the twist angle $ \theta $ is increased between the graphene layers.
		
	In the neighbourhood of Dirac point of L1, $ \bmk = -\Delta\bm{K}/2 + \bmq$, the Hamiltonian $ H_{\bmk} $ can be written as $ H_{\bmk = -\Delta\bm{K}/2 + \bmq} = H_{-\Delta\bm{K}/2} + V_{\bmq} $, where the Hamiltonian $ H_{-\Delta\bm{K}/2} $ contains the interlayer tunnelling matrices,
	\begin{equation}
		H_{-\Delta\bm{K}/2} = 
		\pmqty{0 & S_{1} & S_{2} & S_{3} \\
			S^{\dagger}_{1}  & h_{1} & 0 & 0 \\
			S^{\dagger}_{2} & 0 & h_{2} & 0 \\
			S^{\dagger}_{3} & 0 & 0 & h_{3}};
		\quad
		V_{\bmq} = \pmqty{V_{-} & 0 & 0 & 0 \\
			0  & V_{+} & 0 & 0 \\
			0 & 0 & V_{+} & 0 \\
			0 & 0 & 0 & V_{+}}
	\end{equation}
	where 
	\bea h_1 & = &  -\hbar v_{F}\bm{\sigma}_{\theta/2}\vdot\Delta\bm{K};~ h_{2} = -\hbar v_{f}\bm{\sigma}_{\theta/2}\vdot\left(\Delta\bm{K}+\bm{b}^{c}_{1}\right) \nonumber \\
	h_{3}  & = & -\hbar v_{F}\bm{\sigma}_{\theta/2}\vdot\left(\Delta\bm{K}+\bm{b}^{c}_{1} + \bm{b}^{c}_{2}\right) \label{Hconti3} \eea
	and 	$\bmq$- dependent part $ V_{\bmq} $ is linear in $ \bmq $, where 
	\beq V_{-} = \hbar v_{F}\bm{\sigma}_{-\theta/2}\vdot\bmq;~\text{and} ~V_{+} = \hbar v_{F}\bm{\sigma}_{\theta/2}\vdot\bmq \label{Hconti4} \eeq 
	
	\begin{table}
		\centering
		\caption{\label{table:inttunnmat} The most dominant Fourier amplitudes for (\ref{FTTC}) .}
		\begin{ruledtabular}
			\begin{tabular}{cccc}
				$ \bm{b}^{c} $ & $ \bm{0} $ & $ \bm{b}^{c}_{1} $ & $ \bm{b}^{c}_{1} + \bm{b}^{c}_{2} $ \\
				\hline
				$ \tilde{t}(\bm{b}^{c}) $  & 
				$ \pmqty{\tilde{t}_{\perp} & \tilde{t}_{\perp} \\ \tilde{t}_{\perp} & \tilde{t}_{\perp}} $ &
				$ \pmqty{e^{i2\pi/3}\tilde{t}_{\perp} & \tilde{t}_{\perp} \\ e^{-i2\pi/3}\tilde{t}_{\perp} & \tilde{t}_{\perp}e^{i2\pi/3}} $ &
				$ \pmqty{e^{-i2\pi/3}\tilde{t}_{\perp} & \tilde{t}_{\perp} \\ e^{i2\pi/3}\tilde{t}_{\perp} & e^{-i2\pi/3}\tilde{t}_{\perp}} $ \\
			\end{tabular}
		\end{ruledtabular}
	\end{table}

	In order to gain addition insight into the low energy behaviour predicted by the Hamiltonian (\ref{Hconti2}), for the smaller magnitude of $ \bmq $, $ V_{\bmq} $ can be treated as a perturbation to $ H_{-\Delta\bm{K}/2} $ and one can derive an effective Hamiltonian in the space of zero-energy doublet of $ H_{-\Delta\bm{K}/2} $.
	It is implicitly assumed that such a doublet exists and will allow us to compare the low-energy behaviour of this continuum theory with that of SLG in a direct way. This will also help us to compare this continuum theory with the more successful BM model that will be analysed in section \ref{BMmodel}. 
	One can explicitly obtain the zero-energy doublet of $ H_{-\Delta\bm{K}/2} $ by considering that the $ H_{-\Delta\bm{K}/2} $ acts on an 8-component column vector $ \Psi = \pmqty{\psi_{0} & \psi_{1} & \psi_{2} & \psi_{3}}^{T} $, where the different components in $ \Psi $ satisfies
	\bse
	\begin{align}
		\psi_{n} &= -h_{n}^{-1}S^{\dagger}_{n}\psi_{0} \quad \forall \quad n=1,2,3 \\
		\sum_{n=1}^{3} & S_{n}h^{-1}_{n}S^{\dagger}_{n}\psi_{0} = 0
	\end{align}
	\ese
	Corresponding to two linearly-independent choices for $ \psi_{0} $, the two zero-energy doublet are denoted as $ \Psi^{(1)} $ and $ \Psi^{(2)} $. Normalizing the column vectors $ \Psi^{(1)} $ and $ \Psi^{(2)} $ as
	\begin{equation}
		\Psi^{(i)\dagger}\Psi^{(j)} = \psi^{(i\dagger)}_{0}\psi^{(j)}_{0} + \psi^{(i\dagger)}_{0} \left(\sum_{n=1}^{3} S_{n} h^{-1}_{n}h^{-1}_{n} S^{\dagger}_{n}\right)\psi^{(j)}_{0}
	\end{equation}
	where $ \sum_{n=1}^{3} S_{n} h^{-1}_{n}h^{-1}_{n} S^{\dagger}_{n} = 6\tilde{t}^{2}_{\perp}/(\Delta K^{2}(\hbar v_{F})^2)\mathcal{I}_{2} $, where $ \mathcal{I}_2 $ is an identity matrix of second order. With the help of (\ref{Hconti4}) the matrix representation of $ V_{\bmq} $ in the zero-energy doublet $ \Psi^{(1)} $ and $ \Psi^{(2)} $ can be obtained as
	\begin{equation}
		V_{ij} = \mel{\Psi^{(i)}}{V_{\bmq}}{\Psi^{(j)}}
		= \psi^{(i)\dagger}_{0}V_{-}\psi^{(j)}_{0} + \sum_{n=1}^{3}  \psi^{(i)\dagger}_{n}V_{+}\psi^{(j)}_{n}
		\label{eqn:99}
	\end{equation}
	where the second term in (\ref{eqn:99}) comes out to be
	\begin{equation}
		\sum_{n=1}^{3} \psi^{(i)\dagger}_{n}V_{+}\psi^{(j)}_{n} = \sum_{n=1}^{3}
		\psi^{(i)\dagger}_{0}S_{n}\,h^{-1}_{n}\,V_{+}\,h^{-1}_{n}\,S^{\dagger}_{n}\,\psi^{(j)}_{0}. 
	\end{equation} 
	A straightforward but lengthy algebra gives $ \sum_{n=1}^{3} S_{n}\,h^{-1}_{n}\,V_{+}\,h^{-1}_{n}\,S^{\dagger}_{n} = -3\left(\tilde{t}_{\perp}^2/\Delta K^2 \hbar v_{F}\right) \bm{\sigma}_{-\theta/2}\vdot\bmq $. Therefore, the matrix element $ V_{ij} $ becomes
	\begin{equation}
		V_{ij} = \hbar v_{F}\psi^{(i)\dagger}_{0}\left[1-3\left(\tilde{t}_{\perp}^2/\Delta K^2 (\hbar v_{F})^{2}\right)\right]\bm{\sigma}_{-\theta/2}\vdot\bmq\,\psi^{(j)}_{0}
	\end{equation}
	
	In the limiting condition of the dimensionless ratio $ \tilde{t}/\hbar v_{F} \Delta K << 1 $, the band velocity of the linear bands near the Fermi level becomes dependent on the twist angle and is given by
	\begin{equation}
		\frac{v_F^{*}}{v_F} = 1 - 9\left(\frac{\tilde{t}_{\perp}}{\hbar v_F \Delta K}\right)^2 \label{velocityrenorm1}
	\end{equation}
	where $ v_F $ is the Fermi velocity in SLG. The Fermi velocity $ v^*_{F} $ decreases as the angle decreases, as shown in Fig.\ref{fig:santos}(e). The above equation breaks down in the limit of small rotation angles due to a failure of the perturbation theory when $ \Delta K \rightarrow 0 $. Even though we do not derive a formal relation due to this breakdown of the perturbation, this relation suggests non-trivial renormalisation of the Fermi velocity in the limit of a vanishing twist angle.
	
	In 2010, E.J.Mele \cite{Mele2010} reported a general continuum theory where the Fourier transform of the position-dependent hopping amplitude is $ t(\bm{G}) $ with $ \bm{G} = p_1\bm{G}^{(1)}_1 + p_2\bm{G}^{(1)}_2 + p'_1 \bm{G}^{(2)}_1 + p'_2 \bm{G}^{(2)}_2 $, where $ p_1,p_2,p'_1,p'_2 $ are integers and the superscripts $ (1) $ and $ (2) $ denotes the layer-1 and layer-2. The momentum conserving conditions occurs when
	\begin{equation}
		\bm{K}^{\theta/2} - \bm{K}^{-\theta/2} = \bm{G} \label{MeleEq}
	\end{equation}
	The essential feature of Mele's theory is the existence of a $ \bm{G} = 0 $ term in the effective inter-layer tunnelling Hamiltonian. This indicates that the electron states in the two layers that have the same crystal momentum modulo $ \bm{G} $ are coupled through the inter-layer Hamiltonian. Also, the continuum theory of Ref.\cite{Santos2007} is recovered by only the $ \bm{G} = 0 $ term \cite{Santos2012}. Mele showed that the terms with $ \bm{G} \ne 0 $ lead to different physics and prevent the massless low-energy behaviour. Mele's result indicated the possibility of a low-energy theory that behaves well for arbitrarily small twist angles and incommensurate structures. In the subsequent discussion, we shall discuss these incommensurate structures.
	
	All other structures which are not commensurate are referred to as incommensurate structures. These structures have no long-range ordering, so the crystalline nature is lost. Consequently, one can not define the unit cell in the same way as in commensurate structures. The position-dependent hopping among two layers in (\ref{periodichopping}) is periodic for commensurate structures, but this is not true for an incommensurate structures. In the following section we shall review the theoretical framework of interlayer interaction in any general bilayers followed by the continuum model by Bistritzer and MacDonald \cite{Bistritzer2011} which provides the electronic structure for arbitrary small twist angles.

	\subsection{Incommensurate structures: A general theoretical framework for interlayer interaction}\label{ICTBLG}
\begin{figure*}
		\centering
		\includegraphics[scale=0.5]{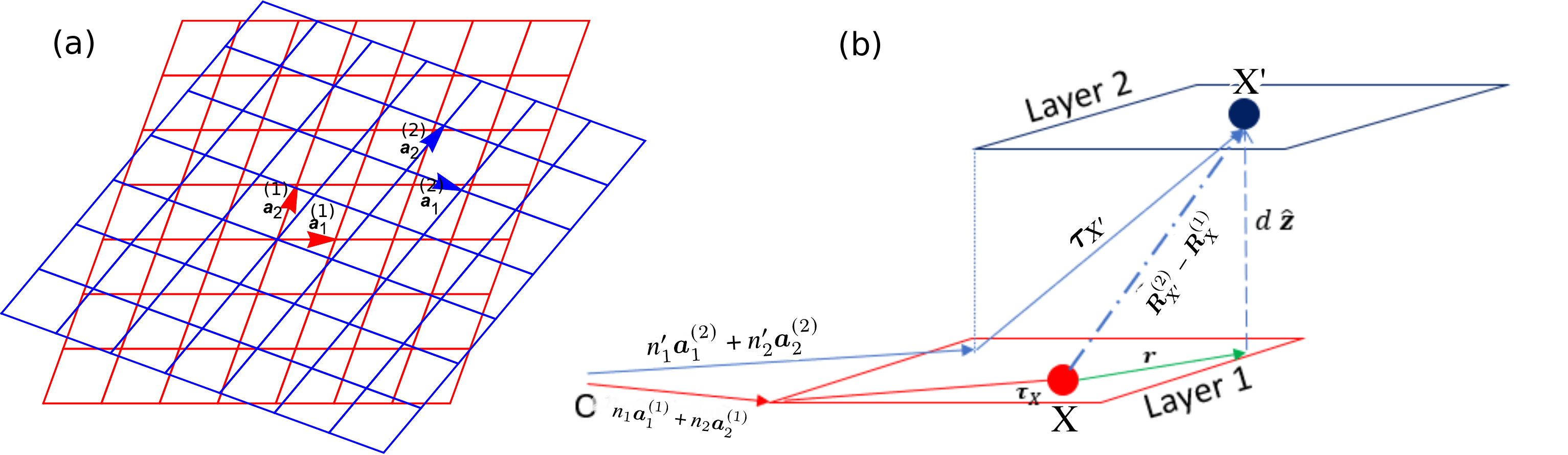}
		\caption{\label{fig:genbl} (a) The figure shows the layer 1 (red) and layer 2 (blue) with different primitive lattice vectors stacked over one another. (b) The figure shows the projection of the atom $ X' $ (blue) of an arbitrarily chosen cell of layer 2 on the layer 1. Also, the in-plane distance $ \bmr = \bm{R}_{X'} - \bm{R}_{X} - d \hat{\bm{z}}$ (green) between the atom $ X $ (red) of a cell of layer 1 and atom $ X' $.}
	\end{figure*}

	In general, the stacking of 2D layers over one another results in an incommensurate structure due to the different crystal structures or the misorientation between the constituting layers. Graphene on hBN hetero-structures discussed in section \ref{GhBN}, as well as TBLG structures at all angles other than the commensurate 
	one discussed in section \ref{CTBLG}, fall under this category. Particularly in TBLG, the incommensuration is entirely due to the misorientation between the layers. 
	We begin the discussion with a general theoretical framework provided by M. Koshino {\it et al.}\cite{Koshino2015jpsj, Koshino2015iop} to describe the interlayer interaction effect in general bilayer systems the lattice vectors of the adjacent layers may have an arbitrary choice of crystal structures and relative orientations. The structure of the interlayer coupling between the stacked layers is very much dependent on the geometrical properties of the individual layers, such as the misorientation between them. The considered structure is shown in Fig.\ref{fig:genbl}(a), and we look for the dependence of the interlayer coupling dependence on various geometrical factors. The Hamiltonian of the composite system of two bilayers can be written as \cite{Santos2007,Bistritzer2011,Koshino2015jpsj}
	\begin{equation}
		H = H_1 + H_2 + H_T
	\end{equation}
	where $ H_1 $ and $ H_2 $ are the intralayer Hamiltonians and $ H_T $ denotes the interlayer coupling between the layers.
	We write the direct and reciprocal space primitive lattice vectors for L1 as $ \bm{a}^{(1)}_i $ and $ \bm{G}^{(1)}_j $ (i,j = 1,2) and $ \bm{a}^{(2)}_i $ and $ \bm{G}^{(2)}_j $ (i,j = 1,2) for L2 which are all in XY-plane, and satisfy
	\begin{equation}
		\bm{a}^{(1)}_i\cdot\bm{G}^{(1)}_j = \bm{a}^{(2)}_i\cdot\bm{G}^{(2)}_j = 2\pi \delta_{ij}
	\end{equation}
	The area of the unit cell is given by $ A_{1} = \abs{\bm{a}^{(1)}_1 \times \bm{a}^{(1)}_2}$ and $ A_{2} = \abs{\bm{a}^{(2)}_1 \times \bm{a}^{(2)}_2}$ for layers 1 and 2, respectively.
	
	In the following, we calculate the matrix elements for $ H_{T} $ using the tight-binding model.  We consider a unit cell in each layer contains several atomic orbitals, specified by $ X=A,B,\dots $ for layer  (L1) and $ X'=A',B',\dots $ for layer 2(L2). The sublattice positions are given by $ \bm{R}^{(1)}_{X} = n_1\,\bm{a}_1 + n_2\,\bm{a}_2 + \bm{\tau}_{X} $ for L1 and $ \bm{R}^{(2)}_{X'} = n'_1\,\bm{a}^{(2)}_1 + n'_2\,\bm{a}^{(2)}_2 + \bm{\tau}_{X'} $ for layer-2, where $ n_i $ and $ n'_i $ are integers, and $ \bm{\tau}_{X} $ and $ \bm{\tau}_{X'} $ are the sublattice position inside the unit cell, which can have in-plane and out-of-plane components such that $ \bm{\tau}_{X} \cdot \hat{\bm{z}} = 0 $ and $ \bm{\tau}_{X'} \cdot \hat{\bm{z}} = d $ for L2, where $ \hat{\bm{z}} $ is the unit vector in $ z$-direction.We define 
	\bea \langle \bmr | \bm{R}^{(1)}_{X} \rangle & = & \phi_{X}(\bmr - \bm{R}^{(1)}_{X}) \nonumber \\ 
	 \langle \bmr | \bm{R}^{(2)}_{X'} \rangle & = & \phi_{X'}(\bmr - \bm{R}^{(2)}_{X'}) \label{Wannier} \eea 
	  as the atomic orbitals of the sublattice $ X $ localized at $ \bm{R}^{(1)}_{X} $ and of the sublattice $ X' $ localized at $ \bm{R}^{(2)}_{X'} $. Both the layers have their own Bloch states because of the periodicity. These Bloch states can be represented in terms of $ \ket{\bm{R}^{(1)}_{X}} $ and $ \ket{\bm{R}^{(2)}_{X'}} $ as,
	\bse
	\begin{align}
		\ket{\bmk_{1},X} = \frac{1}{\sqrt{N}_{1}} \sum_{\bm{R}^{(1)}_{X}} e^{i\bmk_{1}\vdot\bm{R}^{(1)}_{X}}\,\ket{\bm{R}^{(1)}_{X}} \\
		\ket{\bmk_{2},X'} = \frac{1}{\sqrt{N}_{2}} \sum_{\bm{R}^{(2)}_{X'}} e^{i\bmk_{2}\vdot\bm{R}^{(2)}_{X'}}\,\ket{\bm{R}^{(2)}_{X'}} \label{Bloch}
	\end{align}
	\ese
	where $ \ket{\bmk_{1},X} $ is the Bloch state corresponding to sublattice $ X $ in layer-1 and $ \ket{\bmk_{2},X'} $ is the Bloch state corresponding to sublattice $ X' $ in layer-2. One can represent the interlayer Hamiltonian $ H_T $ in the momentum space as
	\begin{equation}
		H_T = \sum_{\bmk_{1},\bmk_{2}} T^{XX'}_{\bmk_{1}\bmk_{2}} \ket{\bmk_{1},X}\,\bra{\bmk_{2},X'} + \text{H.c.}
	\end{equation}
	where $ T^{XX'}_{\bmk_{1},\bmk_{2}} $ is the corresponding matrix element of $ H_T $ over the two Bloch states $ \ket{\bmk_{1},X} $ and $ \ket{\bmk_{2},X'} $ and is given by through (\ref{Bloch}) as 
	\begin{equation}
		T^{XX'}_{\bmk_{1}\bmk_{2}} = \frac{1}{\sqrt{N_{1}N_{2}}} \sum_{\bm{R}^{(1)}_{X},\bm{R}^{(2)}_{X'}} e^{i\left(\bmk_{2}\vdot\bm{R}^{(2)}_{X'}-\bmk_{1}\vdot\bm{R}^{(1)}_{X}\right)}
		\mel{\bm{R}^{(1)}_{X}}{H_{T}}{\bm{R}^{(2)}_{X'}}
		\label{eqn:genmatel}
	\end{equation}
	where the matrix element 
	\beq \mel{\bm{R}^{(1)}_{X}}{H_{T}}{\bm{R}^{(2)}_{X'}} = t_{XX'}(\bm{R}^{(1)}_{X}, \bm{R}^{(2)}_{X'}) \nonumber \eeq is the transfer integral from the site $ \bm{R}^{(2)}_{X'} $ of layer-2 to site $ \bm{R}^{(1)}_{X} $ of layer-1 which also depends on the kind of atomic orbitals of $ X $ and $ X' $. Using (\ref{Wannier}) the integral form of
	$t_{XX'}(\bm{R}^{(1)}_{X}, \bm{R}^{(2)}_{X'}) $ can be given by
	\begin{equation}
		t_{XX'}(\bm{R}^{(1)}_{X}, \bm{R}^{(2)}_{X'}) = \int \dd[3]{\bmr}\, \phi^{*}_{X}(\bmr - \bm{R}^{(1)}_{X})\, H_{T}\, \phi_{X'}(\bmr - \bm{R}^{(2)}_{X'})
		\label{eqn:trandintTBLG}
	\end{equation}
	In the \emph{two-center} approximation, the transfer integral $ t_{XX'} $ depends on the displacement vector $ \bm{R}^{(2)}_{X'}-\bm{R}^{(1)}_{X} $ instead of depending explicitly on $ \bm{R}^{(1)}_{X} $ and $ \bm{R}^{(2)}_{X'} $. It may be noted that the general form of $ H_T $ can be written as 
	\begin{widetext}
	\begin{equation}
		H_T = \sum_{\bm{R}^{(1)}_{X}} U(\bm{R}^{(1)}_{X}) + \sum_{\bm{R}^{(2)}_{X'}} U(\bm{R}^{(2)}_{X'}) + \frac{1}{2}\sum_{\bm{R}^{(1)}_{X}, \bm{R}^{(2)}_{X'}} U(\bm{R}^{(2)}_{X'} - \bm{R}^{(1)}_X) + \\
		\frac{1}{6}\sum_{\bm{R}^{(1)}_X, \bm{R}^{(2)}_{\bar{X}}, \bm{R}^{(2)}_{\bar{X}''}} 
		U(\bm{R}^{(2)}_{X'} - \bm{R}^{(1)}_X, \bm{R}^{(2)}_{X''} - \bm{R}^{(1)}_X, \phi) + ...
	\end{equation}
	where $ \phi = \phi_{\bm{R}^{(2)}_{X'}, \bm{R}^{(2)}_{X''}, \bm{R}^{(1)}_{X}}$ where the third term can further be expanded in terms of Legendre polynomials as,
	\begin{equation}
		U(\bm{R}^{(2)}_{X'} - \bm{R}^{(1)}_X, \bm{R}^{(2)}_{X''} - \bm{R}^{(1)}_X, \phi) = \\
		\sum_{l} C_l\,f_l(\bm{R}^{(2)}_{X'} - \bm{R}^{(1)}_X, \bm{R}^{(2)}_{\bar{X}''} - \bm{R}^{(1)}_X)\,P_l(\cos(\phi))
	\end{equation}
	\end{widetext}
	In the two-center approximation, we ignore this and higher order terms. 
	With this approximation the matrix element in Eq.(\ref{eqn:genmatel}) becomes
	\begin{multline}
		T^{XX'}_{\bmk_{1}\bmk_{2}} = \frac{1}{\sqrt{N_{1}N_{2}}} \sum_{\bm{R}^{(2)}_{X'}} e^{i\bmk_{2}\vdot \bm{R}^{(2)}_{X'}}\\
		\times\sum_{\bm{R}^{(1)}_{X}} t_{XX'} (\bm{R}^{(2)}_{X'} - \bm{R}^{(1)}_{X})\,e^{-i\bmk_{1}\vdot\bm{R}^{(1)}_{X}}
		\label{eqn:transintTBLG}
	\end{multline}
	Since, $ \bm{\tau}_{X}\vdot\hat{\bm{z}} = 0 $ and $ \bm{\tau}_{X'}\vdot\hat{\bm{z}} = d $, one can write the argument of $ t_{X'X} $ into two seperate vectors, a planar vector $ \bmr $ and the component along z-axis as shown in Fig.\ref{fig:genbl}(b), that is $ t_{XX'}(\bm{R}^{(2)}_{X'}-\bm{R}^{(1)}_{X}) = t_{XX'}(\bmr + z_{X'X}\hat{\bm{z}}) $, where $ z_{X'X} = \left(\bm{\tau}_{X'}-\bm{\tau}_{X}\right)\vdot\hat{\bm{z}} $, then the in-plane inverse Fourier transform of the transfer integral is defined by
	\begin{equation}
		t_{XX'} (\bmr + z_{X'X} \hat{\bm{z}}) = \int \frac{d^{2}\bmq}{\left(2\pi\right)^{2}}~\tilde{t}_{XX'}(\bmq)~e^{-i\bmq \vdot \bmr}
		\label{eqn:transfintq}
	\end{equation}
	By inserting the above expression (\ref{eqn:transfintq}) into the second summation of (\ref{eqn:transintTBLG}), it is transformed as
	\begin{multline}
		T^{XX'}_{\bmk_{1}\bmk_{2}} = \frac{1}{\sqrt{N_{1}N_{2}}} \sum_{\bm{R}^{(2)}_{X'}} e^{i\bmk_{2}\vdot \bm{R}^{(2)}_{X'}}
		\int\frac{d^{2}\bmq}{(2\pi)^{2}}\,\tilde{t}_{XX'}(\bmq) \\
		\times
		e^{-i\bmq\vdot \bm{R}^{(2)}_{X'}}~
		e^{i\left(\bmq - \bmk_{1}\right) \vdot \bm{\tau}_{X}} 
		\sum_{n_1,n_2} e^{i\left(\bmq - \bmk_{1}\right) \vdot \left( n_1\bm{a}^{(1)}_1 + n_2\bm{a}^{(1)}_2 \right)}
		\label{eqn:transintTBLG1}
	\end{multline}
	We now replace the integral $ \int \dd[2]{\bmq}/(2\pi)^{2} \rightarrow (1/\sqrt{\mathcal{A}_{1}\mathcal{A}_{2}}) \sum_{\bmq} $, where $ \mathcal{A}_{1} = N_{1}A_{1} $ is the total area of the L1 and $ \mathcal{A}=N_{2}A_{2} $ is the total area of the L2. We now use the following identity for the reciprocal lattice for L1  \cite{Ashcroft}, i.e.,
	\begin{equation}
		\sum_{n_1,n_2} e^{i\left(\bmq - \bmk_{1}\right) \vdot \left( n_1\bm{a}^{(1)}_1 + n_2 \bm{a}^{(1)}_2 \right)} = N_{1} \sum_{\bm{G}^{(1)}} \delta_{\bmq - \bmk_{1}, \bm{G}^{(1)}} \label{Reciprocal} 
	\end{equation}
	The matrix element in (\ref{eqn:transintTBLG1}) can now be written as 
	\begin{multline}
		T^{XX'}_{\bmk_{1}\bmk_{2}} = \sqrt{\frac{N_{1}}{N_{2}}}\frac{1}{\sqrt{\mathcal{A}_{1}\mathcal{A}_{2}}} \sum_{\bm{G}^{(1)}}\tilde{t}_{XX'}(\bmk_{1}+\bm{G}^{(1)}) e^{i\bm{G}^{(1)}\vdot \bm{\tau}_{X}} \\
		\times  e^{i\left(\bmk_{2}-\bmk_{1}-\bm{G}^{(1)}\right)\vdot\bm{\tau}_{X'}}
		\sum_{n'_1,n'_2} e^{i\left(\bmk_{2}-\bmk_{1}-\bm{G}^{(1)} \right)\vdot \left( n'_1\bm{a}^{(2)}_1 + n'_2 \bm{a}^{(2)}_2 \right)}
	\end{multline}
	Repeating the identity (\ref{Reciprocal}) once more, \emph{i.e.,} 
	\beq  \sum_{n'_1,n'_2} e^{i\left( \bmk_{2} - \bmk_{1} - \bm{G}^{(1)} \right)\cdot \left(n'_1\bm{a}^{(2)}_1 + n'_2\bm{a}^{(2)}_2\right)} = N_{2} \sum_{\bm{G}^{(2)}} \delta_{\bmk_{2} - \bmk_{1} - \bm{G}^{(1)}, -\bm{G}^{(2)}}, \nonumber \eeq 
	the matrix element finally becomes
	\begin{multline}
		T^{XX'}_{\bmk_{1}\bmk_{2}} = \frac{1}{\sqrt{A_{1}A_{2}}} \sum_{\bm{G}^{(1)},\bm{G}^{(2)}}\tilde{t}_{XX'}(\bmk_{1}+\bm{G}^{(1)}) \\
		\times e^{i\left(\bm{G}^{(1)} \vdot \bm{\tau}_{X} - \bm{G}^{(2)} \vdot \bm{\tau}_{X'}\right)} \delta_{\bmk_{1} + \bm{G}^{(1)}, \bmk_{2} + \bm{G}^{(2)}}. 
		\label{eqn:genintlaymatel}
	\end{multline}
	The above form of the matrix element $ T^{XX'}_{\bmk_{1}\bmk_{2}} $ suggests that it can be non-zero only when an electronic state with a Bloch wave vector $ \bmk_{1} $ in layer-1 and one with $ \bmk_{2} $ in layer-2 are coupled such that
	\begin{equation}
		\bmk_{1} + \bm{G}^{(1)} = \bmk_{2} + \bm{G}^{(2)}
		\label{eqn:genmomentcoup}
	\end{equation}
	since the reciprocal lattice vectors of layer-1 and layer-2 are $ \bm{G}^{(1)} = m_1\,\bm{G}^{(1)}_1 + m_2\,\bm{G}^{(1)}_2 $ and $ \bm{G}^{(2)} = m'_1\,\bm{G}^{(2)}_1 + m'_2\,\bm{G}^{(2)}_2 $ respectively, the condition (\ref{eqn:genmomentcoup}) can be interpreted as a generalized \emph{Umklapp} process \cite{Koshino2015,Koshino2015iop} between arbitrary misoriented layers.

	The significance of the relation (\ref{eqn:genmomentcoup}) can be understood from an analogy with scattering by a periodic potential within the first Born approximation. In a periodic potential, the scattering matrix element of the potential between the two states with wave vectors $ \bmk_{1},\bmk_{2} $ is non-zero only when the difference between them is a reciprocal lattice vector of the underlying periodic lattice. Here the presence of interlayer coupling $ H_T $ between two layers gives rise to the scattering between Bloch states from layer-1 and layer-2. The matrix element is non-zero only when the difference between their wave vectors equals the difference between the reciprocal lattice vectors of layer-1 and layer-2. The set of vectors $ \left\{\bm{G}^{(1)}_{1}, \bm{G}^{(1)}_{2}\right\} $ forms the BZ of L1 and similarly the set of vectors $ \left\{\bm{G}^{(2)}_{1}, \bm{G}^{(2)}_{2}\right\} $ forms the BZ of L2, but their difference may or may not belong to a BZ.
	
	If the BZ does not exist, one can not obtain the electronic bands using the standard band theory of electrons in a periodic potential\cite{Ashcroft}. In a pioneering work in 2011, Bistritzer and MacDonald \cite{Bistritzer2011} showed that the incommensurate structure of TBLG can be successfully mapped to a periodic lattice in momentum space and showed the existence of the moir\'e Brillouin zone (mBZ) with the corresponding electronic bands being referred to as \emph{moir\'e} bands. In the next section \ref{BMmodel}, we shall discuss in detail in this work by dealing with the condition when the relation (\ref{eqn:genmomentcoup}) is applied to two graphene layers which are rotated with respect to each other by an arbitrary angle.
	
	\begin{figure*}
		\centering
		\includegraphics[scale=0.6]{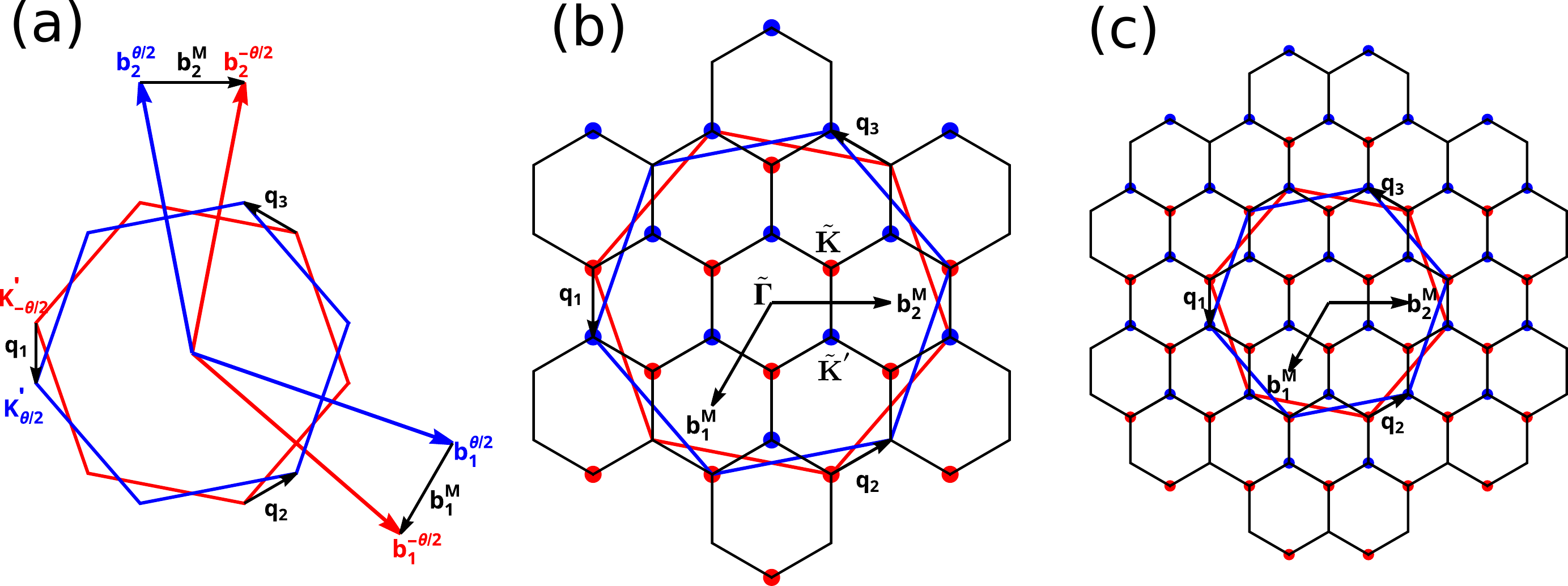}
		\caption{\label{fig:TBLGmbz} (a) It shows the BZ of individual layers (red and blue) and their reciprocal lattice vectors. The vectors $ \bm{b}^{M}_1 = \bm{b}^{-\theta/2}_{1} - \bm{b}^{\theta/2}_{1} $ and $ \bm{b}^{M}_2 = \bm{b}^{-\theta/2}_{2} - \bm{b}^{\theta/2}_{2} $ are the moir\'e reciprocal lattice vectors. $ \bmq_{1}, \bmq_{2} $ and $ \bmq_{3} $ connect the Dirac points of one layer to another. (b) The moir\'e Brillouin zone (mBZ) created for the momenta states such that their magnitude is less than $ 2\bm{b}^{M}_{1} $. (c) The moir\'e Brillouin zone for the momenta states with magnitude is less than $ 4\bm{b}^{M}_{1} $.}
	\end{figure*}

	\section{Bistritzer-MacDonald (BM) model}\label{BMmodel} 
	In section \ref{CTBLG}, we have considered the commensurate TBLG, which gives an exact one-to-one mapping between the commensurate structure in real space and the corresponding BZ in momentum space. On the other hand, at an angle other than a commensurate angle, there is no such exact periodic structure in real space. Instead the approximate periodicity comes into picture due to the emerging moir\'e pattern. The BM model shows that using the momentum-coupling condition (\ref{eqn:genmomentcoup}) for the incommensurate graphene bilayers, one can still define a BZ for such an incommensurate structure for an arbitrary angle under some conditions. This paved the way for insightful theoretical modelling of the band structure for incommensurate structure and predicted interesting physics that was subsequently verified by experiments \cite{Cao2018one, Cao2018two}. 
	
	We begin with the atomic positions in layer-1 and layer-2 respectively as 
	\bse
	\begin{align}
		\bm{R}^{(1)}_{\alpha} = \mathcal{R}(-\theta/2)\left(n_1\bm{a}_1 + n_2\bm{a}_2 + \bm{\tau}_{\alpha}\right) \\
		\bm{R}^{(2)}_{\beta} = \mathcal{R}(\theta/2)\left(n_1\bm{a}_1 + n_2\bm{a}_2 + \bm{\tau}_{\beta}\right), 
	\end{align}
	\ese
	where $ \mathcal{R} $ is the 2D rotation matrix about z-axis. The vectors $ \bm{a}_{1} $ and $ \bm{a}_{2} $ are the primitive lattice vectors of SLG as shown in Fig.\ref{fig:slgbasic}(a). The $ \alpha = A_1, B_1 $ and $ \beta = A_2, B_2 $ are sublattice indices respectively in L1 and L2, and $ \bm{\tau}_{\alpha}~\left(\bm{\tau}_{A_1} = 0, \bm{\tau}_{B_1} = \bm{\tau}\right)  $ and $ \bm{\tau}_{\beta}$  are the corresponding sublattice position in the unit cell of SLG.
	
\begin{figure}
		\centering
		\includegraphics[scale=0.6]{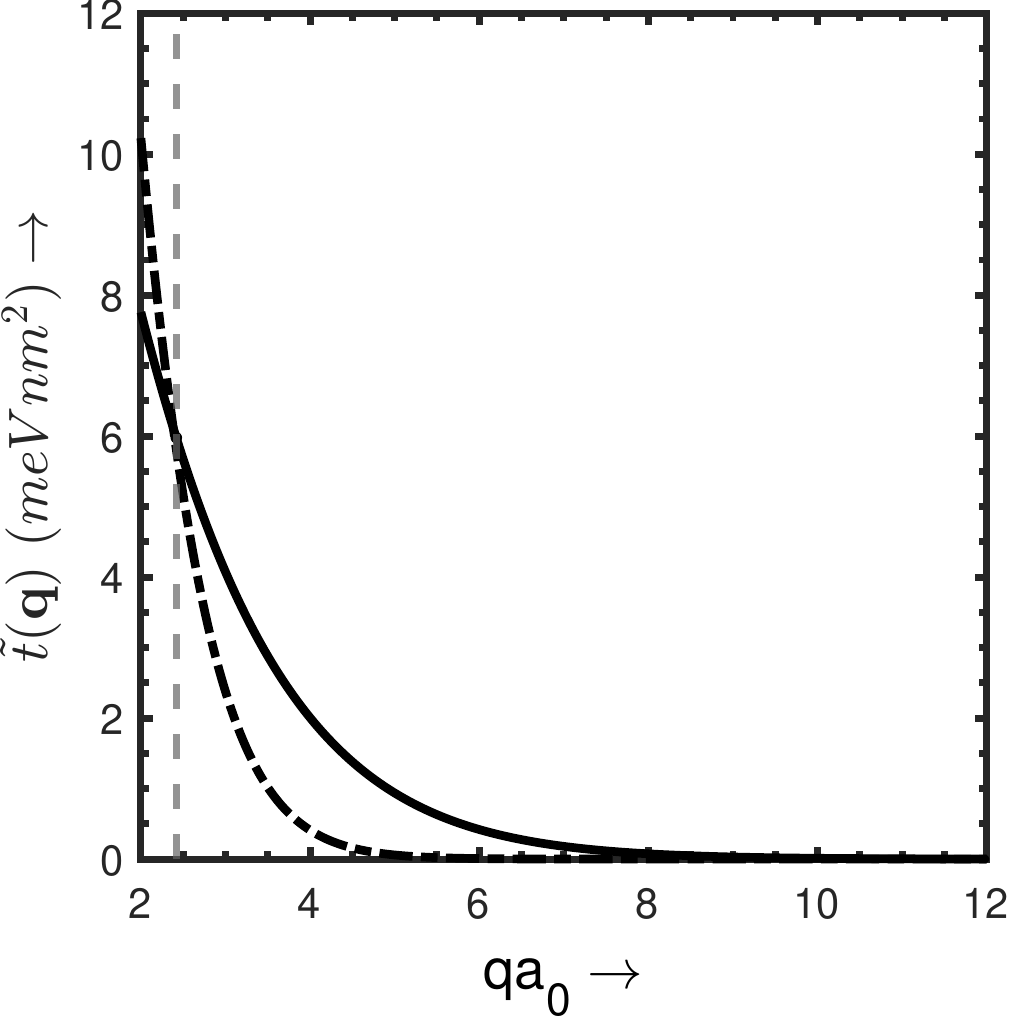}
		\caption{\label{fig:ilhoppingpar}It shows the rapid decay of the Fourier amplitude $ t_{\bmq} $ of the interlayer coupling. The black solid line is plotted for the functional form of $ V_{\text{pp}\pi} $ and $ V_{\text{pp}\sigma} $ given in \cite{Moon2013}, while the black (dashed-dot) line shows the plot for ansatz in (\ref{eqn:BMansatz}). The vertical dashed gray line parallel to y-axis cross the x-axis at $ K_{D}a_{0} = 4\pi/3\sqrt{3} $.}
	\end{figure}

	The orbitals at each sublattice position are identical, and therefore the sublattice indices in the transfer integral $ t\left(\bm{R}^{(2)}_{\beta}-\bm{R}^{(1)}_{\alpha}\right)$ such as one appeared in eq. (\ref{eqn:transintTBLG}), is dropped. The behaviour of this transfer integral can be understood by writing the spatial dependence 
	of $ t $ in terms of the Slater-Koster parameterization as \cite{Slater1954}
	\begin{equation}
		t(\bm{R}) = -V_{pp\pi} \left( 1 - n_z^2 \right) - V_{pp\sigma} n_z^2
	\end{equation}
	where $ n_z = \bm{R}\cdot\bm{\hat{z}}/\abs{\bm{R}} $ is the projection of unit vector of $ \bm{R} $ on z-axis. $ V_{pp\pi} = V^0_{pp\pi}\,e^{-\left(R-a/\sqrt{3}\right)/r_{0}} $ and $ V_{pp\sigma} = V^0_{pp\sigma}\,e^{-\left(R-d\right)/r_{0}} $ are distance dependent parameters \cite{Trambly2010,Moon2013}. 
	The constants $ a\approx0.246\,nm $ is the lattice constant of SLG, $d\approx0.335\,nm $ is the interlayer spacing, $ V^0_{pp\pi}\approx-2.7\,eV $ is the transfer integral between the nearest-neighbour atoms of SLG, and $ V^0_{pp\sigma} \approx 0.48\,\si{\electronvolt} $ is the integral between vertically located atoms on the neighbouring layers. $ r_0 \approx 0.184\,a $ is the decay length of the transfer integral. Setting $ \bm{R} = \sqrt{r^2 + d^2}$
yields 	
\begin{equation}
		t(\bm{R}) = -V^{0}_{\text{pp}\pi} e^{-\frac{\left(\abs{\bm{R}}-a_{0}\right)}{r_{0}}}
		\frac{r^{2}}{r^2 + d^2}
		-V^{0}_{\text{pp}\sigma} e^{-\frac{\left(\abs{\bm{R}}-d\right)}{r_{0}}}
		\frac{d^{2}}{r^2 + d^2}. 
	\end{equation}
	The in-plane FT for $t(\bm{R})$ using the plane-polar coordinates gives 
	\begin{equation}
		\tilde{t}(\bmq) =\int_{0}^{\infty}\dd{r}\,r\,t(\bm{R}) \int_{0}^{2\pi}\dd{\theta}\,e^{-iq\,r\,\cos(\theta)}
	\end{equation}	
	The above integration is obtained numerically and has been plotted in Fig.(\ref{fig:ilhoppingpar}) along with an \emph{ansatz} which has been fitted directly for $ t_{\bmq} $ \cite{Bistritzer2010} 
	and is given as 
	\begin{equation}
		\tilde{t}(\bmq)_{{\it ansatz}} = t_{0}\,e^{-\alpha\left(q\,d\right)^{\gamma}}. 
		\label{eqn:BMansatz}
	\end{equation}
	Here $ t_{0} = 0.02\,eV\,\si{\nano\meter}^{2} $, $ \alpha = 0.13 $, $ \gamma = 1.25 $ and $ d = 0.335\,\si{\nano\meter} $ is the interlayer distance. As a result, the $ \tilde{t}_{\bm{q}} $ (2D-Fourier transform of $ t $) comes out to be a rapidly decaying function in $ \bmq $, which leads to the short-rangedness of $ \tilde{t}_{\bmq} $ in momentum space.
	
	For the case of TBLG, the matrix element of interlayer Hamiltonian $ H_{T} $ in over the Bloch states $ \ket{\bmk_{1},\alpha} $ and $ \ket{\bmk_{2},\beta} $ given in (\ref{eqn:genintlaymatel}) can be written as 	
	\begin{multline}
		T^{\alpha\beta}_{\bmk_{1}\bmk_{2}} = \sum_{\bm{G}^{(1)},\bm{G}^{(2)}}
		\frac{\tilde{t}(\bmk_{1}+\bm{G}^{(1)})}{\Omega} e^{i\left(\bm{G}^{(1)}\vdot\bm{\tau}_{\alpha}-\bm{G}^{(2)}\vdot\bm{\tau}_{\beta}\right)}
		\\
		\times \delta_{\bm{k}_1+\bm{G}^{(1)},\bm{k}_2+\bm{G}^{(2)}} 
		\label{eqn:intlaymel}
	\end{multline}
	where $ \Omega $ is the unit cell area. The reciprocal lattice vectors of L1 and L2 for the TBLG  are $ \bm{G}^{(1)} = \mathcal{R}(-\theta/2)\left(m_1\bm{
	b}_{1} + m_2\bm{b}_{2}\right) $ and $ \bm{G}^{(2)} = \mathcal{R}(\theta/2)\left(m_1\bm{
	b}_{1} + m_2\bm{b}_{2}\right) $ respectively, with $ \bm{b} $ are the reciprocal lattice vectors of SLG. The wave-vectors $ \bmk_{1} $ and $ \bmk_{2} $ are measured from the center $ \Gamma $-point of the BZ of layer-1 and layer-2 respectively. Replacing the wave-vectors $ \bmk_{1} \rightarrow \bm{k}_1 + \bm{K}^{(1)} $ and $ \bmk_{2} \rightarrow \bm{k}_{2} + \bm{K}^{(2)} $, where $ \bm{K}^{(1)} $ and $ \bm{K}^{(2)} $ of the Dirac points of respective layers one gets the coupling condition in (\ref{eqn:genmomentcoup}) for the TBLG as 
	\begin{equation}
		\bm{k}_1 = \bm{k}_{2} + \bm{K}^{(2)} - \bm{K}^{(1)} +\bm{G}^{(2)} - \bm{G}^{(1)}. 
		\label{eqn:coupcond1}
	\end{equation}
	Here the $ \bmk_{1} $ and $ \bmk_{2} $ are measured from the Dirac points of respective layers. Inserting the expressions of $ \bm{G}^{(1)} $ and $ \bm{G}^{(2)} $ in (\ref{eqn:coupcond1}) one gets
	\begin{equation}
		\bmk_{1} = \bmk_{2} +\Delta\bm{K} + m_1\left(\bm{b}^{\theta/2}_{1} -\bm{b}^{-\theta/2}_{1}\right) + m_2 \left(\bm{b}^{\theta/2}_{2} -\bm{b}^{-\theta/2}_{2}\right)
	\end{equation}
	where $ \Delta\bm{K} = \bm{K}^{(2)} - \bm{K}^{(1)} $, $ \bm{b}^{\theta/2}_{j} = \mathcal{R}(\theta/2)\bm{b}_{j} $ and $ \bm{b}^{-\theta/2}_{j} = \mathcal{R}(-\theta/2)\bm{b}_{j} $ for $ j=1,2 $. It provides the moir\'e reciprocal lattice vectors $ \bm{b}^{M}_{1} = \bm{b}_{1}^{-\theta/2}-\bm{b}_{1}^{\theta/2} $ and $ \bm{b}^{M}_{2} = \bm{b}_{2}^{-\theta/2}-\bm{b}_{2}^{\theta/2} $ as can be noted from (\ref{eqn:moiregenrec}). This is shown in Fig.\ref{fig:TBLGmbz}(a). The mBZ is constructed using the reciprocal lattice vectors $ \bm{b}^{M}_{1} $ and $ \bm{b}^{M}_{2} $ as shown in Figs.\ref{fig:TBLGmbz}(b)-(c).
	
	The Fourier amplitude $ \tilde{t}(\bmq)$ of the interlayer coupling is given as $ \tilde{t}\left(\bmk_{1} + \bm{K}^{(1)} + \bm{G}^{(1)}\right) $. Due to the rapid decay of $ \tilde{t} $ in momentum space, a significant value results when the reciprocal lattice vector $ \bm{G}^{(1)} $ takes one of the values in $ \left\{0,-\bm{b}^{-\theta/2}_{1},-\bm{b}^{-\theta/2}_{1}-\bm{b}^{-\theta/2}_{2}\right\} $ and the value comes out to be $ \tilde{t}\left(\bmk_{1} + \bm{K}^{(1)} + \bm{G}^{(1)}\right)/\Omega = w \approx 110~\si{\milli\electronvolt} $ \cite{Bistritzer2011}. The vectors that connect the nearest Dirac points of both the layers for left valley $ \xi = -1 $ are 
	\bea \bm{q}_1 & = & \Delta\bm{K} = \bm{K}'_{\theta/2} - \bm{K}'_{-\theta/2}, \nonumber \\
	\bm{q}_2  & = & \Delta\bm{K} - \bm{b}^{M}_{1} \nonumber \\ 
	\text{and},~ \bm{q}_3 & = & \Delta\bm{K} - \bm{b}^{M}_{1} - \bm{b}^{M}_{2} \label{mreciprocal} \eea
	as shown in Fig.\ref{fig:TBLGmbz}(a). The interlayer tunnelling matrix element defined in (\ref{eqn:intlaymel}) can now be written in terms of them and can be denoted as $T^{\alpha \beta}(\bs{q})$. This 
	depends upon the initial stacking configuration of graphene layers. For the case of Bernal stacked bilayer graphene, the interlayer tunnelling matrices $T^{\alpha\beta}(\bm{q}_{1}) , T^{\alpha \beta}(\bm{q}_{2})$ and $T^{\alpha \beta} \bm{q}_{3}$ are listed in Table-(\ref{table:inttunnmatBM}).
	For brevity, we shall refer to them respectively as $T_{1}, T_{2}, T_{3}$
	\begin{table}
		\centering
		\caption{\label{table:inttunnmatBM} The most dominant Fourier amplitudes for the Bernal-stacked BLG where $ \phi=2\pi/3 $. $T^{\alpha \beta}(\bmq)$ is defined in the text.}
		\begin{ruledtabular}
			\begin{tabular}{cccc}
				$ \bm{q} $ & $ \bm{q}_1 $ & $ \bm{q}_2 $ & $ \bm{q}_{3} $ \\
				\hline
				$ T^{\alpha \beta}(\bmq) $  & 
				$ w\begin{bmatrix}
					1&1 \\
					1&1\\
				\end{bmatrix} $ &
				$ w\begin{bmatrix}
					e^{i\phi}&1\\
					e^{-i\phi}&e^{i\phi}
				\end{bmatrix} $ &
				$ w\begin{bmatrix}
					e^{-i\phi}&1\\
					e^{i\phi}&e^{-i\phi}
				\end{bmatrix} $ \\
			\end{tabular}
		\end{ruledtabular}
	\end{table}
	\begin{figure*}
		\centering
		\includegraphics[scale=0.35]{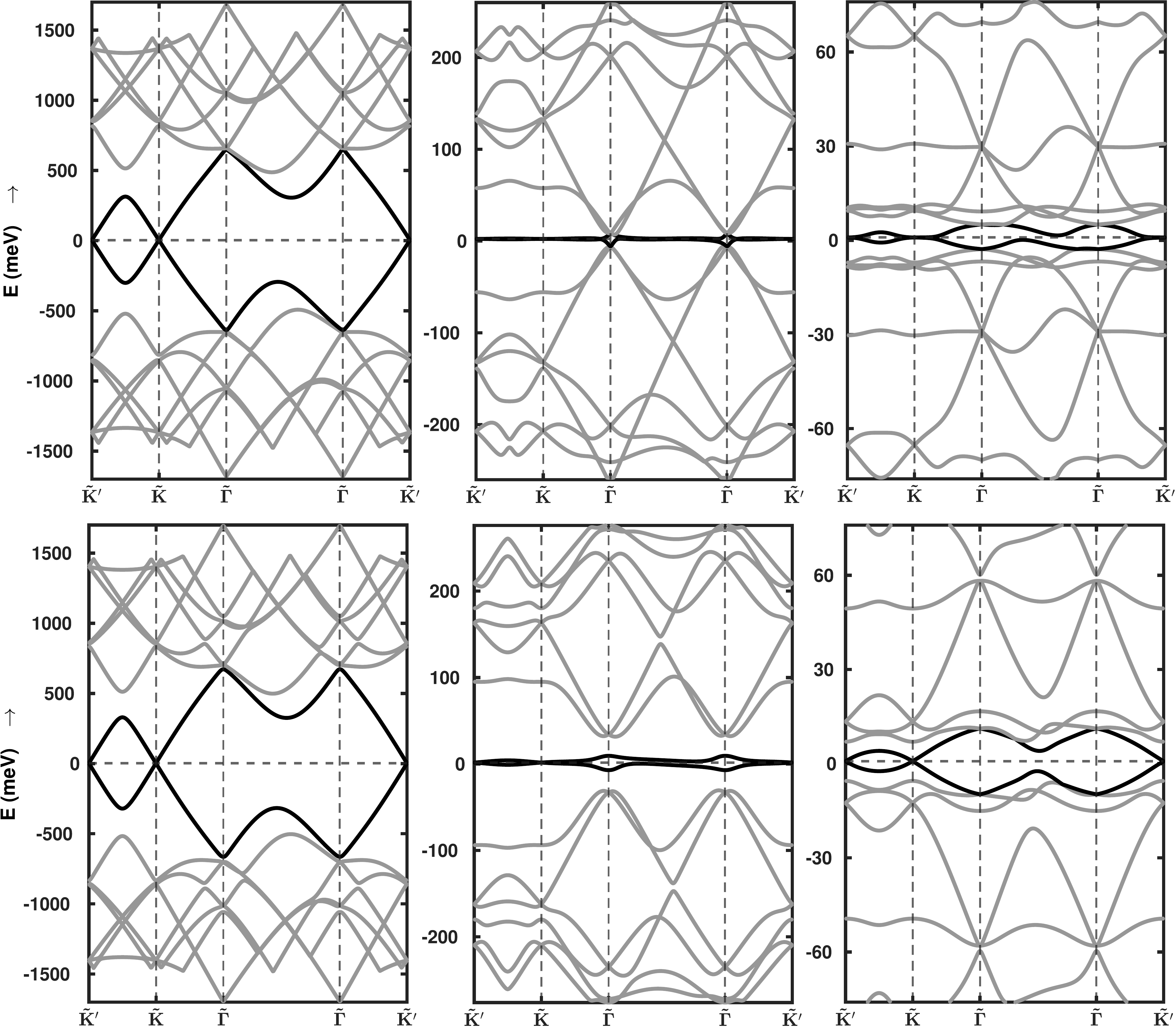}
		\caption{(Top) The Band structure of TBLG is plotted against the high-symmetry path $ \tilde{\bm{K}'}\rightarrow \tilde{\bm{K}} \rightarrow \tilde{\Gamma} \rightarrow \tilde{\Gamma} \rightarrow \tilde{\bm{K}'} $ at twist angles $ \theta = 5^{\circ }$, $ \theta = \ang{1.05} $ and $ \theta = \ang{0.5} $. The interlayer hopping amplitude $ w_{AA} = \tilde{t}(\bmq)/\Omega $ in AA-rich region and $ w_{AB} = \tilde{t}(\bmq)/\Omega $ in AB-rich region are considered identical and is equal to $ 110~\si{\milli\electronvolt} $ \cite{Bistritzer2011}. The gray dashed line at $ E = 0 $ is the Fermi level and the two bands nearest to the Fermi level is shown in solid black lines. (Bottom) The Band structure of TBLG is plotted against the same high-symmetry path as in top figure at twist angles $ \theta = 5^{\circ }$, $ \theta = \ang{1.05} $ and $ \theta = \ang{0.5} $. The interlayer hopping amplitude $ w_{AA} = \tilde{t}(\bmq)/\Omega $ in AA-rich region and $ w_{AB} = \tilde{t}(\bmq)/\Omega $ in AB-rich region are considered to be different and are equal to $ w_{AA} = 79.7~\si{\milli\electronvolt} $ and $ w_{AB} = 97.5~\si{\milli\electronvolt} $. \cite{Koshino2018}}
		\label{fig:TBLGbands}
	\end{figure*}

\begin{figure}
		\centering
		\includegraphics[scale=0.38]{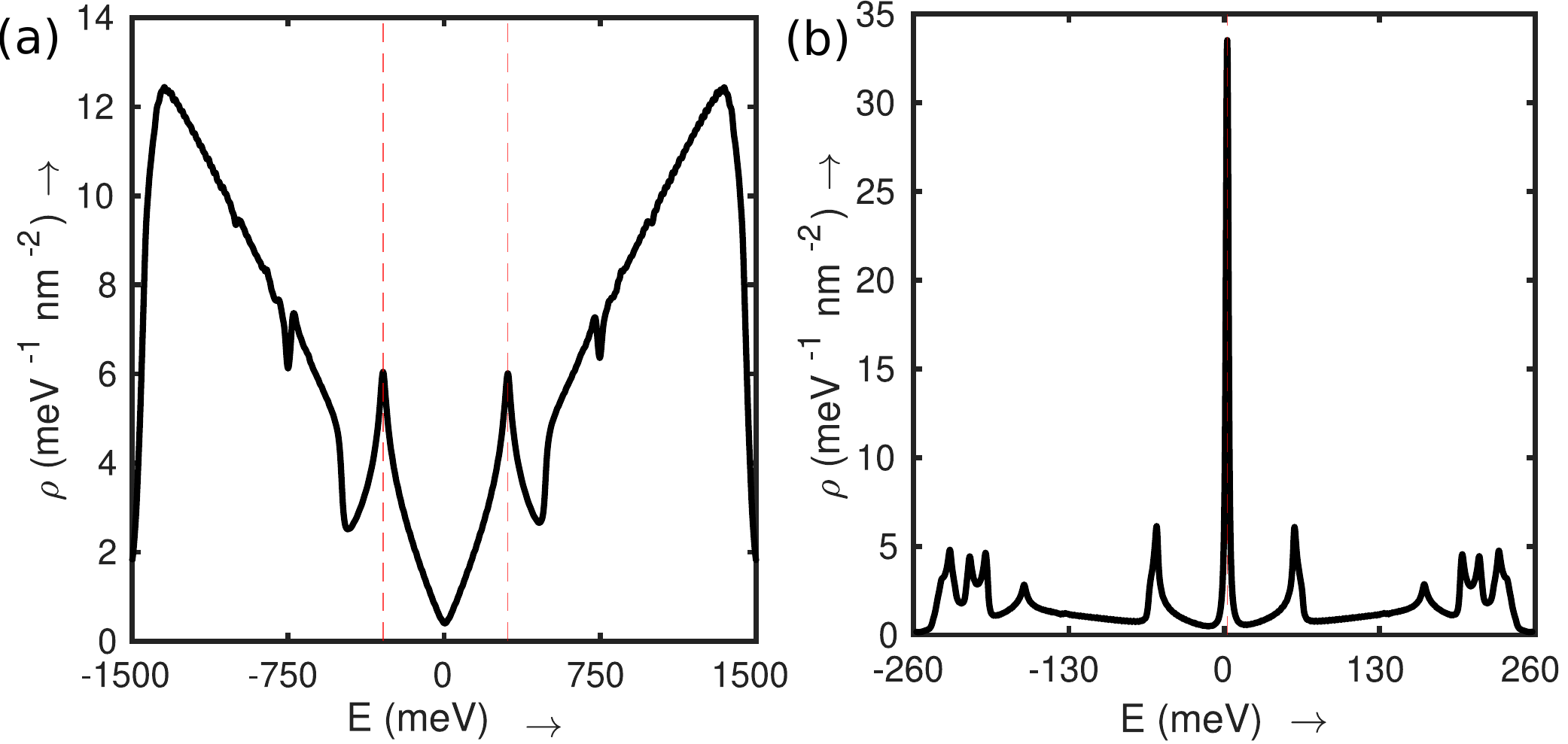}
		\caption{\label{fig:TBLGdos}(a) It shows the DOS plot for TBLG at twist angle $ \theta = \ang{5} $. The locations of VHS are marked using red dashed lines parallel to y-axis. (a) The DOS plot at the first magic angle $ \theta = \ang{1.05} $. The locations of VHS are marked using red dashed lines parallel to y-axis.}
	\end{figure}

	The $ \bmk $-space Hamiltonian for the system is $H=H_1+H_2+H_{\perp}$. In the tight-binding approximation, the Bloch function is,
	\begin{equation}
		|\psi_{\bmk} \rangle=\sum_{\alpha,\beta}c^{(1)}_{\alpha}(\bmk)|\psi^{(1)}_{\bmk,\alpha}\rangle + c^{(2)}_{\beta}(\bmk)|\psi^{(2)}_{\bmk,\beta}\rangle
	\end{equation}
	As written above, the matrix element of the interlayer term is $ \langle \psi^{(1)}_{\bmk_{1},\alpha}|H_{\perp}|\psi^{(2)}_{\bmk_{2},\beta}\rangle $, the matrix elements corresponding to the intralayer Hamiltonian are
	\bse
	\begin{align}
		\langle \psi^{(1)}_{\bmk_{1},\alpha}|H_1|\psi^{(1)}_{\bmk_{1}',\alpha'}\rangle & =h^{\alpha,\alpha'}_{\bmk_{1}}\left(\frac{\theta}{2}\right)\delta_{\bmk_{1},\bmk'_{1}} \\
		\langle \psi^{(2)}_{\bmk_{2},\beta}|H_2|\psi^{(2)}_{\bmk'_{2},\beta'}\rangle & =h^{\beta,\beta'}_{\bmk_{2}}\left(-\frac{\theta}{2}\right)\delta_{\bmk_{2},\bmk'_{2}}
	\end{align}
	\ese
	\begin{widetext}
	Therefore, the $ 8\times8 $ Hamiltonian can be written as
\begin{equation}
H(\bm{k})=
\begin{pmatrix}
h_{ \bm{k}}(\frac{\theta}{2})&T_{1}&T_{2}&T_{3}\\
T^{\dagger}_{1}&h_{\bm{k}+\bm{q}_{1}}(\frac{-\theta}{2})&0&0\\
T^{\dagger}_{2}&0&h_{\bm{k}+\bm{q}_{2}}(\frac{-\theta}{2})&0\\
T^{\dagger}_{3}&0&0&h_{ \bm{k}+\bm{q}_{3}}(\frac{-\theta}{2})
\end{pmatrix}
\label{eqn:TBLGminham}
\end{equation}
	\end{widetext}
	This procedure can be carried out by considering a large number of plane-wave states to obtain a Hamiltonian of higher dimension which captures the bands with vanishing Fermi velocity in a certain region. The band structures at two different twist angles are shown in Fig.(\ref{fig:TBLGbands}). The DOS corresponding to two twist angles $ \theta = \ang{5} $ and $ \theta = \ang{1.05} $ is shown in Figs.\ref{fig:TBLGdos}(a)-(b). The emergent symmetries in the band structure of TBLG in the limit of small twist angles and the corresponding tight-binding model were discussed in detail in Ref.\cite{Zou2018}. A detailed discussion of the same though is outside the scope of this pedagogical review. In section \ref{2BM} we shall elaborate on the treatment given in the BM work by which they calculated the first magic angle analytically. Namely, we shall project the system on to the two lowest bands at the Fermi energy to analytically obtain the magic twist angle at which the bands flatten and the Fermi velocity, which is the salient finding of the BM model. 
	To obtain the twist-angle at which these bands become flat, we isolate these two bands as is given in the next section.
	
	\subsection{Effective two-bands model}\label{2BM}
	In the vicinity of the Dirac point, one can derive an effective two-bands Hamiltonian by projecting the $ 8 \times 8 $ Hamiltonian in the space of zero-energy doublet of $ H $ with $ \bmk = 0 $ \cite{Bistritzer2011}. This can be done by writing the Hamiltonian as a sum of the $ \bmk = 0 $ term $ H_{\bmk = 0} = H^{(0)} $ and $ \bmk \ne 0 $ term $ H_{\bmk \ne 0} = H^{(1)}$, namely 
	\begin{equation}
		H_{\bmk} = H^{(0)} + H^{(1)}. 
	\end{equation}
	Here the unperturbed part $ H^{(0)} $ is given by
	\begin{equation}
		H^{(0)} = 
		\begin{pmatrix}
			O_2 & T_1 & T_2 & T_3 \\
			T_1^{\dagger} & h_{\bmq_1}(-\theta/2) & O_2 & O_2 \\
			T_2^{\dagger} & O_2 & h_{\bmq_2}(-\theta/2) & O_2 \\
			T_3^{\dagger} & O_2 & O_2 & h_{\bmq_3}(-\theta/2)
		\end{pmatrix}
	\end{equation}
	In the neighbourhood of the Dirac points $ \tilde{\bm{K}'} $ or $ \tilde{\bm{K}} $, the Hamiltonian $ H^{(1)} $ is treated as the perturbation. The basic idea is to write the representation of the perturbing part in the degenerate zero energy subspace of the $H^{0}$ to get an analytical form of the effective Hamiltonian in a form analogous to that of the charge carriers of SLG. One can then get the renormalised Fermi velocity from the same in an analytic form. The subsequent treatment is similar to the one done with Hamiltonian (\ref{Hconti3}) and (\ref{Hconti4}) in section \ref{contimod}. However, in this case, the problem of divergence that appears in the limit of arbitrarily small twist through the expression (\ref{velocityrenorm1}) does not appear. Instead, one gets a finite modification of the Fermi velocity for any arbitrarily small twist that shows the existence of the magic angle.
	 
As earlier, we identify the zero-energy doublet of $ H^{(0)} $, from the Schr\"{o}dinger equation, $ H^{(0)}\Psi = E~\Psi $, where $ \Psi = \Pmqty{\psi_0 & \psi_1 & \psi_2 & \psi_3}^{T} $ is an 8-component column vector with each $ \psi_j $ being a two-component column vector. 
 For $E=0$, the first diagonal element of the Hamiltonian (\ref{eqn:TBLGminham}) at $\bmk=0$ gives $ h_{\bmk = 0}(\theta/2)\psi_0 = 0 $. 
 In terms of the components, the other equations from $ H^{(0)}\Psi = E~\Psi $ yield
	\bse
	\begin{align}
		\sum_{i=1}^{3} T_i~\psi_i & = E~\psi_0 \\
		T^{\dagger}_{j}\,\psi_{0} + h_{\bmq_{j}}(-\theta/2)\,\psi_{j} & = E\,\psi_{j} \quad \forall \quad j =1,2,3
		\label{eqn:psij}
	\end{align}
	\ese
	The eigenvalue equation  for $ \psi_{0} $ gives two independent solutions for $ \psi_0 $ which we denote as $ \psi_0^{(1)} $ and $ \psi_0^{(2)} $. Once the $ \psi_{0} $ is chosen, the other components $ \psi_{j} $, for $ j=1,2,3 $ can be determined in terms of $ \psi_0 $ using (\ref{eqn:psij}) for a given energy $ E $. The same holds for $E=0$. Therefore, the different components of zero-energy eigenstates satisfy
	\bse
	\begin{align}
		\psi_{j} & = -h^{-1}_{\bmq_{j}}(-\theta/2)~T^{\dagger}_{j}~\psi_0 \quad \forall \quad j = 1,2,3 \label{eqn:psij1} \\
		\sum_{i=1}^{3} & T_i~h^{-1}_{\bmq_{i}}(-\theta/2)~T^{\dagger}_{i}~\psi_0 = 0 \label{eqn:psij2}
	\end{align}
	\ese
	
	Corresponding to the two linearly independent solutions of $ \psi_0 $, that is $ \psi_0^{(1)} $ and $ \psi_0^{(2)} $, we denote the zero-energy eigenstates of the full $ H^{(0)} $ as $ \Psi^{(1)} $ and $ \Psi^{(2)} $, respectively. We now choose $ \psi^{(1)}_0 = \mqty(1 & 0)^{T} $ and $ \psi^{(2)}_0 = \mqty(0 & 1)^{T} $. Then the different components $ \psi^{(1)}_{j} $ can be obtained using (\ref{eqn:psij1}), 
	\begin{equation}
		\psi^{(1)}_{j}= \frac{w}{\hbar v_F\,\abs{\bmq_{j}}}
		\begin{pmatrix}
			e^{i\left[\theta_{\bmq_{j}} + \theta/2\right]} \\
			e^{-i\left[\theta_{\bmq_{j}} + \theta/2 - \left(j-1\right)\phi \right]}
		\end{pmatrix}
		\quad \forall \quad j=1,2,3. 
	\end{equation}
	Even though $ \psi^{(1)}_{0} $ is chosen to be normalized, the other components in terms of $ \psi_{0} $ that make $ \Psi^{(1)} $ do not necessarily make $ \Psi^{(1)} $ normalized. The norm of $ \Psi^{(1)} $ is
	\begin{equation}
		\norm{\Psi^{(1)}} = \sqrt{\Psi^{(1)\dagger}~\Psi^{(1)}} = \sqrt{1+6\alpha^2}
	\end{equation}
	where $ \alpha^2 = w^2/\left(4~\hbar^2~v_F^2~K_D^2~\sin[2](\theta/2)\right) $. Thus the normalized $ \Psi^{(1)} $ become
	\begin{equation}
		\Psi^{(1)} = \frac{1}{\sqrt{1+6\alpha^2}}
		\begin{pmatrix}
			\psi_0^{(1)}  &	\psi_1^{(1)} & \psi_2^{(1)} & \psi_3^{(1)}
		\end{pmatrix}^{T}
	\end{equation}
	Similarly from $ \psi^{(2)}_{0}$, we can determine the the normalized $ \Psi^{(1)} $. These two degenerate zero-energy eigenstates $ \Psi^{(1)} $ and $ \Psi^{(2)} $ of $ H^{(0)} $ correspond to a vanishing eigenvalue. We can write the perturbing Hamiltonian in this degenerate subspace of zero-energy eigenvalue, namely as $ H^{(1)}_{ij} = \mel{\Psi^{(i)}}{H^{(1)}}{\Psi^{(j)}} $. 
	
	To do that, we first find the action of the operator $ H^{(1)} $ on $ \ket{\Psi^{(j)}} $ yielding 
	\begin{equation}
		H^{(1)} \ket{\Psi^{(j)}} = \frac{1}{\sqrt{1+6\alpha^2}}
		\begin{pmatrix}
			h_{\bmk}(\theta/2) \psi^{(j)}_0 + \sum_{\alpha=1}^{3} T_{\alpha} \psi^{(j)}_{\alpha} \\
			T^{\dagger}_1 \psi^{(j)}_0 + h_{\bmk_1}(-\theta/2) \psi^{(j)}_1 \\
			T^{\dagger}_2 \psi^{(j)}_0 + h_{\bmk_2}(-\theta/2) \psi^{(j)}_2 \\
			T^{\dagger}_3 \psi^{(j)}_0 + h_{\bmk_3}(-\theta/2) \psi^{(j)}_3		
		\end{pmatrix}
	\end{equation}
	Using the relation in (\ref{eqn:psij1}) between different components of the eigenket, therefore the matrix element becomes
	\begin{widetext}
	\begin{multline}
		H^{(1)}_{ij} = \frac{1}{1+6\alpha^2} \Bigg[
		\psi^{(i)\dagger}_0~h_{\bmk}(\theta/2)~\psi^{(j)}_0 - 2\, \psi^{(i)\dagger}_0~\sum_{\alpha=1}^{3} T_{\alpha}~h^{-1}_{\bmq_{\alpha}}(-\theta/2)~T^{\dagger}_{\alpha}~\psi^{(j)}_{0} + \\ \sum_{\alpha=1}^{3}\psi^{(i)\dagger}_0~T_{\alpha}~h^{-1\dagger}_{\bmq_{\alpha}}(-\theta/2)~h_{\bmk_{\alpha}}(-\theta/2)~h^{-1}_{\bmq_{\alpha}}(-\theta/2)~T^{\dagger}_{\alpha}~\psi^{(j)}_0 \Bigg]
	\end{multline}
	The middle term vanishes because of eq. (\ref{eqn:psij2}), leading to 
	\begin{equation}
		H^{(1)}_{ij} = \frac{1}{1+6\alpha^2}~
		\psi^{(i)\dagger}_0 \left[ h_{\bmk}(\theta/2) +	
		\sum_{\alpha=1}^{3} T_{\alpha}~h^{-1\dagger}_{\bmq_{\alpha}}(-\theta/2)~h_{\bmk_{\alpha}}(-\theta/2)~h^{-1}_{\bmq_{\alpha}}(-\theta/2)~T^{\dagger}_{\alpha}
		\right] \psi^{(j)}_0
	\end{equation}	
	\end{widetext}
	The second term can be simplified as shown in Appendix-(\ref{appendix:a}). Therefore, one gets
	\begin{equation}
		\sum_{\alpha=1}^{3} T_{\alpha}~h^{-1\dagger}_{\bmq_{\alpha}}~h_{\bmk_{\alpha}}~h^{-1}_{\bmq_{\alpha}}~T^{\dagger}_{\alpha} = \frac{3~w^2}{\hbar^2~v_F^2~q^2} \bm{\sigma} \cdot \bmk = 3~\alpha^2~\bm{\sigma} \cdot \bmk
	\end{equation}
	Finally, the matrix element becomes
	\begin{equation}
		\mel{\Psi^{(i)}}{H^{(1)}}{\Psi^{(j)}} = -\frac{1-3~\alpha^2}{1+6~\alpha^2}~\hbar~v_F~\psi^{(i)\dagger}_0~\bm{\sigma} \cdot \bmk~\psi^{(j)}_0 \label{magicangle}
	\end{equation}
	Further using the solutions for $ \psi_0 $, the perturbation matrix comes out to be
	\begin{equation}
		H^{(1)} = -\hbar~v^*~\bm{\sigma} \cdot \bmk
	\end{equation}
	Eq. (\ref{magicangle}) shows that at $\alpha = \frac{1}{\sqrt{3}}$, the Fermi velocity vanishes at the $K$ and $K'$ points and the bands become flat. The corresponding twist angle is called the magic angle. Detailed numerical calculation shows a set of angles at which the Fermi velocity of the charge carrier vanishes. At each magic angle, the Fermi velocity at the Dirac point vanishes and the two-lowest electronic bands get flattered in neighbourhood region around it. However, the bandwidth between the lowest two bands at the centre of the mBZ does not vanish and is around $ 7-10~\si{\milli\electronvolt} $. Theoretically, in \textbf{a} later work, it was shown that the interlayer hopping amplitude could be tweaked to give perfectly flat bands as shown \cite{Tarnopolsky2019}. Their model not only produces the perfectly flat bands since the bandwidth at $ \Gamma $-point vanishes but also provides the origin of magic angles at which the Fermi velocity of the Dirac point vanishes. In the subsequent section \ref{PFB}, we review this model to understand the origin such perfectly flat bands and the magic angles.

	\section{Perfectly flat bands}\label{PFB}
	The Hamiltonian in the BM model (\ref{eqn:TBLGminham}) when written in the real space (for a detailed derivation see Appendix-(\ref{appendix:b})), appears as 
	\begin{equation}
		H = \Pmqty{-i\hbar v_{F}\bm{\sigma}_{\theta/2}\vdot\bm{\nabla} & T(\bmr) \\
		T^{\dagger}(\bmr) & -i\hbar v_{F}\bm{\sigma}_{-\theta/2}\vdot\bm{\nabla}}
		\label{eqn:BMham2}
	\end{equation}
	Here the rotated Pauli matrices are $ \bm{\sigma}_{\theta} = e^{-i\theta\sigma_z/2}\left(\sigma_x,\sigma_y\right)e^{i\theta\sigma_z/2}$, $ v_{F} $ is again the Fermi velocity in SLG, and the position dependent interlayer tunnelling matrix $T(\bmr)$ is given as 
	\bse
	\begin{align}
		& T(\bmr) = \sum_{n=1}^{3} T_{n}\,e^{-i\bmq_{n}\vdot\bmr}\label{eqn:Tndef}
		\\
		T_{n} = w_{0}\sigma_{0} + w_{1} & \left[\sigma_{x}\,\cos(n-1)\phi + \sigma_y\,\sin(n-1)\phi\right]
	\end{align}
	\ese
	with $ n=1,2,3 $. The vectors $ \bmq_{1},\bmq_{2} $ and $ \bmq_{3} $ are defined in (\ref{mreciprocal}), and have equal magnitude whose value is $ k_{\theta} $.
	
	In the BM model, the interlayer hopping across the AA-regions ($ w_{0} $) and AB/BA-regions ($ w_{1} $) are considered identical $ (w_{0} = w_{1} = w) $. This consequently results in the isolation of the two bands nearest to the Fermi level from other higher bands except at the $ \Gamma $-point \cite{Bistritzer2010} as shown in Fig.(\ref{fig:TBLGbands}). It was illustrated that the interlayer hopping parameters affects the bandwidth at the $ \Gamma $-point. A detailed analysis of the effect of the hopping parameter can be found in \cite{Guinea2019}. It was also observed that the interlayer hopping amplitudes across the two regions AA- and AB/BA- are not identical. Instead, they differ in such a way so that the ratio of the hopping amplitudes in AA-region to AB/BA $ \left(\kappa = w_{0}/w_{1}\right) $ is around $ 0.7-0.8 $. The different hopping amplitudes in each region isolate the two lowest bands also at the $ \Gamma $-point. This isolation increases as the hopping amplitude in AA-region is further decreased. Experimentally, the interlayer coupling of TBLG is tuned by molecular adsorption \cite{Meng2014}.
	
	In subsequent work, G. Tarnopolsky {\it et al.} \cite{Tarnopolsky2019} showed that the perfectly flat bands can be engineered by switching off the interlayer hopping across the AA-regions. In addition to the perfectly flat bands, this work also showed that the Hamiltonian with $ w_{0} = 0 $ turns out to be chirally symmetric and leads to the particle-hole symmetry due to which the bands are symmetric about the Fermi level at $ E = 0 $. This chirally symmetric Hamiltonian is also used to explain the origin of the series of magic angles in TBLG.
	\begin{figure*}
		\centering
		\includegraphics[scale=0.55]{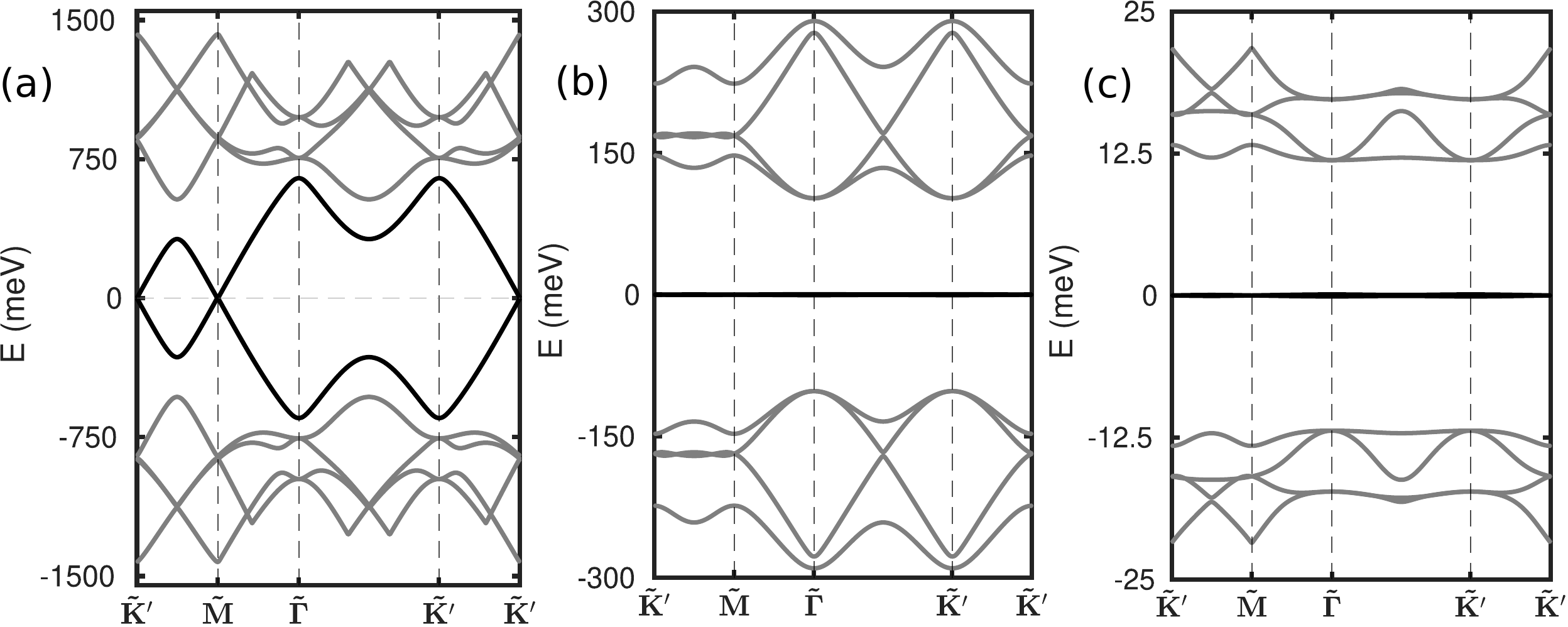}
		\caption{\label{fig:AVbands} (a) Band structure of TBLG without the interlayer coupling in AA-rich regions at twist angle $ \theta = \ang{5} $. The dashed Gray line marks the Fermi level and the two solid black lines shows the two lowest bands which are isolated from the higher bands. (b) The band structure at the first largest magic angle of approximately $ \theta = \ang{1.086}$. The two-lowest bands form the perfectly flat bands. (c) The band structure at the second magic angle of approximately $ \theta = \ang{0.29} $.}
	\end{figure*}
	
	Setting the interlayer coupling off in AA-rich regions $ \left(w_0 \rightarrow 0\right) $ is a crucial assumption in this scheme of calculation to generate perfectly flat bands. In this way, the term responsible for the diagonal matrix elements in interlayer hopping matrices vanishes. As a result the hopping matrices $ T_1, T_2 $ and $ T_3 $ becomes purely off-diagonal and are given as,
	\begin{equation}
		T_{n} = w_{1}\left[\sigma_{x}\,\cos(n-1)\phi + \sigma_{y}\,\sin(n-1)\phi\right]
		\label{eqn:hopingmod}
	\end{equation}
	The corresponding band structures at twist angle $ \theta = \ang{5} $ is shown in Fig.\ref{fig:AVbands}(a), where the two-lowest bands are isolated from the other higher-energy bands. The band structures at the first and second largest magic angles $ \theta = \ang{1.086} $ and $ \theta = \ang{0.29} $ are shown in  Fig.\ref{fig:AVbands}(b) and Fig.\ref{fig:AVbands}(c), respectively. The two bands nearest to Fermi energy form the perfectly flat bands. For each case, the bands are symmetric about $ E=0 $ line and evidently shows the presence of particle-hole (PH) symmetry in contrast to the BM model where the PH symmetry is not explicitly manifested. The PH symmetry emerges as the limit $ w_{0} \rightarrow 0 $ makes the Hamiltonian (\ref{eqn:BMham2}) manifestly chiral. To show that we first rewrite the $ T(\bmr) $ with modified $ T_{n} $ in (\ref{eqn:hopingmod}) as
	\begin{equation}
		T(\bmr) = \Pmqty{0 & w_{1}\,U(\bmr) \\ w_{1}U^{*}(-\bmr) & 0}
	\end{equation}
	where the off-diagonal term $ U(\bmr) = \sum_{n=1}^{3} e^{-i(n-1)\phi}\,e^{-i\bmq_{n}\vdot\bmr} $. Now Substituting $T(\bmr)$ back into the Hamiltonian (\ref{eqn:BMham2}), and using the similarity transformation on the Hamiltonian with $ W = \text{diag}\left(e^{-i\sigma_{z}\theta/2},e^{i\sigma_{z}\theta/2}\right) $  to removes the $ \theta $-dependence from $ H $, we finally get 
	\begin{widetext}
	\begin{equation}
		\mathcal{H} = \hbar v_{F} k_{\theta} 
		\pmqty{0 & \frac{1}{k_{\theta}}\left(-i\partial_x - \partial_y\right) & 0 & \alpha U(\bmr) \\
		\frac{1}{k_{\theta}}\left(-i\partial_x+\partial_y\right) & 0 & \alpha U^{*}(-\bmr) & 0 \\
		0 &\alpha U(-\bmr) & 0 & \frac{1}{k_{\theta}}\left(-i\partial_x-\partial_y\right)\\
		\alpha U^{*}(\bmr) & 0 & \frac{1}{k_{\theta}}\left(-i\partial_x+\partial_y\right) & 0}. 
		\label{eqn:BMham3}
	\end{equation}
	\end{widetext}
	 The Hamiltonian (\ref{eqn:BMham3}) acts on a four-component spinor $ \Psi $, where the upper two-components $ \pmqty{\psi_{1}(\bmr) & \chi_{1}(\bmr)}^{T} = u^{(1)}_{\bmk}(\bmr)\,e^{i\left(\bmk-\bm{K}_{1}\right)\vdot\bmr} $ are the probability amplitudes corresponding to L1 with the Dirac point at $\bm{K}_{1}$, and the lower two-components $ \pmqty{\psi_{2}(\bmr) & \chi_{2}(\bmr)}^{T} = u^{(2)}_{\bmk}(\bmr)\,e^{i\left(\bmk-\bm{K}_{1}+\bm{q}_{1}\right)\vdot\bmr} $ are of L2. Further shuffling the two-components $ \chi_{1} $ and $ \psi_{2} $ in $ \Psi $ also transforms the Hamiltonian as
	 \begin{widetext}
	 	\begin{equation}
	 		\mathcal{H} =
	 		\pmqty{0 & 0 & \frac{1}{k_{\theta}}\left(-i\partial_x - \partial_y\right) & \alpha U(\bmr) \\
	 			0 & 0 & \alpha U(-\bmr) & \frac{1}{k_{\theta}}\left(-i\partial_x-\partial_y\right)\\
	 			\frac{1}{k_{\theta}}\left(-i\partial_x+\partial_y\right) &\alpha U^{*}(-\bmr) & 0 & 0 \\
	 			\alpha U^{*}(\bmr) & \frac{1}{k_{\theta}}\left(-i\partial_x+\partial_y\right) & 0 & 0}
	 		\label{eqn:BMham4}
	 	\end{equation}
	 \end{widetext}
 	The scale of the energy is considered to be $ \hbar v_{F} k_{\theta} $and the Hamiltonian now operates on the spinor $ \Psi = \pmqty{\psi_{1} & \psi_{2} & \chi_{1} & \chi_{2}}^{T} $. Now, for some complex number $ z = x + iy$, one can write $ x = \left(z + \bar{z}\right)/2 $ and $ y = (z-\bar{z})/2i $, therefore
 	\begin{equation}
 		\pdv{\bar{z}} = \pdv{x}{\bar{z}} \pdv{x} + \pdv{y}{\bar{z}}\pdv{y} = \frac{1}{2}\left(\partial_x +i \partial_y\right) \equiv \bar{\partial}
 	\end{equation}
 	Therefore, one can define an operator $ \mathcal{D}(\bmr) $ which contains only antiholomorphic derivatives in its main diagonal, \emph{i.e.,}
 	\begin{equation}
 		\mathcal{D}(\bmr) = \Pmqty{\frac{-2i}{k_{\theta}} \bar{\partial} & \alpha U^{*}(-\bmr) \\
 		\alpha U^{*}(\bmr) & \frac{-2i}{k_{\theta}} \bar{\partial}}
 		\label{eqn:Doprtr}
 	\end{equation}
	In terms of this operator (\ref{eqn:Doprtr}), the Hamiltonian $ H $ in (\ref{eqn:BMham4}) becomes
	\begin{equation}
		\mathcal{H} = \Pmqty{0 & \mathcal{D}^{*}(-\bmr) \\ \mathcal{D}(\bmr) & 0}. 
		\label{eqn:chiralHam}
	\end{equation}
	The Hamiltonian $H$ in (\ref{eqn:chiralHam}) is manifestly a chiral Hamiltonian as it satisfies the anticommutation relation $ \left\{\mathcal{I}_{2}\times\sigma_{z},\mathcal{H}\right\} = 0 $.
	
	In the subsequent subsection, the chiral Hamiltonian in (\ref{eqn:chiralHam}) is used to explain the origin of the magic angles.
	
	\subsection{The origin of magic angles}\label{MA}
	It may be noted that the $ \mathcal{D} $ operator in (\ref{eqn:Doprtr}) is symmetric up to a phase factor when a counter-clockwise rotation of $ \phi = 2\pi/3 $ is applied, \tcog{\emph{i.e.,}}
	\begin{equation}
		\mathcal{D}(\mathcal{R}(\phi)\bmr) = \Pmqty{\frac{-2i}{k_{\theta}} \bar{\partial} & \alpha U^{*}(-\mathcal{R}(\phi)\bmr) \\
			\alpha U^{*}(\mathcal{R}(\phi)\bmr) & \frac{-2i}{k_{\theta}} \bar{\partial}}. 
	\end{equation}
	Given $ U^{*}(\mathcal{R}(\phi)\bmr) = e^{i\phi} U^{*}(\bmr) $, therefore $ \mathcal{D}(\mathcal{R}(\phi)\bmr)  = e^{i\phi} \mathcal{D}(\bmr)$. As this symmetry feature holds for all $ \alpha $ turning on $ \alpha > 0 $ gradually preserves symmetry.  For $ \alpha = 0 $, the interlayer tunnelling is absent, and the two graphene layers become uncoupled, and the Hamiltonian in (\ref{eqn:BMham4}) has four zeros modes: two from each Dirac points $ \bm{K} $ and $ \bm{K}' $. Since, there are always zero modes at some points in the moir\'e BZ, the appearance of the perfectly flat band at the set of magic angles implies that the zero-energy equation with $ \psi_{\bmk} = \pmqty{\psi_{1} & \psi_{2}}^{T} $,
	\begin{equation}
		\mathcal{D}(\bmr)\psi_{\bmk}(\bmr) = 0
		\label{eqn:Dpsi}
	\end{equation}
	has solutions for arbitrary momenta $ \bmk \in $ moir\'{e} BZ. The function $ \psi_{\bmk} $ should be periodic with primtive translations vectors $ \bm{a}^{M}_{1} $ and $ \bm{a}^{M}_{2} $, that is,
	\begin{equation}
		\psi_{\bmk}(\bmr + \bm{a}^{M}_{1,2}) =
		\pmqty{u^{(1)}_{\bmk}(\bmr + \bm{a}^{M}_{1,2}) e^{i\left(\bmk-\bm{K}_{1}\right)\vdot\left(\bmr + \bm{a}^{M}_{1,2}\right)} \\
		u^{(2)}_{\bmk}(\bmr + \bm{a}^{M}_{1,2}) e^{i\left(\bmk-\bm{K}_{1}\right)\vdot\left(\bmr + \bm{a}^{M}_{1,2}\right)} e^{i\bmq_{1}\vdot\left(\bmr + \bm{a}^{M}_{1,2}\right)}}
	\end{equation}
	where the form of $ \psi_{\bmk} $ is given in Appendix-(\ref{appendix:b}) and can be written as
	\begin{equation}
		\psi_{\bmk}(\bmr + \bm{a}^{M}_{1,2}) = e^{i\bmk\vdot\bm{a}^{M}_{1,2}}
		\pmqty{1 & 0 \\ 0 & e^{i\bmq_{1}\vdot\bm{a}^{M}_{1,2}}}
		\psi_{\bmk}(\bmr)
	\end{equation}
	since $ \bmq_{1}\vdot\bm{a}^{M}_{1,2} = (2\bm{b}^{M}_{1} + \bm{b}^{M}_{2})\vdot\bm{a}^{M}_{1,2}/3 = 4\pi/3 ~\text{or}~ 2\pi/3 $.
	
	The Eq. (\ref{eqn:Dpsi}) ensures that  the zero-mode solution $ \psi_{\bm{K}}(\bmr) $ at the Dirac point where $ \bmk = \bm{K} $ is guaranteed. Therefore $ \mathcal{D}(\bmr)\psi_{\bm{K}}(\bmr) = 0 $ with the property
	\begin{equation}
		 \psi_{\bm{K}}(\bmr + \bm{a}^{M}_{1,2}) =
		 \pmqty{1 & 0 \\ 0 & e^{i\bmq_{1}\vdot\bm{a}^{M}_{1,2}}}\psi_{\bm{K}}(\bmr). 
	\end{equation}
	Now, since $ \mathcal{D}(\bmr) $ contains the antiholomorphic derivative $ \bar{\partial} = \partial/\partial \bar{z} $ in the main diagonal, therefore the derivative is with respect to $\bar{z}$ only. Consequently any complex function $f_{\bmk}(z)$ at a particular $ \bmk $ can be multiplied with $ \psi_{\bm{K}}(\bmr) $ such that
	\begin{equation}
		\mathcal{D}(\bmr)\left[f_{\bmk}(z)\psi_{\bm{K}}(\bmr)\right] =
		f_{\bmk}(z) \mathcal{D}(\bmr) \psi_{\bm{K}}(\bmr) = 0. 
	\end{equation}
	If one is able to find a function $ f_{\bmk}(z) $ which is also periodic with the moir\'e primitive vectors $ \bm{a}^{M}_{1,2} $, \emph{i.e.,}
	\begin{equation}
		f_{\bmk}(z + a^{M}_{1,2}) = e^{i\bmk\vdot\bm{a}^{M}_{1,2}} f_{\bmk}(z)
		\label{eqn:blochperiodfz}
	\end{equation}
	where $ a^{M}_{j} = (\bm{a}^{M}_{j})_{x} + i(\bm{a}^{M}_{j})_{y} $ for $ j=1,2 $ then the other zero-mode solutions can be found just by multiplying $ \psi_{\bm{K}}(\bmr) $ by $f(z)$, such that $ \mathcal{D}(\bmr)\left[f_{\bmk}(z)\psi_{\bm{K}}(\bmr)\right] = 0 $.
	
	Now, Liouville's theorem says that any entire or holomorphic function $ f(z) $ is either constant or blows up at infinity, and hence this state cannot be a Bloch state at a momentum other than $ \bm{K}$ \cite{Ledwith2021}. This leads to the choice of $ f(z) $ to be a meromorphic function. Although such a function would necessarily have poles at certain positions in the moir\'e cell that should be precisely cancelled by zeros in $ \psi_{\bm{K}}(\bmr) $. Note, since this is a spinor wavefunction, we will need both components of $ \psi_{\bm{K}}(\bmr) $ to simultaneously vanish at that location in the unit cell. We can explore if such a condition is satisfied by varying the angle. In fact, the magic angles are precisely those angles for which $ \psi_{\bm{K}} $ has a zero in both of its components at some $ \bmr $.
	
	At the locations of AB/BA-stacking which are given by $ \pm\bmr_{0} = \pm \left(-\bm{a}^{M}_{1} + 2\bm{a}^{M}_{2}\right)/3 $, these stacking points are distinguished by $ C_{3} $ (counterclockwise rotation of $ 2\pi/3$ about the centre of the moir\'e cell). This is because they map themselves up to a lattice vector, \emph{i.e.,} $ C_{3}\bm{r}_{0} = \bm{r}_{0} - \bm{a}^{M}_{2}$ (see Fig.\ref{fig:TBLG1}(a)). This implies that if $ \psi_{\bm{K}}(\bmr) $ is a solution to the equation $ \mathcal{D}(\bmr) \psi_{\bm{K}}(\bmr) = 0 $, then $ \psi_{\bm{K}}(\mathcal{R}(\phi)\bmr) $ is also a solution because $ \mathcal{D}(\bmr) $ is symmetric under a $ \phi = 2\pi/3 $ rotation. This in turn implies that at arbitrary $\alpha$, the following relations,
	\bse
	\begin{align}
		\psi_{\bm{K},1}(\mathcal{R}(\phi)\bm{r} \pm \bm{r}_{0}) &= \psi_{\bm{K},1}(\bmr \pm \bm{r}_{0}), \\
		\psi_{\bm{K},2}(\mathcal{R}(\phi)\bm{r} \pm \bm{r}_{0}) &=
		e^{\pm i \phi} \psi_{\bm{K},1}(\bmr \pm \bm{r}_{0}).
	\end{align}
	\ese
	Due to the form of the Bloch states as given in Appendix-(\ref{appendix:b}), the second component $ \psi_{\bm{K},2} $ acquires a phase under translations by $ \pm \bm{r}_{0} $. The second relation relation implies that $ \psi_{\bm{K},2}(\bmr) $ always vanishes at $ \bmr =\pm \bmr_{0} $ for all $ \alpha $ and $ \psi_{\bm{K},1} $ is in general non-zero.

	Therefore, to generate the other states, one need to choose a meromorphic function $ F_{\bmk}(z) $ which is Bloch periodic. The function $ F_{\bmk}(z) $ is written as $ F_{\bmk}(z) = f_{\bmk}(z)/g(z) $ where both $ f_{\bmk}(z) $ and $ g(z) $ are holomorphic functions and
	\begin{equation}
		g(z_{0} + n_{1}a^{M}_{1} + n_{2}a^{M}_{2}) = 0. 
	\end{equation}
	Here $ z_{0} = \bm{r}_{0x} + i\bmr_{0y} $ and the complex number $ a^{M}_{1} $ and $ a^{M}_{2} $ are defined in (\ref{eqn:blochperiodfz}).
	The function $ g(z) $ is the Jacobi theta function,
	\begin{equation}
		g(z) = \vartheta_{1} \left(\frac{z-z_{0}}{a^{M}_{1}}|\omega\right)
	\end{equation}
	where $ \omega = e^{i\phi} $ with $ \phi = 2\pi/3 $, and $ \vartheta_{a,b}(z|\tau) $ is the Jacobi theta function of the first kind of particular period is defined as
	\begin{equation}
		\vartheta_{1}(z|\tau) = -i \sum_{n=-\infty}^{\infty} (-1)^{n} e^{i\pi\tau\left(n+1/2\right)^2}e^{i\pi \left(2n+1\right)z}
		\label{eqn:gz}
\end{equation}
Also, $ \vartheta_{1}(0|\tau) = 0 $ such that $ g(z) $ has the desired zeros. One can see from (\ref{eqn:gz}) that the translation $ z \rightarrow z + a^{M}_{1} $ shifts argument of theta function by $ 1 $ and the translation by $ z \rightarrow z + a^{M}_{2} $ shifts the argument by $ \omega = e^{i\phi} $. Under translations the Jacobi theta functions satisfy
\begin{align}
	\vartheta_{1}(z+1|\tau) & = \vartheta_{1}(z|\tau) e^{i(2n+1)\pi} = -\vartheta_{1}(z|\tau) \\
	\vartheta_{1}(z+\tau|\tau) & = -e^{-i\pi\tau-2i\pi z}\vartheta_{1}(z|\tau). 
\end{align}
The function $ f_{\bmk}(z) $ can be created by shifting the arguments of theta functions and is given as
\begin{equation}
	f_{\bmk}(z) = e^{2\pi i kz/a^{M}_{1}} \vartheta_{1}\left(\frac{z-z_{0}}{a^{M}_{1}} - \frac{k}{b^{M}_{2}}|\omega\right)
\end{equation}
where $ k = k_{x} + i k_{y} $ measured from the $ K- $point. Therefore, the function $ \psi_{\bmk}(\bmr) $ becomes
\begin{equation}
	\psi_{\bmk}(\bmr) =  e^{2\pi i kz/a^{M}_{1}} \vartheta_{1}\left(\frac{z-z_{0}}{a^{M}_{1}} - \frac{k}{b^{M}_{2}}|\omega\right)
	\frac{\psi_{\bm{K}}(\bmr)}{\vartheta_{1} \left(\frac{z-z_{0}}{a^{M}_{1}}|\omega\right)}
\end{equation}
It has been shown that the zero Fermi velocity is also connected to the zeros of wave functions $ \psi_{\bm{K}}(\bmr) $ \cite{Tarnopolsky2019}. Now, the function $ \psi_{\bm{K}}(\bmr) $ is taken as
\begin{equation}
	\psi_{\bm{K}}(\bmr) = \pmqty{1 + \alpha^{2}u_{2} + \dots \\ \alpha u_{1} + \dots}
\end{equation}
Using the above $ \psi_{\bm{K}} $ in the zero-mode solution $ \mathcal{D}(\bmr)\psi_{\bm{K}}(\bmr) = 0 $, one obtains
\begin{align}
	u_{1}(\bmr) & = -i\left(e^{i\bmq_{1}\vdot\bmr} + e^{i\bmq_{2}\vdot\bmr} + e^{i\bmq_{3}\vdot\bmr}\right) \\
	u_{2}(\bmr) & = \frac{i}{\sqrt{3}} e^{-i\phi}\left(e^{-i\bm{b}^{M}_{1}\vdot\bmr} + e^{i\bm{b}^{M}_{2}\vdot\bmr} + e^{i\left(\bm{b}^{M}_{1}-\bm{b}^{M}_{2}\right)\vdot\bmr}\right) + \text{c.c}
\end{align}
Therefore, the function $ \psi_{\bm{K}} $ at $ \bmr = \bmr_{0} $ is obtained by calculating $ u_{1}(\bmr_{0}) = 0 $, and $ u_{2}(\bmr_{0}) = -3 $ implies that the spinor $ \psi_{\bm{K}} $ vanishes when $ \alpha \approx 1/\sqrt{3} $, which is very close to the first magic angle in TBLG. The function $ \psi_{\bm{K}} $ vanishes for infinitely many $ \alpha $ with the quasiperiodicity $ \alpha_{n} \approx \alpha_{n-1} + 1.5 $ holding for large $ \alpha $ \cite{Tarnopolsky2019}. The chiral model was thus able to provide analytical reasoning for the occurrence of a series of magic angles in TBLG from which a perfect flat band emerges. 

\section{Extended Hamiltonian of TBLG}
Although the BM model in section \ref{BMmodel} or the chiral model in section \ref{PFB} predicts the existence of a series of magic angles, only the largest magic angle $ (\approx \ang{1.1}) $ has been observed experimentally \cite{Cao2018one,YankowitzTBG2019,Kerelsky2019,Utama2021}. In the case of an unrelaxed lattice, the size of the moir\'{e} unit cell becomes larger in size with an increase in the size of the local AA regions as the twist angle decreases. Realistically, a significant atomic relaxation occurs in small-angle TBLG systems, resulting in the development of strain fields. These strain fields minimize the diameter of the energetically less stable AA-stacking regions (since the atoms of two layers are perfectly aligned) and lead to the formation of triangular domain patterns in the moir\'{e} pattern \cite{Lamparski2020}. This domain network pushes the small-angle TBLG into the soliton regime \cite{JonathanAlden2013,Zhang2018,Gargiulo2018,Lin2018,Yoo2019,Efimkin2018,Lamparski2020,Gadelha2021}. Owing to the different stacking network in small-angle TBLG systems compared to the large-angle ones, this atomic reconstruction separates the TBLG systems into two large- and small-angle classes with different electronic properties \cite{Nguyen2021}.

The formation of a network of domain walls separating AB and BA stacking regions in small-angle TBLG facilitates opposite Chern numbers in these regions when inversion symmetry is broken with an external electric field. This leads to the emergence of topologically protected helical states that appears on the domain walls \cite{Jose2013,JonathanAlden2013,Efimkin2018,Shengqiang2018,Rickhaus2018}.
	The reconstructed stacking structure in small-angle TBLG has been shown to modulate the electronic properties \cite{Sunku2018}, vibrational properties \cite{Jiang2016}. This also produces changes in the behaviour of electron–phonon coupling \cite{Eliel2018} and to the observation of strong correlations and superconductivity \cite{Cao2018one}.

The more general and exact Hamiltonian that accounts for the reconstruction due to lattice relaxation is given as \cite{Carr2019},
	\begin{widetext}
	\begin{equation}
		H = 
		\begin{pmatrix}
			-i\hbar v_{F}\bm{\sigma}_{\theta/2}\vdot\bm{\nabla} & 0 \\
			0 & -i\hbar v_{F}\bm{\sigma}_{-\theta/2}\vdot\bm{\nabla}
		\end{pmatrix}
		+
		\begin{pmatrix}
			V_{1}(\bmr) + A_{1}(\bmr) & T(\bmr) + \left\{M^{\dagger}_{+}(\bmr),\hat{k}_{-}\right\} + \left\{M^{\dagger}_{-}(\bmr),\hat{k}_{+}\right\} \\
			T^{\dagger}(\bmr) + \left\{M_{+}(\bmr),\hat{k}_{+}\right\} + \left\{M_{-}(\bmr),\hat{k}_{-}\right\} & V_{2}(\bmr) + A_{2}(\bmr)
		\end{pmatrix}
	\label{eqn:extHamTBLG}
	\end{equation}
	\end{widetext}
where $ V_{i}(\bmr) $ for $ i=1,2 $ is the external potential for each individual layer that can include an electric gating potential, sublattice mass terms, or a potential from doping or charge redistribution. The interlayer hopping matrices $ T(\bmr) $ is defined in (\ref{eqn:Tndef}). The field $ A_{i}(\bmr) $ for $ i=1,2 $ coupled to the Dirac electron is generated due to the geometric deformation and strain in each layer. This field is periodic with the primitive lattice vectors of the moir\'{e} pattern and therefore its Fourier expansion can be written as
\begin{equation}
	A_{i}(\bmr) = \sum_{\bm{b}^{M}} A^{i}_{\bm{b}^{M}} e^{i\bm{b}^{M}\vdot\bmr}
\end{equation}
where $ \bm{b}^{M} $ are the reciprocal lattice vectors of the moir\'{e} Brillouin zone. The remaining off-diagonal terms with the interlayer hopping matrix is due to the $ \bmk $-dependent scattering between the Bloch states with $ \hat{k}_{\pm} = \hat{k}_{x} \pm i \hat{k}_{y} $ in relaxed TBLG\tcog{;} the scattering matrix element between the Bloch states $ \ket{\Psi^{(2)}_{\bmk + \bmq_{j},\beta}} $ and $ \ket{\Psi^{(1)}_{\bmk,\alpha}} $ not only depends upon the wave-vector $ \bm{q}_{j} $, but also the wave-vector $ \bmk $. For a more detailed explanation of each term in the extended Hamiltonian in (\ref{eqn:extHamTBLG}), one can refer to \cite{Fang2019}. The lattice deformation significantly enhances the Fermi velocity in contrast to the suppressed Fermi velocity in non-relaxed system \cite{NNTnam2017}. Other similar DFT calculations, which fully account for the atomic reconstruction or in-plane and out-of-plane relaxations, have not been able to predict the existence of smaller magic angles $ ( < \ang{1.1}) $ \cite{Walet2020}. Till date, no experimental observations of the other magic angles are extant.

The flat band physics is quite interesting as the electron-electron interaction among the electrons dominates over their kinetic energy, and the only scale of energy is the interaction energy scale. Consequently, the behaviour of electrons is controlled purely by the interactions between them, therefore their motion is correlated to a greater extent. The predicted many-body phases like superconductivity and correlated-insulator at the half-fillings of the flat bands manifest themselves in TBLG because of the presence of flat bands. 
One of the most exotic possibilities that the flat bands in magic-angle twisted bilayer graphene occurs is the flat-band-assisted super-conductivity, and has been revisited in correlated electron systems \cite{Balents2020}, where the inter-band scattering between the dispersive and flat bands plays an essential role. In particular, this mechanism is thought of as one of the possible origins of enhancement of $ T_c $ in a TBLG with so called “magic angles”. There it has been pointed out that the preferable band structure for such mechanism is (i) the flat band is located slightly above or below the Fermi level, and (ii) the dispersive band has a large density of states (DOS) near the flat band. Tomonari {\it et al.} \cite{Tomonari2019} proposed a simple method to tune the energy of flat bands without losing the exact flatness of bands. The main idea is to add farther-neighbour hopping to the existing NN models with flat band(s) and it has two prominent advantages : (i) the flat bands retain exact flatness after the modulation of the Hamiltonian, and (ii) a few parameters are needed to control a flat-band energy.

\section{Topics not covered in the review}
This brings us almost to the end of this review. There has been an explosion in the experimental and theoretical work in twisted graphene structures, particularly after the seminal experimental discovery of strongly correlated phases in MATBLG \cite{Cao2018one, Cao2018two} by the MIT group. The current review aimed at providing a detailed overview of all the basic theoretical tools to beginners interested in entering this field. We have therefore omitted a huge chunk of the latest developments in this field by design. Instead, we provide a brief list of some of the most exciting theoretical and experimental works in the field of twistronics. We hasten to add that this list is by no means exhaustive, and many equally important works are left out. It shall be helpful though to get a flavour of these recent developments once the basic theory of this field is understood with the help of preceding sections. 
	
In the BM model, the calculation of the overlap matrix elements is carried out entirely in momentum space. A general real-space continuum model was proposed by L. Balents \cite{Balents2019} more recently. This continuum model has been proposed for bilayer graphene with smooth lattice deformation where the deformations have been parameterized by the small displacement gradients. It has been shown that any deformation described by an inhomogeneous small gradient displacement field in each layer can be treated with an accuracy equal to that of the case of a uniform small-angle twist, and therefore the BM model becomes a special case of this more general model \cite{Balents2019}. Since the derivation is entirely in real space, one obtains a complete real-space BM model. The model also generalizes the other extensions to the BM model, such as the flat bands designing in TBLG and bilayer transition metal dichalcogenide (TMDC) systems by the heterostrain \cite{Zhen2019}. Along this line, a generic topological criterion has recently been reported to find flat bands in a wide range of 2D materials which is not limited to moir\'{e} bilayers \cite{Parhizkar2023}.
		
 In this review, we confined ourselves to studying the TBLG in the absence of electrostatic interactions among the charge carriers. Owing to the fact that TBLG hosts the flat bands at the magic angles where the interactions among the electrons dominate over their kinetic energy, the $e-e$ interactions play a crucial role. In fact, the exciting, strongly correlated phases in MATBLG demand a complete understanding of the $e-e$ interactions specially in the flat band region. The effects of long-range electrostatic interactions using Hartree-Fock approximations is considered in \cite{TomassoCea2021,TommasoCea2022}. It has been shown that the electron-electron interaction get maximized at the magic angle \cite{Kerelsky2019}. The role of $e-e$ interaction in plasmon modes in TBLG was discussed in \cite{Ding2022}, and the plasmonic bands in small-angle twisted bilayer graphene is analysed in \cite{Stauber2016}. RPA calculations of the low energy excitations in TBLG have also been considered extensively in refs \cite{Pizarro2019,Cyprian2019,Fengcheng2020,GSharma2020,Khalaf2020}. For other many-body calculations in TBLG, one can look at the ref. \cite{Romanova2022}.
		
The emergence of superconductivity and correlated insulators in MATBLG has raised the intriguing possibility that its pairing mechanism is distinct from that of conventional superconductors \cite{Cao2018one,Cao2018two,Lu2019,YankowitzTBG2019,Saito2020,Stepanov2020,Choi2021,Oh2021,Sujay2019}, as described by the Bardeen–Cooper–Schrieffer (BCS) theory. After the discovery of unconventional superconductivity there are many theoretical papers aiming to understand the pairing interaction responsible for superconductivity. The attractive $e-e$ interaction mediated by electron-phonon coupling is considered in \cite{Peltonen2018,Wu2018,Isobe2018,Choi2018,Liu2018,Lian2019,Wu2019,Schrodi2020}. The electron-phonon superconductivity and strong correlations in moir\'e-flat bands were considered by Balents {\it et al.} \cite{Balents2020}.
		
Other strongly correlated phases in TBLG are an equally exciting topics of research. The TBLG goes through several exciting phases as the electronic density is tuned by gating. In this context, Sharpe et al. \cite{Sharpe2019} showed that the MATBLG becomes magnetic at a particular electronic density. More specifically, they observed emergent ferromagnetic hysteresis with a giant anomalous Hall effect near three-quarters $ (3/4) $ filling of the moir\'e conduction band. Orbital Magnetism \cite{Lu2019} was also explored. Theoretically, the existence of magnetic order in TBLG may be possible due to interactions that lift the spin and valley degeneracies \cite{Ochi2018,Dodaro2018,Thomson2018,Venderbos2018,Kang2019,Seo2019}. Electrical controllable magnetism  was also analysed in Ref.\cite{Gonzalez2017}.
		
TBLG under a magnetic field is naturally an exciting topic given the moir\'{e} pattern has a large lattice constant. The fractal Hofstadter spectrum is a canonical example of electronic structure in a system with incommensurate length scales. In this direction, the pioneering early work by Bistritzer and MacDonald \cite{Bistritzerbutterfly2011} showed the existence of moir\'e butterflies in TBLG in the presence of an external magnetic field. The Landau-levels have also been observed in TBLG in the presence of magnetic field \cite{Uri2020,Hejazi}. Recently, a unified flavor polarization mechanism is proposed to understand the intricate interplay of topology, interactions and symmetry breaking as a function of density and applied magnetic field in MATBLG \cite{Yu2022}. 

If flat bands emerge in a condensed matter system, the quantum Hall effect and Chern Insulator cannot be far behind. Theoretically, the quantum Hall effect (QHE) in TBLG has been studied for various twist angles and for different strengths of magnetic field in \cite{Lee2011,Koshino2012}. Experimentally, the Quantum anomalous Hall-effect is reported in TBLG on hBN \cite{Shi2021}. The existence of the Fractional Chern Insulators (FCI) in MATBLG has been recently reported \cite{Xie2021}.
		
Interesting single particle effects were also explored. Similar to the chiral tunnelling of massless Dirac fermions in SLG, the chiral tunnelling in TBLG has also been reported \cite{Hechtunnn2013}. The transport properties of moir\'e electrons is recently studied in \cite{Padhi2020}, where they look at phenomena like 
moir\'e-tunneling. 

TBLG is a new addition in the expanding horizon of topological condensed matter systems. The existence of a series of magic angles in TBLG has been predicted to have a strong connection with the topology of the system. That the magic angles in TBLG are all topological was theoretically studied in 2019 by Song et al \cite{Song2019}. The edge states in TBLG have also been studied, and the fact that these edge states have a topological connection to the quantum pumping has been reported recently by Fujimoto et al \cite{Fujimoto2021}. Topologically protected zero-modes in TBLG \cite{deGail2011} were also studied. 

In the 1950s it was shown that the the liquid $^{3}$He can solidify on heating and this effect is referred to as Pomeranchuk effect. The same effect has also been seen in MATBLG \cite{Rozen2021}. Thermal transport phenomena, such as thermopower, are sensitive to the PH asymmetry and the emergent highly PH asymmetric electronic structure in TBLG can show giant thermopower peaks \cite{Paul2022}. The cross-plane thermoelectricity in TBLG controllable via misorientation \cite{Mahapatra2020} and experimentally, the breakdown of semiclassical description of thermoelectricity near-magic angle TBLG were observed by Ghawri {\it et al.} \cite{Ghawri2022}. The highly tunable Josephson junctions in TBLG were studied in refs \cite{Rodan-Legrain2021, deVries2021}.

Another interesting and related structure is  twisted double bilayer-graphene. Namely, two Bernal (or AB-) stacked bilayer graphene are stacked over one another followed by a rotation of one bilayer with respect to another gives rise to twisted double bilayer graphene (TDBG). The continuum approximation of both the bilayers is massive and having parabolic dispersion in contrast to the case of linear dispersion in case of TBLG. The existence of flat bands was also reported in such a system by Haddadi {\it et al.} \cite{Haddadi2020} and tunable multi-bands by Zhu {\it et al.} \cite{Zhu2022}. Due to the existence of flat bands in TDBG, there have been a lot of work on TDBG to understand the existence of strongly correlated phases and some of the important experimental breakthroughs in TDBG are spin-triplet superconductivity \cite{Lee2019}, correlated states in TDBG \cite{Shen2020}, tunable metal-insulator transition in double-layer graphene hetero-structures \cite{Ponomarenko2011}, correlated electron-hole state \cite{Rickhaus2021}, tunable correlated states, and spin-polarized phases in twisted bilayer--bilayer graphene \cite{Cao2020}. The manifestation of dielectric screening has also been studied in by Mukai et al. \cite{Mukai2021}. Some other interesting works include Floquet engineering in TDBG \cite{Vega2020} , symmetry breaking in TDBG \cite{He2021}.

What about twisted trilayer graphene? The flat bands in twisted trilayer graphene (TTG) has been studied by \cite{Mora2019} and the tunable superconductivity in TTG has also been observed \cite{Park2021,ZeyuHao2021}.
One may of course ask about $n$-layer generalizations of TBLG as well. To name a few exciting efforts in the direction of twisted multilayer graphene (tMLG) such as topological superconductivity in tMLG \cite{XuLeon2018}, the quantum Hall effect and orbital magnetism in tMLG \cite{Liu2019}. Correlated electronic phases in twisted bilayer transition metal dichalcogenides \cite{Wang2020} were also explored.

Some other important works include the comparison between various tight-binding models on TBLG \cite{Po2019}, charge order and broken rotational symmetry in MATBLG \cite{Jiang2019}, symmetry-adapted maximally localized Wannier states for the lowest four-bands in TBLG \cite{Kang2018}, were also analysed. Owing to the similar hexagonal structure of hBN as of graphene, it has been shown that the flat bands can also emerge in a twisted bilayer hBN \cite{Walet2021}. Commensurate-incommensurate transition in G/BN \cite{Woods2014}, one-dimensional electrical contact to a two-Dimensional material \cite{LWang2013}, bilayer graphene's direct measurement of discrete valley and orbital quantum numbers \cite{Hunt2017}, bilayer graphene's spontaneous chiral symmetry breaking \cite{Zhang20159}, optical properties of massive anisotropic tilted Dirac systems \cite{Mojarro2021}, graphene on hexagonal boron nitride: valley order and loop currents \cite{Uchoa2015}, applications of 2D materials with tunable optical characteristics \cite{Ma2021}, engineering of band structures in graphene and silicene induced by the moire potential \cite{Zhao2021}, are some of other related works.

The above list, though not exhaustive, provides a glimpse of the explosion in the interest in the field of twistronics, begotten by the discovery of flat bands at a magic angle in TBLG, but branching out in innumerable ways, enriching physics, chemistry, material science, technology and engineering. We hope our pedagogical introduction to the theory of this field will be helpful to a large variety of researchers in this direction.

We thank Disha Arora and Raghav Chaturvedi for helpful discussion at various points of this work. We particularly thank Prof. G. Murthy for sharing with us some of his unpublished notes on the BM model. 

The work of SG is supported by the project MTR/2021/000513 funded by SERB, DST, Govt. of India. The work of DA is supported by a UGC ( Govt. of India) fellowship.

\appendix
\section{Simplification of the term in velocity renormalization}\label{appvelo}
\label{appendix:a}
From (\ref{eqn:slgHamxi}), the rotated SLG Hamiltonian for valley $ \xi = -1 $ is given by
\begin{equation}
	h_{\bmk}(\theta) = -\hbar v_{F} \abs{\bmk}\pmqty{0 & e^{i\left(\theta_{\bmk}-\theta\right)} \\ e^{-i\left(\theta_{\bmk}-\theta\right)} & 0}
\end{equation}
The matrix $ h_{\bmk}(\theta) $ is invertible and therefore its inverse is given by
\begin{equation}
	h^{-1}_{\bmk}(\theta) = -\frac{1}{\hbar v_{F}\abs{\bmk}}
	\pmqty{0 & e^{i\left(\theta_{\bmk}-\theta\right)} \\ e^{-i\left(\theta_{\bmk}-\theta\right)} & 0}
\end{equation}
Also, the hermitian adjoint of the inverse matrix $ (h^{-1}_{\bmk}(\theta))^{\dagger} = h^{-1}_{\bmk}(\theta) $. Therfore the term in (133) becomes
\begin{equation}
	\sum_{\alpha=1}^{3} T_{\alpha}~h^{-1}_{\bmq_{\alpha}}(-\frac{\theta}{2})~h_{\bmk_{\alpha}}(-\frac{\theta}{2})~h^{-1}_{\bmq_{\alpha}}(-\frac{\theta}{2})~T^{\dagger}_{\alpha}
\end{equation}
Before proceeding further, one can note that the term $ h_{\bmk_{\alpha}}(-\frac{\theta}{2}) $ using (18) can be rewritten as
\begin{equation}
	h_{\bmk_{\alpha}}(-\frac{\theta}{2}) = h_{\bmk}(-\frac{\theta}{2}) + h_{\bmq_{\alpha}}(-\frac{\theta}{2})
\end{equation}
\begin{widetext}
	Using (A4) in (A3) gives two-terms,
	\begin{equation}
		\sum_{\alpha=1}^{3} T_{\alpha}~h^{-1}_{\bmq_{\alpha}}(-\frac{\theta}{2})~h_{\bmk}(-\frac{\theta}{2})~h^{-1}_{\bmq_{\alpha}}(-\frac{\theta}{2})~T^{\dagger}_{\alpha}
		+ \sum_{\alpha=1}^{3} T_{\alpha}~h_{\bmq_{\alpha}}(-\frac{\theta}{2})~T^{\dagger}_{\alpha}
	\end{equation}
	We first simplify the first term in (A5), which is the sum of three terms. Each of the term is given by
	\begin{align}
		T_{1}~h^{-1}_{\bmq_{1}}(-\frac{\theta}{2})~h_{\bmk}(-\frac{\theta}{2})~h^{-1}_{\bmq_{1}}(-\frac{\theta}{2})~T^{\dagger}_{1} & =
		-\frac{w^{2}\abs{\bmk}}{\hbar v_{F}\abs{\bmq_{1}}^2} 
		\left[e^{i\left(\theta_{\bmk}-2\theta_{\bmq_{1}}-\frac{\theta}{2}\right)} + \text{c.c.}\right]
		\pmqty{1 & 1 \\ 1 & 1} \\
		T_{2}~h^{-1}_{\bmq_{2}}(-\frac{\theta}{2})~h_{\bmk}(-\frac{\theta}{2})~h^{-1}_{\bmq_{2}}(-\frac{\theta}{2})~T^{\dagger}_{2} & = -\frac{w^{2}\abs{\bmk}}{\hbar v_{F}\abs{\bmq_{2}}^2}
		\pmqty{e^{i\left(\theta_{\bmk}-2\theta_{\bmq_{2}}-\frac{\theta}{2} + \phi\right)} + \text{c.c.} & e^{i\left(\theta_{\bmk}-2\theta_{\bmq_{2}}-\frac{\theta}{2} - \phi\right)} + e^{-i\left(\theta_{\bmk}-2\theta_{\bmq_{2}}-\frac{\theta}{2}\right)} \\
			e^{i\left(\theta_{\bmk}-2\theta_{\bmq_{2}}-\frac{\theta}{2}\right)} + e^{-i\left(\theta_{\bmk}-2\theta_{\bmq_{2}}-\frac{\theta}{2} - \phi\right)} &
			e^{i\left(\theta_{\bmk}-2\theta_{\bmq_{2}}-\frac{\theta}{2} - 2\phi\right)} + \text{c.c.}} \\
		T_{3}~h^{-1}_{\bmq_{3}}(-\frac{\theta}{2})~h_{\bmk}(-\frac{\theta}{2})~h^{-1}_{\bmq_{3}}(-\frac{\theta}{2})~T^{\dagger}_{3} & = -\frac{w^{2}\abs{\bmk}}{\hbar v_{F}\abs{\bmq_{3}}^2}
		\pmqty{e^{i\left(\theta_{\bmk}-2\theta_{\bmq_{3}}-\frac{\theta}{2} - \phi\right)} + \text{c.c.} & e^{i\left(\theta_{\bmk}-2\theta_{\bmq_{3}}-\frac{\theta}{2} + \phi\right)} + e^{-i\left(\theta_{\bmk}-2\theta_{\bmq_{3}}-\frac{\theta}{2}\right)} \\
			e^{i\left(\theta_{\bmk}-2\theta_{\bmq_{3}}-\frac{\theta}{2}\right)} + e^{-i\left(\theta_{\bmk}-2\theta_{\bmq_{3}}-\frac{\theta}{2} + \phi\right)} &
			e^{i\left(\theta_{\bmk}-2\theta_{\bmq_{3}}-\frac{\theta}{2} + 2\phi\right)} + \text{c.c.}}
	\end{align}
	Since, $ \abs{\bmq_{1}} = \abs{\bmq_{2}} = \abs{\bmq_{3}} = k_{\theta} $, therefore, the first of the two diagonal terms gives
	\begin{equation}
		-\frac{w^{2} \abs{\bmk}}{\hbar v_{F} k^{2}_{\theta}}
		\left[ 
		e^{i\left(\theta_{\bmk}-\frac{\theta}{2}\right)}
		\left(e^{i\left(-2\theta_{\bmq_{1}}\right)}
		+ e^{i\left(-2\theta_{\bmq_{2}} + \phi\right)}
		+ e^{i\left(-2\theta_{\bmq_{3}} - \phi\right)}\right) + \text{c.c.}\right]
	\end{equation}
	Since $ \phi  = 2\pi/3 $ and $ \theta_{\bmq_{1}} = -\pi/2 $, $ \theta_{\bmq_{2}} = \pi/6 $ and $ \theta_{\bmq_{3}} = 5\pi/6 $, therefore,
	\begin{equation}
		e^{i\left(-2\theta_{\bmq_{1}}\right)}
		+ e^{i\left(-2\theta_{\bmq_{2}} + \phi\right)}
		+ e^{i\left(-2\theta_{\bmq_{3}} - \phi\right)}
		= e^{i\pi} + e^{i\pi/3} + e^{-i7\pi/3} = -1 + \left(\frac{1}{2} + i\frac{\sqrt{3}}{2}\right) + \left(\frac{1}{2} - i\frac{\sqrt{3}}{2}\right) = 0
	\end{equation}
	and similarly, the second diagonal term becomes
	\begin{equation}
		-\frac{w^{2} \abs{\bmk}}{\hbar v_{F} k^{2}_{\theta}}
		\left[ 
		e^{i\left(\theta_{\bmk}-\frac{\theta}{2}\right)}
		\left(e^{i\left(-2\theta_{\bmq_{1}}\right)}
		+ e^{i\left(-2\theta_{\bmq_{2}} - 2\phi\right)}
		+ e^{i\left(-2\theta_{\bmq_{3}} + 2\phi\right)}\right) + \text{c.c.}\right]
	\end{equation}
	and
	\begin{equation}
		e^{i\left(-2\theta_{\bmq_{1}}\right)}
		+ e^{i\left(-2\theta_{\bmq_{2}} - 2\phi\right)}
		+ e^{i\left(-2\theta_{\bmq_{3}} + 2\phi\right)}
		= e^{i\pi} + e^{-i5\pi/3} + e^{-i\pi/3} = -1 + \left(\frac{1}{2} + i\frac{\sqrt{3}}{2}\right) + \left(\frac{1}{2} - i\frac{\sqrt{3}}{2}\right) = 0
	\end{equation}
	hence both of the diagonal terms vanishes. The only thing remain is to shows the off-diagonal terms, the first off-diagonal term is 
	\begin{equation}
		\left[ 
		e^{i\left(\theta_{\bmk}-\frac{\theta}{2}\right)}
		\left(e^{i\left(-2\theta_{\bmq_{1}}\right)}
		+ e^{i\left(-2\theta_{\bmq_{2}} - \phi\right)}
		+ e^{i\left(-2\theta_{\bmq_{3}} + \phi\right)}\right) + 
		e^{-i\left(\theta_{\bmk}-\frac{\theta}{2}\right)}
		e^{2i\left(\theta_{\bmq_{1}}+\theta_{\bmq_{2}}+\theta_{\bmq_{3}}\right)}\right] = -3\,e^{i\left(\theta_{\bmk}-\frac{\theta}{2}\right)} - e^{-i\left(\theta_{\bmk}-\frac{\theta}{2}\right)}
	\end{equation}
	Similarly, the second off-diagonal term becomes
	\begin{equation}
		\left[ 
		e^{i\left(\theta_{\bmk}-\frac{\theta}{2}\right)}
		e^{-2i\left(\theta_{\bmq_{1}}+\theta_{\bmq_{2}}+\theta_{\bmq_{3}}\right)}
		+ 
		e^{-i\left(\theta_{\bmk}-\frac{\theta}{2}\right)}
		\left(e^{i\left(2\theta_{\bmq_{1}}\right)}
		+ e^{i\left(2\theta_{\bmq_{2}} + \phi\right)}
		+ e^{i\left(2\theta_{\bmq_{3}} - \phi\right)}\right)
		\right] = -e^{i\left(\theta_{\bmk}-\frac{\theta}{2}\right)}-3\,e^{-i\left(\theta_{\bmk}-\frac{\theta}{2}\right)}
	\end{equation}
	Finally, the first term in (A5) becomes
	\begin{equation}
		\sum_{\alpha=1}^{3} T_{\alpha}~h^{-1}_{\bmq_{\alpha}}(-\frac{\theta}{2})~h_{\bmk}(-\frac{\theta}{2})~h^{-1}_{\bmq_{\alpha}}(-\frac{\theta}{2})~T^{\dagger}_{\alpha}
		= \frac{w^{2} \abs{\bmk}}{\hbar v_{F} k^{2}_{\theta}}
		\pmqty{0 & 3\,e^{i\left(\theta_{\bmk}-\frac{\theta}{2}\right)} + e^{-i\left(\theta_{\bmk}-\frac{\theta}{2}\right)} \\
			e^{i\left(\theta_{\bmk}-\frac{\theta}{2}\right)}+3\,e^{-i\left(\theta_{\bmk}-\frac{\theta}{2}\right)} & 0}
	\end{equation}
	In the similar way, the second term in (A5) becomes
	\begin{equation}
		\sum_{\alpha=1}^{3} T_{\alpha}~h^{-1}_{\bmq_{\alpha}}(-\frac{\theta}{2})~T^{\dagger}_{\alpha}
		= w^{2} \hbar v_{F} k_{\theta}
		\pmqty{0 & 2\sin(\theta/2) \\ 2\sin(\theta/2) & 0}
	\end{equation}
\end{widetext}

\section{The alternative form of Hamiltonian in BM model}
\label{appendix:b}
Here, we write an alternative form of TBLG Hamiltonian in the BM paper. If the top layer-1 is rotated anti-clockwise by an angle $ \theta/2 $ and the bottom layer-2 is rotated clockwise by the same angle $ \theta/2 $ such that the relative rotational misalignment between them is $ \theta $, the block Hamiltonian can also be described as
\begin{equation}
	\mathcal{H}(\bmr) =
	\begin{pmatrix}
		H_{11} & H_{12} \\
		H_{21} & H_{22}
	\end{pmatrix}
	\label{eqn:rsham1}
\end{equation}
where, the form of the each term in (\ref{eqn:rsham1}) is
\bse
\begin{align}
	H_{11} & = \hbar v_{F} \left(-i\bm{\nabla}-\bm{K}_{1}\right)\vdot \bm{\sigma}_{\theta/2}, \\
	H_{22} & = \hbar v_{F} \left(-i\bm{\nabla}-\bm{K}_{2}\right)\vdot \bm{\sigma}_{-\theta/2},\qq{and}\\
	H_{12} & = H^{\dagger}_{21} = T_{0}(\bmr)\,\sigma_{0} + T_{AB}(\bmr)\,\sigma_{+} + T_{BA}(\bmr)\,\sigma_{-}
\end{align}
\ese
where $ \bm{K}_{1} $ and $ \bm{K}_{2} $ are the rotated wave-vectors of the Dirac point in layer-1 and layer-2 respectively. $ \sigma_{0} $ is a $ 2\times 2 $ identity matrix and $ \sigma_{+} = \frac{1}{2}\left(\sigma_{x} + i\sigma_{y}\right) = \spmqty{0 & 1 \\ 0 & 0} $ and $ \sigma_{-} = \frac{1}{2}\left(\sigma_{x} - i\sigma_{y}\right) = \spmqty{0 & 0 \\ 1 & 0} $.

The Fig.(\ref{fig:TBLG1}(a)) shows the real-space moir\'e pattern emerging as a result of the relative twist among the layers. The resulting moir\'e pattern consists of three different local regions AA or BB, AB and BA of different sizes. The tunnelling, in the AA or BB regions, can be expressed as
\begin{multline}
	T_{0}(\bmr) \approx \sum_{n_1,n_2} \delta\left(\bmr- n_1\,\bm{a}^{M}_{1} - n_2\,\bm{a}^{M}_{2} \right) \\
	= \frac{1}{A_{M}}\sum_{m_1,m_2} e^{-i\left(m_1\,\bm{b}^{M}_1 + m_2\,\bm{b}^{M}_2\right)\cdot\bmr}
\end{multline}
where $ n_1\bm{a}^{M}_1 + n_2\,\bm{a}^{M}_2 $ is an arbitrary moir\'e direct-space translation vector with $ \bm{a}^{M}_1 $ and $ \bm{a}^{M}_2 $ are shown in Fig.(\ref{fig:TBLG1}(a)), and $ A_{M} = \abs{\bm{a}^{M}_1 \times \bm{a}^{M}_2} $ is the area of the moire unit cell. The moir\'e reciprocal space translation vectors $ \bm{b}^{M} = m_1\bm{b}^{M}_1 + m_2\bm{b}^{M}_2 $ are defined such that $ \bm{a}^{M}_{i}\vdot\bm{b}^{M}_{j} = 2\pi\delta_{ij} $. Similarly, the interlayer tunneling in the AB and BA regions are expressed as
\bse
\begin{align}
	T_{AB}(\bmr) & \approx \sum_{n_1, n_2} \delta\left(\bmr- n_1\,\bm{a}^{M}_{1} - n_2\,\bm{a}^{M}_2 - \bmr_{A}\right) \nn \\
	& = \frac{1}{A_{M}}\sum_{m} e^{-i\bm{b}^{M}\vdot\bmr_{A}} e^{-i\bm{b}^{M}\vdot\bmr}
\end{align}
\begin{align}
	T_{BA}(\bmr) & \approx \sum_{n_1, n_2} \delta\left(\bmr- n_1\,\bm{a}^{M}_{1} - n_2\,\bm{a}^{M}_2 - \bmr_{B}\right) \nn \\
	& = \frac{1}{A_{M}}\sum_{m}	e^{-i\bm{b}^{M}\vdot\bmr_{B}} e^{-i\bm{b}^{M}\vdot\bmr}
\end{align}
\ese
where the vectors $ \bm{r}_{A} = (\bm{a}^{M}_{1} + \bm{a}^{M}_2)/3 $ and $ \bm{r}_B = (2\bm{a}^{M}_{1} - \bm{a}^{M}_2)/3 $ are the locations of AB and BA-regions within a moir\'e unit cell, respectively, also shown in Fig.\ref{fig:TBLG1}(a).

To shift the origin in the top and bottom layer to $ \bm{K}_{1} $ and $ \bm{K}_{2} $, respectively, in the Hamiltonian in (\ref{eqn:rsham1}), it is transformed using the unitary matrix $ W $, such that the transformed Hamiltonian is
\begin{equation}
	H' = W^{\dagger}\,H\,W = \pmqty{e^{-i\textbf{K}_{1}\cdot\textbf{r}}\,H_ {11}\,e^{i\textbf{K}_{1}\vdot\textbf{r}}
		& e^{-i\textbf{K}_{1}\vdot\textbf{r}}\,H_ {12}\,e^{i\textbf{K}_{2}\vdot\textbf{r}} \\
		e^{-i\textbf{K}_{2}\cdot\textbf{r}}\,H_ {21}\,e^{i\textbf{K}_{1}\vdot\textbf{r}} & e^{-i\textbf{K}_{2}\cdot\textbf{r}}\,H_ {22}\,e^{i\textbf{K}_{2}\vdot\textbf{r}}}
\end{equation}
where the matrix $ W $ is given by
\begin{equation}
	W = \Pmqty{\mathcal{I}_2\,e^{i\bm{K}_{1}\cdot\bmr} & \mathcal{O}_2 \\
		\mathcal{O}_2 & \mathcal{I}_2\,e^{i\bm{K}_{2}\cdot\bmr}}
\end{equation}
the diagonal elements of $ H' $ are
\begin{align}
	e^{-i\bm{K}_{1}\cdot\bmr}~H_{11} e^{i\bm{K}_{1}\cdot\bmr}
	= -i\hbar v_{F} \bm{\nabla}\cdot\bm{\sigma}_{\theta/2} \\
	e^{-i\bm{K}_{2}\cdot\bm{r}}\,H_ {BB}\,e^{i\bm{K}_{2}\cdot\bm{r}}
	= -i\hbar v_{F} \bm{\nabla}\cdot\bm{\sigma}_{-\theta/2}
\end{align}
and the off-diagonal matrix elements are
\begin{equation}
	H'_{12} = e^{-i\bm{K}_{1}\cdot\bm{r}}\,\left[T_ 0(\bm{r})\,\sigma_0	+ T_{AB}(\bm{r})\,\sigma_{+} + T_{BA}(\bm{r})\,\sigma_{-}\right] e^{i\bm{K}_{2}\cdot\bm{r}}
	\label{eqn:transTME}
\end{equation}
where the last step is written by explicit calculation of the transformation of the T-matrices as shown below:
\begin{align}
	e^{-i\bm{K}_{1}\cdot\bm{r}}~T_{0}(\bm{r})~& e^{i\bm{K}_{2}\cdot\bm{r}}
	= \frac{1}{A_M}\sum_{\bm{b}^{M}} \tilde{T}_{0}\,(\bm{b}^{M}) e^{-i\left(\bm{K}_{1}-\bm{K}_{2}+\bm{b}^{M}\right)\cdot\bm{r}} \nn \\
	& = w_ 0\left(e^{-i\bmq_{1}\vdot\bm{r}} +
	e^{-i\bmq_{2}\cdot\textbf{r}} + 
	e^{-i\bmq_{3}\cdot\textbf{r}}\right)
\end{align}
In the last expression, only the Fourier components with moir\'e reciprocal lattice vectors $ 0,-\bm{b}^{M}_1,-\bm{b}^{M}_1-\bm{b}^{M}_2 $ significant as explained in BM model. Similarly the other two terms in the Eq.(\ref{eqn:transTME}),
\begin{align}
	e^{-i\bm{K}_{1}\cdot\bm{r}}& ~T_{AB}(\bm{r})~ e^{i\bm{K}_{2}\cdot\bm{r}} \nn \\
	& = \frac{1}{A_M}\sum_{\bm{b}^{M}} \tilde{T}_{AB}\,(\bm{b}^{M})\,e^{-i\bm{b}^{M}\cdot\bmr_{A}} e^{-i\left(\bm{K}_{1}-\bm{K}_{2}+\bm{b}^{M}\right)\cdot\bm{r}} \nn \\
	& = w_1\left(e^{-i\bmq_{1}\cdot\bm{r}} +
	e^{-i\bm{b}^{M}_1\cdot\bmr_A}e^{-i\bmq_{1}\vdot\textbf{r}} + 
	e^{-i\bm{b}^{M}_2\cdot\bmr_A}e^{-i\bmq_{3}\vdot\textbf{r}}\right)
\end{align}
and,
\begin{align}
	e^{-i\bm{K}_{1}\vdot\bm{r}}&~T_{BA}(\bm{r})~e^{i\bm{K}_{2}\vdot\bm{r}} \nn \\
	& = \frac{1}{A_M}\sum_{\bm{b}^{M}} \tilde{T}_{BA}\,(\bm{b}^{M})\,e^{-i\bm{b}^{M}\cdot\bmr_{B}} e^{-i\left(\bm{K}_{1}-\bm{K}_{2}+\bm{b}^{M}\right)\cdot\bm{r}} \nn \\
	& = w_1\left(e^{-i\bmq_{1}\cdot\bm{r}} +
	e^{-i\bm{b}^{M}_1\cdot\bmr_B}e^{-i\bmq_{2}\vdot\textbf{r}} + 
	e^{-i\bm{b}^{M}_2\cdot\bmr_B}e^{-i\bmq_{3}\vdot\textbf{r}}\right)
\end{align}
Since, the scalar products $ \bm{b}^{M}_{1}\vdot\bm{r}_{A} = \bm{b}^{M}_{2}\vdot\bm{r}_{A} = 2\pi/3 $ and $ \bm{b}^{M}_{1}\vdot\bm{r}_{B} = 4\pi/3 $ and $ \bm{b}^{M}_{2}\vdot\bm{r}_{B} = -2\pi/3 $, therefore, the transformed Hamiltonian describing the TBLG in a single valley is given as
\begin{equation}
	H'(\bmr) =
	\Pmqty{-i\hbar v_{F} \bm{\nabla}\cdot\bm{\sigma}_{\theta/2} & T(\bmr) \\ T^{\dagger}(\bmr) & -i\hbar v_{F} \bm{\nabla}\cdot\bm{\sigma}_{-\theta/2}}
	\label{eqn:rsham2}
\end{equation}
where the off-diagonal matrices $ T(\bmr) $ is given as
\begin{equation}
	T(\bmr) = \sum_{n=1}^{3} T_{n} e^{-i\bmq_{n}\vdot\bmr}
\end{equation}
where
\begin{equation}
	T_{n} = w_{0}\sigma_{0} + w_1\left[\sigma_{x} \cos \left(n-1\right)\phi + \sigma_{y} \sin \left(n-1\right)\phi\right]
\end{equation}
where $ \phi = 2\pi/3 $. If the Hamiltonian in Eq.(\ref{eqn:rsham1}) acts on the wavefunction given by
\begin{equation}
	\psi_{\bmk}(\bmr) =
	\Pmqty{u^{(1)}_{\bmk}(\bmr)\,e^{i\bmk\cdot\bmr} \\ u^{(2)}_{\bmk}(\bmr)\,e^{i\bmk\cdot\bmr}}
	\label{eqn:rswfun1}
\end{equation}
where $ u^{(1)}_{\bmk}(\bmr) $ and $ u^{(2)}_{\bmk}(\bmr) $ both are the two-component spinors corresponding to layer-1 and layer-2 respectively, then the transformed wavefunction which is acted upon by the Hamiltonian $ H'(\bmr) $ in Eq.(\ref{eqn:rsham2}) is
\begin{equation}
	\psi'_{\bmk}(\bmr) = W^{\dagger}\psi(\bmr)
	= \Pmqty{u^{(1)}_{\bmk}(\bmr)\,e^{i\left(\bmk-\bm{K}_{1}\right)\cdot\bmr} \\ u^{(2)}_{\bmk}(\bmr)\,e^{i\left(\bmk-\bm{K}_{1} + \bmq_{1}\right)\cdot\bmr}}
	\label{eqn:rswf}
\end{equation}

In order to obtain the $ 8\times 8 $ k-space Hamiltonian, we use (\ref{eqn:rsham2}) and (\ref{eqn:rswf}) in the Schr\"{o}dinger equation $ \left(H'(\bmr)-E\right)\psi'_{\bmk}(\bmr) = 0 $, gives
\begin{widetext}
	\bse
	\begin{align}
		\left(-iv_0 \bm{\nabla}\cdot\bm{\sigma}_{\theta/2}-E\right)u^{(1)}_{\bmk}(\bmr)e^{i\left(\bmk-\bm{K}_{1}\right)\cdot\bmr} + T(\bmr)\,u^{(2)}_{\bmk}(\bmr)\,e^{i\bm{q}_1\cdot\bmr}\,e^{i\left(\bmk-\bm{K}_{1}\right)\cdot\bmr} = 0, \label{eqn:bseqn1a}\\
		T^{\dagger}(\bmr)\,u^{(1)}_{\bmk}(\bmr)\,e^{i\left(\bmk-\bm{K}_{1}\right)\cdot\bmr} + \left(-iv_0 \bm{\nabla}\cdot\bm{\sigma}_{-\theta/2}-E\right)u^{(2)}_{\bmk}(\bmr)\,e^{i\bm{q}_1\cdot\bmr}\,e^{i\left(\bmk-\bm{K}_{1}\right)\cdot\bmr} = 0
		\label{eqn:bseqn1b}
	\end{align}
	\ese
\end{widetext}
since, $ u^{(1)}_{\bmk}(\bmr) $ and $ u^{(2)}_{\bmk}(\bmr) $, both are periodic functions with the periodicity of the moir\'e reciprocal space, and therefore one can expand them as
\begin{align}
	u^{(1)}_{\bmk}(\bmr) &= \sum_{\bm{b}_{M}} C^{(1)}_{\bm{b}^{M}}(\bmk)\, e^{i\bm{b}^{M}\cdot\bmr} \\
	u^{(2)}_{\bmk}(\bmr)\,e^{i\bm{q}_1\cdot\bmr} & = \sum_{\bm{b}^{'M}} C^{(2)}_{\bm{b}^{'M}+\bm{q}_1}(\bmk)\, e^{i\left(\bm{b}^{'M}+\bm{q}_1\right)\cdot\bmr} 
\end{align}
where $ C^{(1)}_{\bm{b}^{M}}(\bmk) $ and $ C^{(2)}_{\bm{b}^{'M}+\bm{q}_1}(\bmk) $ are the expansion coefficients as a function of the wave vector $ \bmk $. For given $\bm{b}^{M}$, $ \bm{b}^{'M} $ and $ j $, therefore, one can write the Eq.(\ref{eqn:bseqn1a}) and Eq.(\ref{eqn:bseqn1b}) as
\begin{widetext}
	\bse
	\begin{align}
		e^{i\left(\bmk-\bm{K}_{1}\right)\cdot\bmr} & 
		\sum_{\bm{b}^{M}}\left\{\left[v_0 \left(\bmk-\bm{K}_{1}+\bm{b}^{M}\right)\cdot\bm{\sigma}_{\theta/2}-E\right]C^{(1)}_{\bm{b}^{M}}(\bmk) +
		\sum_{j=1}^{3} T_{j} \, C^{(2)}_{\bm{b}^{M}+\bm{q}_{j}}(\bmk)\right\} e^{i\bm{b}^{M}\cdot\bmr} = 0,\label{eqn:bseqna}\\
		e^{i\left(\bmk - \bm{K}_{1} + \bm{q}_1\right)\cdot\bmr} & \sum_{\bm{b}'^{M}}\left\{\sum_{j=1}^{3} T^{\dagger}_{j}\,C^{(1)}_{\bm{b}'^{M}+\bm{q}_{1}-\bm{q}_{j}}(\bmk)
		+ \left[v_0\left(\bmk - \bm{K}_{1}+\bm{b}'^{M} + \bm{q}_1\right)
		\cdot\bm{\sigma}_{-\theta/2}-E\right] C^{(2)}_{\bm{b}'^{M}+\bm{q}_1}(\bmk)\right\}
		e^{i\bm{b}'^{M}\cdot\bmr} = 0 \label{eqn:bseqnb}
	\end{align}
	\ese
\end{widetext}
To obtain the minimal set of linear equations, one restricts the expansion coefficients in such a way that only $ C^{(1)}_{\bm{0}} $, $C^{(2)}_{\bm{q}_1}  $, $C^{(2)}_{\bm{q}_2}  $ and $C^{(2)}_{\bm{q}_3}  $ are considered to survive. Therefore, we consider first the Eq.(\ref{eqn:bseqna}) for $ \bm{b}^{M} = \bm{0} $,  which takes the form
\begin{widetext}
	\begin{equation}
		\left[v_0 \left(\bmk-\bm{K}_{1}\right)\cdot\bm{\sigma}_{\theta/2}-E\right]C^{(1)}_{\bm{0}}(\bmk) +	T_{1}~C^{(2)}_{\bm{q}_1}(\bmk) + T_{2}~C^{(2)}_{\bm{q}_{2}}(\bmk)
		+ T_{3}~C^{(2)}_{\bm{q}_{3}}(\bmk) = 0
	\end{equation}
	and then, the Eq.(\ref{eqn:bseqnb}) for $ \bm{b}'^{M} = \bm{0}, \bm{b}^{M}_{1} $ and $ \bm{b}^{M}_{2} $, which gives
	\bse
	\begin{align}
		T^{\dagger}_{1}\,C^{(1)}_{\bm{0}}(\bmk) +
		T^{\dagger}_{2}\,C^{(1)}_{-\bm{b}^{M}_{1}}(\bmk) +
		T^{\dagger}_{3}\,C^{(1)}_{-\bm{b}^{M}_{2}}(\bmk) +
		\left[v_0\left(\bmk - \bm{K}_{1}+\bm{q}_1 \right)\cdot\bm{\sigma}_{-\theta/2}-E\right] C^{(2)}_{\bm{q}_1}(\bmk) = 0 \\
		T^{\dagger}_{1}\,C^{(1)}_{\bm{b}^{M}_{1}}(\bmk) +
		T^{\dagger}_{2}\,C^{(1)}_{\bm{0}}(\bmk) +
		T^{\dagger}_{3}\,C^{(1)}_{\bm{b}^{M}_{1}-\bm{b}^{M}_{2}}(\bmk) +
		\left[v_0\left(\bmk - \bm{K}_{1}+\bm{q}_2 \right)\cdot\bm{\sigma}_{-\theta/2}-E\right] C^{(2)}_{\bm{q}_{2}}(\bmk) = 0 \\
		T^{\dagger}_{1}\,C^{(1)}_{\bm{b}^{M}_{2}}(\bmk) +
		T^{\dagger}_{2}\,C^{(1)}_{\bm{b}^{M}_{2}-\bm{b}^{M}_{1}}(\bmk) +
		T^{\dagger}_{3}\,C^{(1)}_{\bm{0}}(\bmk) +
		\left[v_0\left(\bmk - \bm{K}_{1}+\bm{q}_3 \right)\cdot\bm{\sigma}_{-\theta/2}-E\right] C^{(2)}_{\bm{q}_{3}}(\bmk) = 0
	\end{align}
	\ese
	Writing the above in the matrix form gives
	\begin{equation}
		\Pmqty{h_{\bmk}(\theta/2)-E & T_{1} & T_{2} & T_{3}\\
			T^{\dagger}_{1} & 	h_{\bmk+\bmq_1}(-\theta/2)-E & \mathcal{O} & \mathcal{O}\\
			T^{\dagger}_{2} & \mathcal{O} & h_{\bmk+\bmq_2}(-\theta/2)-E & \mathcal{O}\\
			T^{\dagger}_{3} & \mathcal{O} & \mathcal{O} & h_{\bmk+\bmq_3}(-\theta/2)-E}
		\Pmqty{C^{(1)}_{\bm{0}}(\bmk) \\ C^{(2)}_{\bm{q}_1}(\bmk) \\ C^{(2)}_{\bm{q}_{2}}(\bmk) \\ C^{(2)}_{\bm{q}_{3}}(\bmk)} = 0
		\label{eqn:8ham}
	\end{equation}
	where $ h_{\bmk}(\theta/2) = v_0\left(\bmk-\bm{K}_{1}\right)\cdot\bm{\sigma}_{\theta/2} $. After finding the expansion coefficients (arranged in the column vector in Eq.(\ref{eqn:8ham})), one can write the real space wavefunction as
	\begin{equation}
		\psi'_{\bmk}(\bmr) = \Pmqty{C^{(1)}_{\bm{0}}(\bmk)\\
			\left[C^{(2)}_{\bm{q}_1}(\bmk)\,e^{i\bm{q}_{1}\cdot\bmr} + C^{(2)}_{\bm{q}_{2}}(\bmk)\,e^{i\bm{q}_{2}\cdot\bmr} + C^{(2)}_{\bm{q}_{3}}(\bmk)\,e^{i\bm{q}_{3}\cdot\bmr}\right]}e^{i\left(\bmk-\bm{K}_{1}\right)\cdot\bmr}
		\label{eqn:rswfun2}
	\end{equation}
\end{widetext}

\bibliography{reference}

\end{document}